# Defect engineering over anisotropic brookite towards substrate-specific photo-oxidation of alcohols

S. M. Hossein Hejazi[1], Mahdi Shahrezaei[1], Piotr Błoński[1], Mattia Allieta[2], Polina M. Sheverdyaeva[3], Paolo Moras[3], Zdeněk Baďura[1], Sergii Kalytchuk[1], Elmira Mohammadi[1], Radek Zbořil[1,4], Štěpán Kment[1,4], Michal Otyepka[1,5], Alberto Naldoni[1]*, Paolo Fornasiero[6,7]*

[1]Czech Advanced Technology and Research Institute, Regional Centre of Advanced Technologies and Materials, Palacký University Olomouc, Křížkovského 511/8, 77900 Olomouc, Czech Republic

[2]Ronin Institute Montclair, NJ 07043 USA

[3]Istituto di Struttura della Materia-CNR (ISM-CNR), SS 14, Km 163,5, I-34149, Trieste, Italy

[4]Nanotechnology Centre, Centre of Energy and Environmental Technologies, VŠB– Technical University of Ostrava, 17. listopadu 2172/15, 70800 Ostrava-Poruba, Czech Republic

[5]IT4Innovations, VSB – Technical University of Ostrava, 17. listopadu 2172/15, 708 00 Ostrava-Poruba, Czech Republic

[6]Department of Chemical and Pharmaceutical Sciences, ICCOM-CNR Trieste Research Unit, INSTM-Trieste, University of Trieste, Via L. Giorgieri 1, 34127 Trieste, Italy

[7]Center for Energy, Environment and Transport Giacomo Ciamician - University of Trieste, Italy

*Corresponding authors: alberto.naldoni@upol.cz; pfornasiero@units.it

**Summary**

Generally adopted design strategies for enhancing the photocatalytic activity are aimed at tuning properties such as the visible light response, the exposed crystal facets, and the nanocrystal shape. Here, we present a different approach for designing efficient photocatalysts displaying a substrate-specific reactivity upon defect engineering. The defective anisotropic brookite $TiO_2$ photocatalyst functionalized with Pt nanocrystals are tested for alcohol photoreforming showing up to an 11-fold increase in methanol oxidation rate, compared to the unreduced one, whilst presenting much lower ethanol or isopropanol specific oxidation rates. We demonstrate that the alcohol oxidation and hydrogen evolution reactions are tightly related, and when the substrate-specific alcohol oxidation ability is increased, the hydrogen evolution is significantly boosted. The reduced anisotropic brookite shows up to twenty-six-fold higher specific photoactivity with respect to anatase and brookite with isotropic nanocrystals, reflecting the different type of defective catalytic sites formed depending on the $TiO_2$ polymorph and its crystal shape. Advanced *in-situ* characterizations and theoretical investigations reveal that controlled engineering over oxygen vacancies and lattice strain produces large electron polarons hosting the substrate-specific active sites for alcohol photo-oxidation.

Keywords: selective photocatalysis, oxygen vacancies, DFT calculations, photoreforming, black $TiO_2$

# INTRODUCTION

The visionary idea of a world powered by solar light proposed by G. Ciamician more than a century ago[1] has become a reality, proving that complex organic synthesis[2,3] and production of solar fuels like hydrogen[4–6] and ammonia[7,8] can be performed more and more efficiently. However, these intrinsically sustainable processes, before becoming industrially competitive with existing polluting technologies, need further material design, fine tuning of light absorption properties, charge carriers management, and surface engineering.[9] Over the past decades, photocatalysis for direct conversion of solar energy into molecular fuels has been focused on designing efficient photocatalysts by improving their fundamental properties. The visible light photoactivity can be enhanced by engineering heterojunction, introducing lattice defects into wide bandgap materials like $TiO_2$,[10–12] or choosing semiconductors having small bandgap energy (e.g. $Cu_2O$, $ZnIn_2S_4$, and i.e. $C_3N_4$) and suitable bands position straddling the molecular redox levels of the investigated chemical reaction.[2,13–15] The use of inorganic nanocrystals with well-defined morphology, determined crystal facets, or dimensional anisotropy have been also demonstrated to be beneficial for the charge carrier separation.[16–18] The kinetic competition between charge recombination and surface catalysis is often overcome by the addition of proper metallic co-catalysts.[19] This playground has stimulated the exploration of countless options to prepare, mix, and engineer semiconductor photocatalysts with enhanced opto-electronic properties with the aim of more efficiently driving benchmark photocatalytic reactions such as hydrogen evolution from water splitting and photoreforming of biomass. However, the majority of these studies involve monitoring the products of the reductive cycle, i.e. the evolved hydrogen, while not analyzing the oxidation products.[20–22] When sacrificial biomass substrates are employed, i.e. alcohol photoreforming, analyzing the oxidation pathway and reactivity becomes especially important not only because they provide a kinetic gain, and therefore an improved hydrogen evolution compared to the case of water oxidation, but also because the oxidized sacrificial agents often participate in hydrogen evolution, thus directly regulating its kinetics.[20,23] Furthermore, controlling the oxidation process of sacrificial biomass is particularly relevant since it can lead to its partial oxidation and the synthesis of added value products such as 2,5-furandicarboxylic acid (bioderived polymer that may substitute PET) and diesel fuel precursors.[3,24,25]

We developed an approach to designing anisotropic defective brookite $TiO_2$ nanocrystals that upon high temperature reduction treatment expose precise defect site with substrate-specific photo-reactivity for methanol oxidation, compared to higher alcohols such as ethanol and isopropanol (Fig. 1A). We show that the reaction rates for the photocatalytic alcohol oxidation and the parallel hydrogen evolution reaction are tightly connected and that optimization of the methanol oxidation leads to an increased production of hydrogen. Although the introduction of point defects and structural deformation results in enhanced visible light absorption and reduced charge carrier lifetime, we show that the selective affinity towards methanol oxidation of reduced brookite is the main parameter enabling a higher apparent quantum yield (AQY) for hydrogen evolution. Using a set of *in-situ* characterization aided by DFT calculations, we demonstrate that the substrate-specific activity is regulated by catalytic sites including sub-surface oxygen vacancies within a locally strained lattice environment that generate shallow hole traps responsible for boosting the first steps of methanol photo-oxidation. These photo-oxidation sites are crucial for the enhanced substrate-specific photoreforming activity of reduced brookite, and they form preferentially within anisotropic nanocrystals, which show twenty-six times higher specific hydrogen evolution rate, compared to isotropic ones, where defect sites with different energy are formed.

**RESULTS AND DISCUSSION**

**Synthesis and characterization of brookite nanorods**

To engineer the photocatalytic sites for alcohol photoreforming at the atomic level, we selected brookite $TiO_2$ nanorods—a promising and still poorly investigated $TiO_2$ polymorph—as a model material and grew anisotropic nanostructures exposing the (210) surface on the lateral facets (Fig. 1B) by hydrothermal synthesis (see Supporting Information).[26] We prepared various brookite photocatalysts reduced under a $H_2$ stream at different temperatures, along with reduced anatase and commercial brookite reference samples (see Table S1). Elemental analysis revealed that the obtained nanopowders did not contain significant quantity of non-metals coming from C, H, or N incorporation (Table S2). Reduction of a pristine brookite $TiO_2$ under pure hydrogen stream at 700°C for 1h created defective nanocrystals showing remarkable changes in their structural and electronic properties alongside giving the best photocatalytic performance. This morphology evolved from anisotropic nanostructures with well-defined shape and exposed crystal facets (Fig. 1B and S1A) into more isotropic nanoparticles that displayed irregular shape and aggregation through twin boundaries formation (Fig. 1C and Fig. S1B). Notably, anatase and brookite with isotropic crystal shapes (Fig. S2-S3) did not reveal any crystal reshaping upon high temperature reduction treatment. However, they presented different color variations (Table S1), compared to those observed for the anisotropic brookite, upon increasing the temperature of the hydrogen treatment, thus suggesting a different reducibility behavior with respect to the $TiO_2$ polymorph and crystal shape, as confirmed by UV-vis reflectance spectroscopy, Raman spectroscopy, and photolumiscence spectroscopy mapping (see below). In order to prepare highly active photocatalysts for alcohol photoreforming, we functionalized the samples by photodepositing Pt co-catalyst nanoparticles on their surface. Inductively coupled plasma mass spectrometry (ICP-MS) analysis detected similar Pt loading on both pristine (0.98 wt%) and reduced brookite samples (0.90 wt%). HRTEM and STEM-HAADF micrographs as well as the elemental mapping showed that Pt nanoparticles with an average diameter of 2.5 nm were homogeneously deposited on the pristine brookite nanorods (Fig. 1D and Fig. S4A, S5, S6). Surprisingly, in the case of reduced brookite nanocrystals, we detected the presence of small Pt nanoparticles with similar sizes as well as larger Pt nanoparticle aggregates reaching 10–30 nm in size (Fig. 1E and Fig. S7–S10). This result was confirmed by a HRTEM analysis of three different brookite batches. The larger Pt conglomerates may be less reactive than the smaller ones, thus negatively affecting the photocatalytic activity of the reduced brookite. Moreover, larger metal aggregates may screen the incoming light during photocatalysis, decreasing the light harvesting efficiency. However, as shown in the next section, in the present investigation these parameters did not reduced neither one aspect nor the other for reduced brookite, suggesting that for the considered reaction the defects in $TiO_2$ played a more crucial role than that of Pt nanoparticles. The brookite treatment under hydrogen was accompanied by a decrease in the BET (Brunauer–Emmett–Teller) specific surface area from 67 to 47 $m^2$ $g^{-1}$ after reduction (Fig. S11 and Table S3).

The temperature (reduction)-dependent structural parameters extracted by the Rietveld refinement of X-ray diffraction (XRD) patterns were in agreement with the described morphological evolution (Fig. S12-S18 and Table S3–S6). Notably, the XRD analysis highlighted that the reduction treatment introduced an anisotropic and preferential deformation of the brookite lattice along the c-axis due to the creation of oxygen vacancies (Fig. S19). Their presence was further supported by the blueshift in the main $A_{1g}$ vibrational mode detected by Raman spectroscopy measurements (Fig. S20 and discussion in the Supporting Information) after reduction, which again pointed out to a different reducing behavior dependent on the $TiO_2$ polymorph and shape (Raman shift is 1.3 $cm^{-1}$ for the reduced anisotropic brookite, 7.6 $cm^{-1}$ for the reduced isotropic anatase, and no observed shift for the reduced isotropic brookite). Having performed the bond valence sum analysis, we observed an average depletion of ~0.5% of the Ti valence for the reduced brookite—a typical feature induced by the formation of oxygen vacancies.[11,27] The moderate decrease in the Ti valence upon a $H_2$ treatment at high temperature is an indication of a low tendency toward defects formation in brookite nanorods. This general feature was also reflected by the color change—from white to grey—that brookite underwent after reduction at 700°C, as opposed to the more reducible anatase phase that assumed a darker color already at lower temperatures (Table S1). The reduced brookite nanorods showed an optical bandgap energy of ~3.38 eV, this making no significant difference from the value retrieved for the as-synthesized sample (Fig. S21 and Table S7). The same results were observed for both the isotropic brookite and anatase samples (Fig. S22-S23 and Table S8-S9). Further analysis of the absorption spectra highlighted an increased visible light absorption and an Urbach tail that ranged for the anisotropic brookite from 69 (for the as-synthesized brookite nanorods) to 115 meV (for nanorods reduced at 700°C). For the isotropic anatase, the Urbach tail increased from 115 (for the as-synthesized anatase nanoparticles) to 205 mV (for anatase reduced at 500°C). This supports the scenario of a phase- and shape-dependent increase in the population of oxygen vacancies after the reduction treatment (see Supporting Information for further discussion).

**Alcohol photoreforming with reduced brookite**

The photocatalytic activity of the platinized brookite nanorods was tested for methanol photoreforming under a simulated AM 1.5G spectrum at one-sun intensity producing $H_2$ and oxidation products. As previously reported by others groups, the increased photocatalytic activity of reduced $TiO_2$ nanomaterials could be ascribed to the co-catalyst role in $H_2$ evolution played by oxygen vacancies.[12,28] In contrast, in order to use the oxygen vacancies as catalytic sites in the photocatalytic oxidation reaction, we photodeposited Pt nanoparticles over the optimized photocatalysts. Following this procedure, $H_2$ generation occurred on the Pt surface as the photogenerated electrons were separated and stabilized into the Pt nanoparticles by the Schottky barrier formation at the Pt–TiO$_2$ interface, while photo-oxidation occurred on the TiO$_2$ surface.[19,29]

In contrast to common practice, where only a H$_2$ production rate is detected, we designed specific experiments to follow the kinetics of methanol photo-oxidation using a solution including deuterated methanol (i.e, CD$_3$OD) and a small aliquot of methanol (i.e. CH$_3$OH), whose corresponding consumption was followed by an $^1$H-NMR analysis of the liquid phase (Fig. S24). Table S10 reports the NMR signal integration of CH$_3$OH relative to the adopted internal standard (DMSO), which highlight no significant difference between the blank measurement containing no photocatalyst and the one at time zero, i.e. after 30 min adsorption/desorption equilibrium in the dark in the presence of the photocatalysts. From this data is clear that the methanol adsorption in the dark did not affect the photocatalytic performance of the investigated photocatalysts. Fig. 1F shows the amount of methanol oxidized over 24h of illumination illustrating the remarkable oxidation activity of the reduced brookite over the pristine material. The corresponding specific methanol consumption rates, computed by using the BET surface area of each sample, for the platinized brookite nanorod samples (Fig. 1G, left) evidenced that the pristine brookite drove the photo-oxidation reaction with a specific rate of 27 µmol h$^{-1}$ m$^{-2}$, whereas for the reduced one, it was 99 µmol h$^{-1}$ m$^{-2}$. We obtained the photoactivity values by considering methanol consumption after 24 h of reaction, which resulted in a 3.7-fold enhancement in favor of the reduced brookite. Notably, if we consider kinetic data after 5 h (Fig. 1F), the reduced brookite performed methanol photo-oxidation up to 11 times faster than the pristine sample. This suggests on the one hand that, in the early stage of reaction, methanol was oxidized faster until the available surface reaction sites were fully occupied and a steady state was reached, which ensured a higher methanol consumption rate even after 24 h of reaction, as evidenced by the divergence of the reaction kinetics curves (Fig. 1F). On the other hand, this behavior can be due to a partial aggregation of the colloidal photocatalysts after several hours of irradiation (see dynamic light scattering measurements in Fig. S25), thus producing a reduced available surface for the methanol oxidation reaction to occur.

The amount of evolved hydrogen determined by gas chromatography analysis followed a linear increase with time (Fig. S26) for both the pristine and the reduced brookite, corresponding to optimized specific $H_2$ production rates of 26 and 88 µmol $h^{-1}$ $m^{-2}$, respectively, with reduced brookite that evolved $H_2$ 3.4 times faster (Fig. 1G, right). The reduced brookite showed a 13% decrease activity after 5 photocatalytic runs (Fig. S27). Notably, the activity decrease appeared almost constant after each recycling test, thus suggesting that it can be due to the loss of catalyst during the tests, which can happen during the centrifugation/re-suspension of the photocatalyst and/or can be due to the loss of material attaching onto the reactor walls. Another aspect that can produce this slight decrease in activity is the increased hydrodynamic diameter of the brookite nanocrystals in suspension, as revealed by dynamic light scattering measurements after 24 h of illumination (Fig. S25). Moreover, two more aspects may produce the observed photocatalytic activity decrease after several recycling cycles. On the one hand, the catalyst surface may be partially passivated by the presence of reaction intermediates, as we did not wash the catalyst before subsequent tests. On the other hand, a partial modification of the surface population of defects (as evidenced by the resonant PES valence band measurements on B700 after catalysis, see Fig. S43) may induce a partial reactivity change. Interestingly, the $H_2$ production rates followed a reactivity trend closely resembling the one observed for the methanol photo-oxidation activity, which suggests that hydrogen production is strictly regulated by the alcohol oxidation and it can be used as reporter figures of merit for tracking the reactivity of the pristine and the reduced brookite for alcohol photo-oxidation. Following this principle, we tested our platinized samples for the photoreforming of ethanol and isopropanol and discovered that the reactivity of the reduced brookite was markedly more pronounced and substrate-specific toward methanol in comparison with the other tested alcohols (Fig. 1G, right). In the case of ethanol photoreforming, both reduced and pristine anisotropic brookites showed a specific $H_2$ production rate of 54 µmol h$^{-1}$ m$^{-2}$, suggesting that they have a similar affinity toward its photo-oxidation. On the other hand, in the case of isopropanol photoreforming, the reduced anisotropic brookite presented a 1.7-fold higher specific $H_2$ production rate (41 µmol h$^{-1}$ m$^{-2}$) when compared to the pristine sample, denoting a higher photo-oxidation ability yet still much lower than that shown for methanol. This observation confirmed that the $H_2$ evolution activity of the reduced brookite was regulated by a substrate-specific reactivity toward alcohol photo-oxidation. Such a stark photo-reactivity toward methanol oxidation was observed only for the brookite nanorods, while platinized reference samples made by reduced spherical anatase nanocrystals or reduced isotropic brookite nanoparticles displayed ~1.8–1.9 times higher specific photocatalytic rates in comparison with the untreated materials (Fig. S28). Notably, the reduced brookite nanorods loaded with Pt showed a remarkably higher specific $H_2$ evolution rate with respect to both the reduced anatase/Pt (19 times) and the reduced isotropic brookite/Pt (26 times), as shown in Fig. S28. However, this difference is significantly reduced when considering the photocatalytic activity per optimized mass (Fig. S29), with the reduced brookite nanorods (B700/Pt) still showing more than two-times the hydrogen evolution rate observed for the reduced anatase nanocrystals (A500/Pt). These observations suggest that the type of produced defects/catalytic sites varies depending on the selected $TiO_2$ polymorph as well as on the crystal shape, which emphasizes how the exposure of different crystal facets having different interfacial energy and therefore resistance to hydrogen treatment under high temperature regulates the defects formation. This is demonstrated by the different degree of reducibility that each sample exhibited, as supported by previously discussed absorption and Raman spectroscopy measurements.

The apparent quantum yield (AQY) for hydrogen evolution from methanol photoreforming was measured for a pristine and a reduced platinized brookite (Fig. 1H) using different monochromatic light sources. The maximum AQY was reached at 334 nm and was 33.5% for the reduced brookite and 22.2% for the pristine nanorods. These AQY values can be further increased by optimizing the methanol concentration, metal loading, metal particle size, or photoreactor design, which however goes beyond the scope of this study. Interestingly, despite its visible light absorption, the reduced brookite did not show AQY in the visible region, with values of ~0.09 and 0.004% at 386 and 402 nm, respectively (AQY below the detection limit for the pristine brookite at both wavelengths). This result was further confirmed by $H_2$ evolution experiments under one-sun illumination applying a longpass optical filter to cut off $\lambda \geq 380$ nm, i.e. cutting optical excitation above bandgap energy did not lead to detecting any $H_2$ after 24 h of reaction. This result confirms that oxygen vacancies introduced upon the hydrogen treatment at high temperature, and the related optical transitions, did not produce visible light photocatalytic activity. We indeed suggest that the introduced oxygen vacancies enhanced the reactivity toward the methanol oxidation by favoring its activation, as discussed in further detail below.

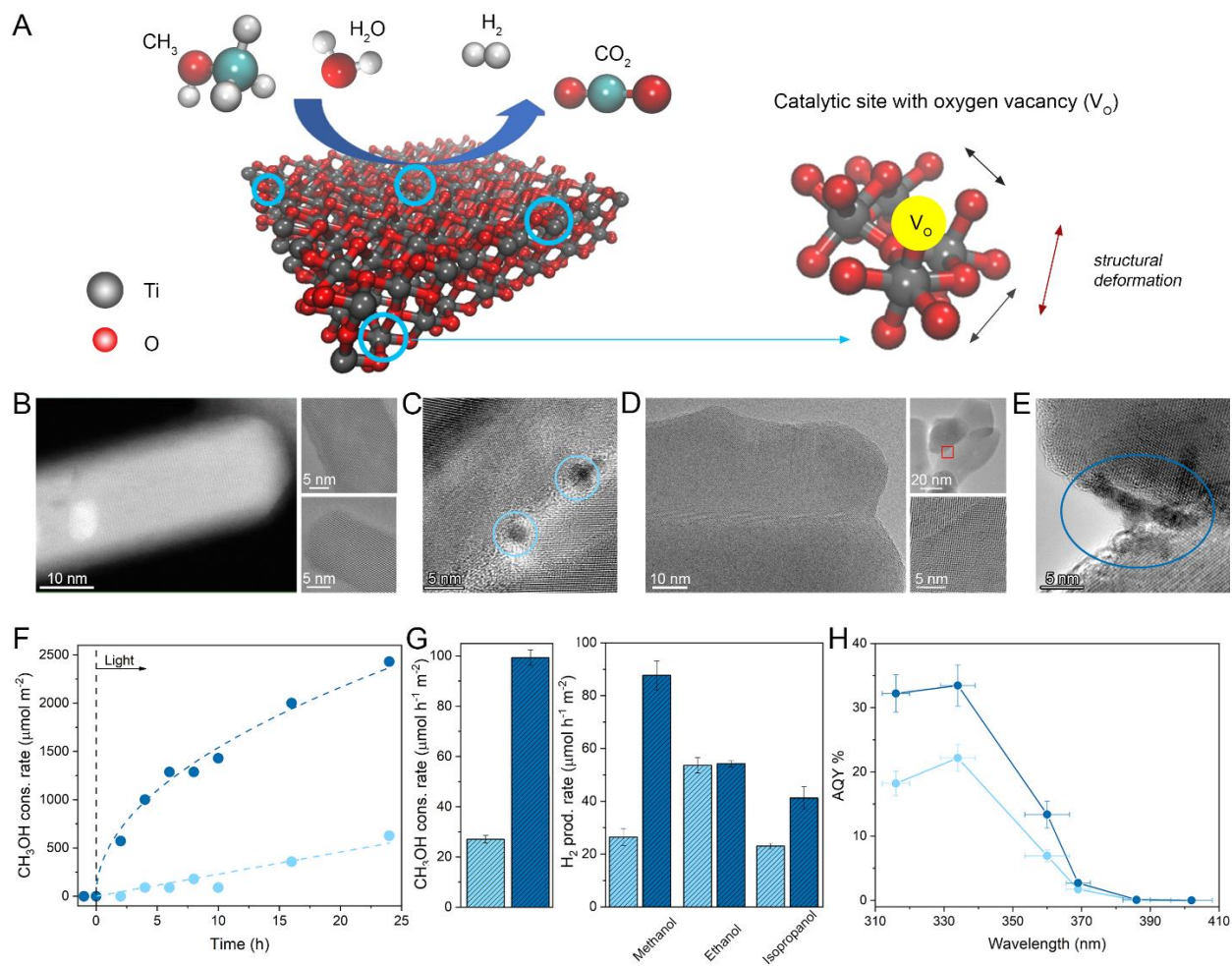

**Fig. 1. Morphology and photocatalytic activity.** (**A**) Schematic representation of defect engineering in reduced brookite showing an exemplary $TiO_2$ surface during photocatalysis and a zoomed view of the catalytic site containing oxygen vacancy ($V_O$) and structural distortions. (**B**) HAADF-STEM (left) and HRTEM (right) micrographs of a single brookite nanorod. (**C**) HRTEM micrograph of isolated Pt nanoparticles deposited on pristine brookite. (**D**) HRTEM of brookite nanorods reduced at 700°C. (**E**) HRTEM micrograph of aggregated Pt nanostructures deposited on reduced brookite. (**F**) Methanol consumption in time for pristine (sky blue) and reduced brookite (dark blue) nanorods. The points before zero time represent the methanol signal before adding the photocatalysts, while time zero was measured once the adsorbtion/desorption equilibrium in the dark was reached. (**G**) Specific methanol consumption rate (left) and specific hydrogen evolution rate (right) during methanol, ethanol, and isopropanol photoreforming for pristine (sky blue) and reduced (dark blue) brookite. Measurements were performed under a simulated AM 1.5G spectrum at one-sun intensity for 24h using a 1:1 vol.% $H_2O$:alcohol mixture. (**H**) Apparent quantum yield for hydrogen evolution from methanol photoreforming for pristine (sky blue) and reduced (dark blue) brookite. In all measurements both pristine and reduced brookite were loaded with 1 wt.% Pt.

In order to study in more detail the methanol photo-oxidation reaction on the reduced brookite, we analyzed the reaction products by both the GC analysis of the gas phase and the NMR analysis of the liquid phase after reaction. Carbon dioxide was the only detected reaction product. Furthermore, we investigated the hydrogen production rate from different possible intermediates of methanol oxidation (e.g. formaldehyde and formic acid) of as-synthesized and reduced brookite nanorods loaded with 1 wt.% Pt. Interestingly, both samples showed similar specific photocatalytic activity in the presence of formaldehyde and formic acid, presenting significant hydrogen production rate of around 25–30 $\mu mol\ h^{-1}\ m^{-2}$ (Fig. S30). This result is far from being trivial, as it has been reported that other $TiO_2$ polymorphs usually oxidize methanol to formaldehyde, thus stopping the methanol photo-oxidation after the first reaction step.[30] Moreover, it repeatedly emphasizes that a reduced brookite demonstrates a substrate-specific oxidation ability toward methanol molecules. The blank test for photolysis of formaldehyde under AM 1.5G 1sun illumination produced a very small hydrogen production rate, namely, ~85 $nmol\ h^{-1}\ m^{-2}$.

Notably, the investigated $TiO_2$ samples showed two order-lower alcohol photoreforming activity without Pt loading; the data on the samples reduced at different temperatures are reported in Fig. S31 and S32. These data are further supported by electron spin resonance spectroscopy (EPR) investigations measured under dark and light conditions both for dried powders and in a water/methanol medium (in situ conditions), demonstrating the increased reactivity of brookite nanorods reduced at 700°C (Fig. S33–S35). For instance, in the case of EPR spectra for dried powders of an anisotropic brookite reduced at different temperatures, the most efficient sample in methanol photoreforming was B700, which indeed gave the highest differential EPR signal (light-dark) in comparison with samples with lower activity (e.g. a pristine brookite and B500). Interestingly, the most active sample (B700) showed the weakest intensity in the EPR powder spectrum among the series (Fig S33 and discussion in the Supporting Information). Therefore, the number of spins recorded by EPR do not directly correlate with the system reactivity and its overall efficiency in the photocatalytic process, all in agreement with previous reports.[12,31]

*In-situ* **photoluminescence spectroscopy**

To understand the nature of the methanol oxidation sites, we measured excitation-dependent photoluminescence (PL) spectra at 80 K, obtaining energy-resolved two-dimensional maps of the radiative recombinations occurring in the pristine and the reduced anisotropic brookite both under inert gas atmosphere ($N_2$) and in the presence of methanol (Fig. 2A). The PL maps in the presence of the latter (i.e., a hole scavenger) showed drastic quenching of the signal, demonstrating that the photogenerated holes trapped within the defect sites, i.e. oxygen vacancies, readily reacted with the surface adsorbed methanol molecules. Moreover, this also provided evidence that such defect sites must be located on either the surface or sub-surface of the brookite nanocrystals, from where they can react with surface adsorbates.[32] This is supported by the synchrotron-based photoemission spectra for the Ti 2p region and valence band (see for details the next section). Comparing the two-dimensional PL maps measured under $N_2$ gas atmosphere for the pristine and the anisotropic brookite reduced at 500, 600, 700, and 800°C (Fig. 2A and Fig. S36), we observed a clear variation in the energy position relative to the radiative recombination centers upon high temperature treatment, eventually underlined by a subtle but significant re-organization of structural defects. Notably, the reference samples (especially the isotropic anatase, which is more reducible than the isotropic brookite) displayed a similar behavior (Fig. S37-S38 and discussion in the Supporting Information).

Focusing on the reduced anisotropic brookite, we observed a significant blue shift in the PL peak after reduction.

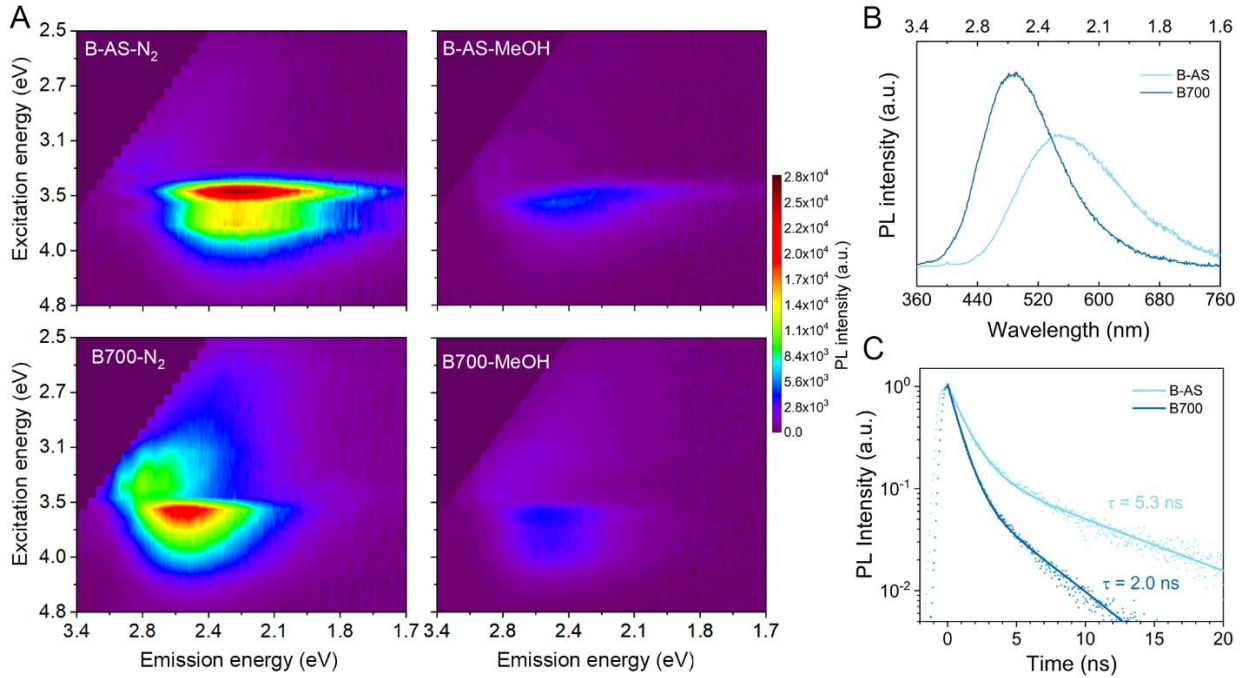

**Fig. 2. Energy distribution of defects-related radiative recombinations and lifetime of photogenerated charge carriers.** (**A**) Excitation-emission color maps under $N_2$ and in the presence of methanol (MeOH) for pristine and reduced brookite. (**B**) PL spectra of pristine (blue sky) and reduced brookite (dark blue) under excitation at 340 nm. (**C**) Time-resolved PL decay curves collected at the corresponding emission maximum of pristine (blue sky) and reduced brookite (dark blue) under excitation at 372 nm.

This is better highlighted in the PL spectra generated using a single excitation wavelength (340 nm) and by analyzing the weight of the deconvoluted components set at 2.75, 2.53, 2.27, and 2.0 eV for all the samples (Fig. 2B and Fig. S39). The dominant radiative recombinations for the as-synthesized anisotropic brookite (B-AS) localize at 2.27 and 2.0 eV, while the components at higher energies are almost negligible. Notably, in the case of the reduced anisotropic brookite (B700), the component at lower energy almost vanished, while the intensity of the radiative recombinations with higher energies (2.75 and 2.53 eV) became dominant, denoting the formation of shallower hole traps upon hydrogen reduction treatment at high temperature. Furthermore, we investigated the lifetime of photogenerated charge carriers measured at PL maximum by time-resolved PL spectroscopy. The charge carriers' lifetime ($\tau$) decreased after the brookite's reduction, similarly to the other TiO$_2$ reference samples, from 5.3 ns to 2.0 ns (Fig. 3C, Fig. S40 and Table S11), suggesting that the enhanced photocatalytic activity of the reduced anisotropic brookite is not related to the enhanced charge separation.

**Synchrotron resonant photoemission spectroscopy**

The investigation by conventional lab scale X-ray photoelectron spectroscopy (XPS) analysis provided similar results for pristine and reduced TiO$_2$ samples (Fig. S41). Therefore, we investigated in more detail the electronic structure of our brookite samples at the VUV-Photoemission beamline (Elettra, Trieste) by synchrotron-based photoemission spectroscopy (PES) for the Ti 2p (Fig. S42), O 1s (Fig. S43, see discussion in the next section), and the valence band (VB) regions. The Ti 2p spectra of both brookite samples contained two components corresponding to the presence of Ti$^{4+}$, due to the coordination of Ti into the stoichiometric lattice, and Ti$^{3+}$ species introduced by the oxygen vacancy formation near or at the TiO$_2$ surface. The presence of low valence Ti ions in the pristine brookite is a common observation, especially in nanocrystals obtained through hydrothermal synthesis and not subjected to a following heat treatment like in the present case. Next, by using soft X-ray photons with energy that is resonant to the Ti absorption edge, it is possible to highlight electronic states even in samples containing a low amount of defects.[33] Fig. 3A shows the VB PES spectra for the pristine and the reduced anisotropic brookite. The main VB edge did not significantly shift upon reduction, while the density of states (DOS) within the bandgap showed a stark difference. The pristine brookite displayed localized mid-gap states peaking at around 1 eV below the Fermi energy. In contrast, the reduced brookite revealed an increased electron density showing an intense VB tailing. We also investigated the reduced brookite after 24 h of photocatalytic reaction B700-AR and compared the result with the spectra of the pristine brookite B-AS and the reduced brookite before reaction B700-BR (Fig. S44). The post-catalytic characterization evidence a spectrum, which features a localize state at around 1 eV below the Fermi level (similarly to B-AS) and an increase density of states –band tailing - at energies closer to the valence band maximum (similarly to B700-BR). The slight modification of the spectrum can be due both to a partial passivation of the surface defects in B700 during photocatalysis and to the adsorption of methoxy groups / reaction intermediates, i.e. the sample was not regenerated after reaction.

**Density functional theory calculations of the photocatalytic sites**

In order to understand the origin of the VB tailing in the electronic structure of the reduced brookite, we calculated the energy band structure using *ab initio* density functional theory (DFT) calculations. Driven by the XRD results, we focused on the (210) surface of the brookite $TiO_2$ introducing two types of structural defects: (1) oxygen vacancies located at different distance from the surface (denoted by V1–V8 in Fig. 3B and Fig. S45), and (2) distortion of $TiO_6$ octahedra (for methods see supplementary materials) by modifying up to ±0.1 Å either the axial or randomly chosen Ti-O distances.

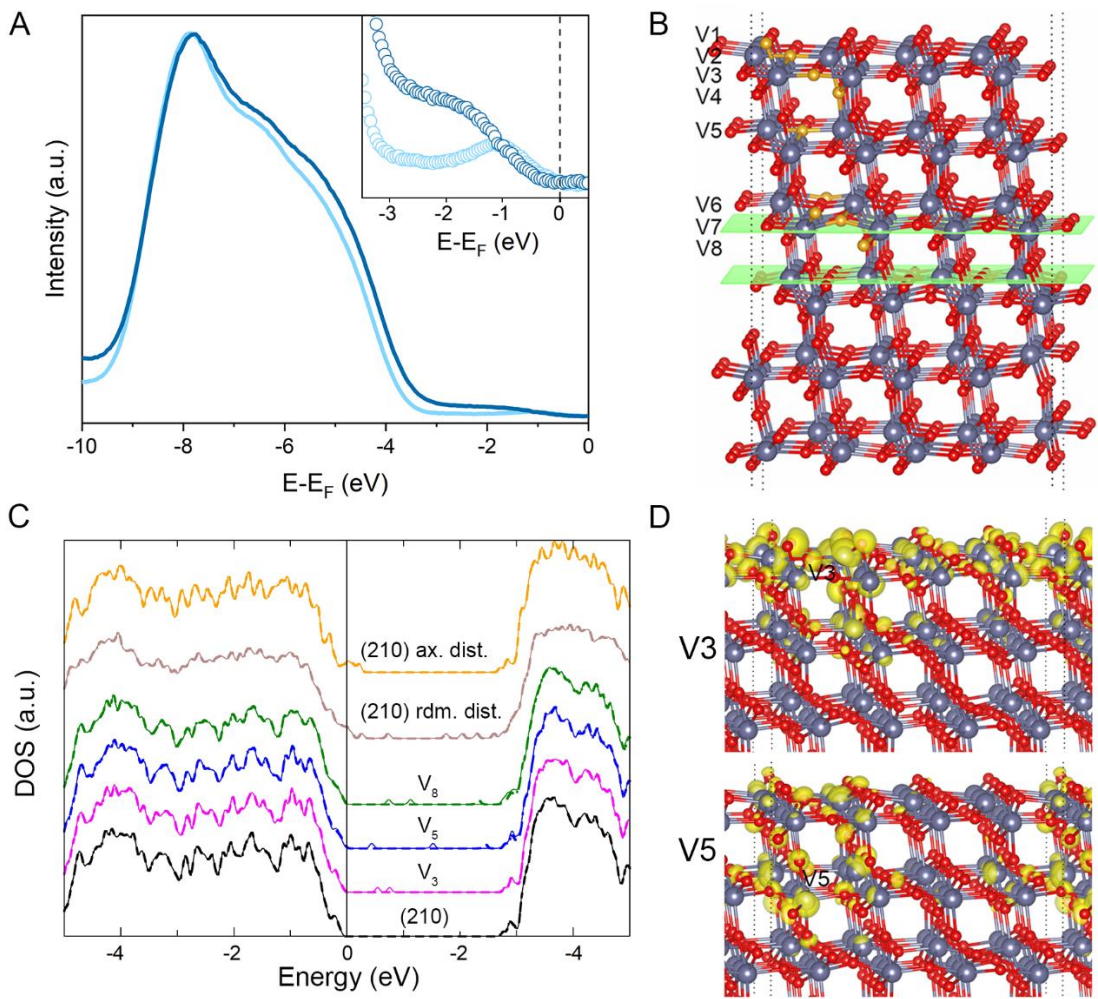

**Fig. 3. Experimental and theoretical determination of the electronic structure of the photocatalytic sites.** (**A**) Synchrotron-based photoemission spectra around the valence band (VB) region for the pristine (light blue) and the reduced (dark blue) brookite. Inset: zoom of the VB region around the Fermi energy. (**B**) Brookite $TiO_2$ supercell employed for the calculations exposing the (210) surface: Ti atoms plotted in grey, O atoms in red, oxygen vacancies in orange. The middle part of the slab corresponds to the bulk region of $TiO_2$ enclosed by green planes, while the supercell's boundaries are marked by the dotted lines. (**C**) Calculated total DOS of the ideal (210) brookite surface, of various defective brookite surfaces with an oxygen vacancy (V in the figure) placed at different locations in the lattice, two distorted brookite surfaces (rdm. and ax. stand for random and axial distortions, respectively). The energy of the VB maximum of the ideal (210) surface is taken to be zero. Spin up/down derived DOS are shown by solid/dashed lines. (**D**) Excess electron density donated by introducing V3 and V5 oxygen vacancies (yellow iso-surface).

The presence of oxygen vacancies introduces localized mid-gap states deriving from the hybridization of O 2p and Ti 3d orbitals (Fig. 3C and Fig. S46) with their energy position that varies with respect to the defect's distance from the surface. In contrast, the primary effect of the expansion of Ti-O axial distances is to produce strong band tailing near the VB edge (Fig. 3C) entering the band gap by ~0.4 eV. When we introduced a lattice disorder by random displacements of both Ti and O atoms from their equilibrium positions, both mid-gap states and VB tailing were seen in the DOS (Fig. 3C). The computational results confirmed that the DOS envelope of the reduced brookite was formed by mid-gap states and VB band tailing due to the combined effect of oxygen vacancies and lattice distortions. The location of oxygen vacancies in the real samples is represented by a statistical distribution of lattice positions. Each of such defect populations produce different DOS and their convolution, alongside the effect from lattice distortions, results in the formation of the VB tailing.

We also examined the differential electron densities due to the introduction of an oxygen vacancy at two different positions in the slab, namely, surface/near-surface (V3) or sub-surface (V5) positions (Fig. 3B and 3D). In both cases, the excess of charge was spread over many lattice sites and accompanied by the relaxation of the lattice atoms by up to 2–4% of the equilibrium Ti-O bond length, thus denoting the generation of a large electron polaron around the oxygen vacancies. We propose that these kind of bound states between oxygen vacancies and large electron polarons represent the substrate-specific photocatalytic active sites for methanol oxidation. The size of this photocatalytic active site and the charge distribution around it make it a well-defined reactive pocket with high specificity toward the methanol molecule rather than to higher alcohols. The finding of Zhang and co-workers support our results, as they recently observed a similar reactivity pattern in Cu-doped $TiO_2$ nanosheets, where the oxygen vacancies within a strained environment enabled strong chemisorption and activation of molecular $N_2$ and water, resulting in high photocatalytic $NH_3$ evolution under visible-light irradiation.[8] Diebold and co-workers recently reported the photo-oxidation mechanism of methanol at the surface of anatase $TiO_2$.[34] Values obtained from DFT calculations, scanning tunneling microscopy, and temperature programmed desorption aided by XPS showed the existence of two different, more favorable pathways for activating methanol adsorbed on $TiO_2$. The methanol molecules are first adsorbed onto the surface $Ti_{5c}$ atoms dissociating into methoxy groups and hydrogen atoms, which are then oxidized to formaldehyde (and eventually to formic acid and carbon dioxide) and molecular hydrogen. Methanol molecules must first dissociate into methoxy groups, and after this step, the hole transfer from $TiO_2$ becomes energetically favorable. Methanol can be activated via two pathways (Fig. S47): (A) by reaction with dissociated $H_2O$ forming terminal $OH^-$ species bound to surface $Ti_{5c}$ atoms, and (B) by reaction with activated adsorbed $O_2$.**Error! Reference source not found.** Mechanism (A) begins with the spontaneous dissociative adsorption of water enabled by the extra charge density due to oxygen vacancies and reflected by the formation of hydroxyl ions.[34,35] Interestingly, brookite $TiO_2$(210) (the same crystallographic direction expressed on the lateral facets of our brookite nanorods) has the same structural building block of anatase $TiO_2$(101), but interatomic distances are slightly shorter and the blocks are arranged in a different way. Selloni and co-workers found that these differences significantly change the reactivity toward adsorption of water (and formic acid), making its dissociation more possible to occur on the brookite surface rather than on the anatase.[36] This may underlie the enhanced specific photocatalytic activity that we observed for the anisotropic brookite over the anatase during methanol photoreforming. This scenario is corroborated by our synchrotron ~~XPS~~ PES of the O 1*s* region that shows a significant, 24% increase in the $OH^-$ species adsorbed on the reduced surface in comparison with the pristine brookite. These species may be derived from the dissociative adsorption of water, which is more favored on the reduced brookite due to the extra electrons provided by the subsurface oxygen vacancies.[32] Mechanism (B) is less probable, as our experiments are performed in the absence of oxygen (under Ar atmosphere). However, it should be noted that some traces of peroxide species were detected by EPR,[12,37] suggesting that mechanism B may occur even at a lower extent than mechanism A. This could also point to a faster decomposition of hydrogen peroxide to water and bridging oxygen dimer (step (iii) → (iv) in mechanism B) in the most photoactive sample (B700) after illumination. Finally, besides the pure A and B mechanisms, an intermediate case can be also considered, in which the OH⁻ formation results from the reaction of coadsorbed $O_2$ and $H_2O$.[38]

## CONCLUSIONS

In summary, we demonstrated the concept of enhancing the photocatalytic activity during alcohol photoreforming by engineering the defect sites in an anisotropic brookite in a way that enables substrate-specific oxidation photo-activity. Synchrotron photoemission spectroscopy and *in situ* photoluminescence investigations aided by DFT calculations showed that creating a low amount of defects (i.e. oxygen vacancies) in well-defined lattice positions produces a kind of bound states between oxygen vacancies and large electron polarons hosting the photocatalytic active sites, which act as shallow hole traps during alcohol photoreforming. Our results also demonstrate that the nature of the produced defects/photocatalytic sites varies with respect to the selected $TiO_2$ polymorph and on its crystal shape. This work highlights the value of analyzing the reaction products of both the reductive and the oxidative pathways during photocatalytic reactions alongside opening new avenues for substrate-selective photocatalytic biomass conversion through the atomic design of the active sites.

**Acknowledgments:** A.N. and R.Z. gratefully acknowledge the support of the Czech Science Foundation (GACR) through the projects no. 20-17636S and 19-27454X. The authors gratefully acknowledge the support by the Operational Programme Research, Development and Education - European Regional Development Fund, project no. CZ.02.1.01/0.0/0.0/15_003/0000416 of the Ministry of Education, Youth and Sports of the Czech Republic. P.F. acknowledge financial support from the European Community (projects H2020 – RIA-CE-NMBP-25 Program – Grant No. 862030 – and H2020-LC-SC3-2019-NZE-RES-CC – Grant No. 884444), INSTM consortium and ICCOM-CNR. P.M.S. and P.M. gratefully acknowledge financial support through the project EUROFEL-ROADMAP ESFRI. The authors gratefully acknowledged G. Zoppellaro and O. Tomanec for EPR discussion and TEM measurements, respectively.

**Author contributions:**

Conceptualization: PF, AN

Methodology: SMHH, MA, EM, PB

Investigation: SMHH, MS, PMS, PM, ZB, SK, EM, PB

Visualization: SMHH, PB, AN

Funding acquisition: RZ, SK, MO, PF, AN

Project administration: AN

Supervision: PF, AN

Writing – original draft: SMHH, AN

Writing – review & editing: SMHH, MA, PM, PB, RZ, MO, PF, AN

**Competing interests:** Authors declare that they have no competing interests.

**Data and materials availability:** All data are available in the main text or the supplementary materials.

# REFERENCES

1. Ciamician, G. (1912). The Photochemistry of the Future. Science *36*, 385–394.

2. Ghosh, I., Khamrai, J., Savateev, A., Shlapakov, N., Antonietti, M., and König, B. (2019). Organic semiconductor photocatalyst can bifunctionalize arenes and heteroarenes. Science *365*, 360–366.

3. Luo, N., Montini, T., Zhang, J., Fornasiero, P., Fonda, E., Hou, T., Nie, W., Lu, J., Liu, J., Heggen, M., et al. (2019). Visible-light-driven coproduction of diesel precursors and hydrogen from lignocellulose-derived methylfurans. Nat Energy *4*, 575–584.

4. Wang, Q., Hisatomi, T., Jia, Q., Tokudome, H., Zhong, M., Wang, C., Pan, Z., Takata, T., Nakabayashi, M., Shibata, N., et al. (2016). Scalable water splitting on particulate photocatalyst sheets with a solar-to-hydrogen energy conversion efficiency exceeding 1%. Nature Mater *15*, 611–615.

5. Kosco, J., Bidwell, M., Cha, H., Martin, T., Howells, C.T., Sachs, M., Anjum, D.H., Gonzalez Lopez, S., Zou, L., Wadsworth, A., et al. (2020). Enhanced photocatalytic hydrogen evolution from organic semiconductor heterojunction nanoparticles. Nat. Mater. *19*, 559–565.

6. Wang, Z., Li, C., and Domen, K. (2019). Recent developments in heterogeneous photocatalysts for solar-driven overall water splitting. Chem. Soc. Rev. *48*, 2109–2125.

7. Zhang, N., Jalil, A., Wu, D., Chen, S., Liu, Y., Gao, C., Ye, W., Qi, Z., Ju, H., Wang, C., et al. (2018). Refining Defect States in $W_{18}O_{49}$ by Mo Doping: A Strategy for Tuning $N_2$ Activation towards Solar-Driven Nitrogen Fixation. J. Am. Chem. Soc. *140*, 9434–9443.

8. Zhao, Y., Zhao, Y., Shi, R., Wang, B., Waterhouse, G.I.N., Wu, L., Tung, C., and Zhang, T. (2019). Tuning Oxygen Vacancies in Ultrathin $TiO_2$ Nanosheets to Boost Photocatalytic Nitrogen Fixation up to 700 nm. Adv. Mater. *31*, 1806482.

9. Takanabe, K. (2017). Photocatalytic Water Splitting: Quantitative Approaches toward Photocatalyst by Design. ACS Catal. *7*, 8006–8022.

10. Chen, X., Liu, L., Yu, P.Y., and Mao, S.S. (2011). Increasing Solar Absorption for Photocatalysis with Black Hydrogenated Titanium Dioxide Nanocrystals. Science *331*, 746–750.

11. Naldoni, A., Allieta, M., Santangelo, S., Marelli, M., Fabbri, F., Cappelli, S., Bianchi, C.L., Psaro, R., and Dal Santo, V. (2012). Effect of Nature and Location of Defects on Bandgap Narrowing in Black TiO2 Nanoparticles. J. Am. Chem. Soc. *134*, 7600–7603.

12. Naldoni, A., Altomare, M., Zoppellaro, G., Liu, N., Kment, Š., Zbořil, R., and Schmuki, P. (2019). Photocatalysis with Reduced $TiO_2$: From Black $TiO_2$ to Cocatalyst-Free Hydrogen Production. ACS Catal. *9*, 345–364.

13. Wang, X., Maeda, K., Thomas, A., Takanabe, K., Xin, G., Carlsson, J.M., Domen, K., and Antonietti, M. (2009). A metal-free polymeric photocatalyst for hydrogen production from water under visible light. Nature Materials *8*, 76–80.

14. Filippini, G., Longobardo, F., Forster, L., Criado, A., Carmine, G.D., Nasi, L., D'Agostino, C., Melchionna, M., Fornasiero, P., and Prato, M. (2020). Light-driven, heterogeneous organocatalysts for C–C bond formation toward valuable perfluoroalkylated intermediates. Science Advances *6*, eabc9923.

15. Li, H., Chen, S., Shang, H., Wang, X., Yang, Z., Ai, Z., and Zhang, L. (2020). Surface hydrogen bond network spatially confined BiOCl oxygen vacancy for photocatalysis. Science Bulletin *65*, 1916–1923.

16. Wu, Y.A., McNulty, I., Liu, C., Lau, K.C., Liu, Q., Paulikas, A.P., Sun, C.-J., Cai, Z., Guest, J.R., Ren, Y., et al. (2019). Facet-dependent active sites of a single $Cu_2O$ particle photocatalyst for $CO_2$ reduction to methanol. Nature Energy *4*, 957–968.

17. Cargnello, M., Montini, T., Smolin, S.Y., Priebe, J.B., Jaén, J.J.D., Doan-Nguyen, V.V.T., McKay, I.S., Schwalbe, J.A., Pohl, M.-M., Gordon, T.R., et al. (2016). Engineering titania nanostructure to tune and improve its photocatalytic activity. PNAS *113*, 3966–3971.

18. Li, R., Zhang, F., Wang, D., Yang, J., Li, M., Zhu, J., Zhou, X., Han, H., and Li, C. (2013). Spatial separation of photogenerated electrons and holes among {010} and {110} crystal facets of $BiVO_4$. Nature Communications *4*, 1432.

19. Chen, X., Shen, S., Guo, L., and Mao, S.S. (2010). Semiconductor-based Photocatalytic Hydrogen Generation. Chem. Rev. *110*, 6503–6570.

20. Kamat, P.V., and Jin, S. (2018). Semiconductor Photocatalysis: "*Tell Us the Complete Story!*." ACS Energy Lett. *3*, 622–623.

21. Schneider, J., and Bahnemann, D.W. (2013). Undesired Role of Sacrificial Reagents in Photocatalysis. J. Phys. Chem. Lett. *4*, 3479–3483.

22. Hainer, A.S., Hodgins, J.S., Sandre, V., Vallieres, M., Lanterna, A.E., and Scaiano, J.C. (2018). Photocatalytic Hydrogen Generation Using Metal-Decorated $TiO_2$: Sacrificial Donors vs True Water Splitting. ACS Energy Lett. *3*, 542–545.

23. Belhadj, H., Hamid, S., Robertson, P.K.J., and Bahnemann, D.W. (2017). Mechanisms of Simultaneous Hydrogen Production and Formaldehyde Oxidation in $H_2O$ and $D_2O$ over Platinized $TiO_2$. ACS Catal. *7*, 4753–4758.

24. Xu, C., Paone, E., Rodríguez-Padrón, D., Luque, R., and Mauriello, F. (2020). Recent catalytic routes for the preparation and the upgrading of biomass derived furfural and 5-hydroxymethylfurfural. Chem. Soc. Rev. *49*, 4273–4306.

25. Wu, X., Li, J., Xie, S., Duan, P., Zhang, H., Feng, J., Zhang, Q., Cheng, J., and Wang, Y. (2020). Selectivity Control in Photocatalytic Valorization of Biomass-Derived Platform Compounds by Surface Engineering of Titanium Oxide. Chem *6*, 3038–3053.

26. Naldoni, A., Montini, T., Malara, F., Mróz, M.M., Beltram, A., Virgili, T., Boldrini, C.L., Marelli, M., Romero-Ocaña, I., Delgado, J.J., et al. (2017). Hot Electron Collection on Brookite Nanorods Lateral Facets for Plasmon-Enhanced Water Oxidation. ACS Catal. *7*, 1270–1278.

27. Di Valentin, C., Pacchioni, G., and Selloni, A. (2009). Reduced and n-Type Doped TiO2: Nature of Ti3+ Species. J. Phys. Chem. C *113*, 20543–20552.

28. Wierzbicka, E., Altomare, M., Wu, M., Liu, N., Yokosawa, T., Fehn, D., Qin, S., Meyer, K., Unruh, T., Spiecker, E., et al. (2021). Reduced grey brookite for noble metal free photocatalytic H2 evolution. J. Mater. Chem. A *9*, 1168–1179.

29. Wang, H., Zhang, L., Chen, Z., Hu, J., Li, S., Wang, Z., Liu, J., and Wang, X. (2014). Semiconductor heterojunction photocatalysts: design, construction, and photocatalytic performances. Chem. Soc. Rev. *43*, 5234–5244.

30. Huang, H., Feng, J., Zhang, S., Zhang, H., Wang, X., Yu, T., Chen, C., Yi, Z., Ye, J., Li, Z., et al. (2020). Molecular-level understanding of the deactivation pathways during methanol photo-reforming on Pt-decorated TiO2. Applied Catalysis B: Environmental *272*, 118980.

31. Bad'ura, Z., Naldoni, A., Qin, S., Bakandritsos, A., Kment, Š., Schmuki, P., and Zoppellaro, G. Light-Induced Migration of Spin Defects in TiO2 Nanosystems and their Contribution to the H2 Evolution Catalysis from Water. ChemSusChem *n/a*.

32. Setvín, M., Aschauer, U., Scheiber, P., Li, Y.-F., Hou, W., Schmid, M., Selloni, A., and Diebold, U. (2013). Reaction of O2 with Subsurface Oxygen Vacancies on TiO2 Anatase (101). Science *341*, 988–991.

33. Naldoni, A., Riboni, F., Marelli, M., Bossola, F., Ulisse, G., Carlo, A.D., Píš, I., Nappini, S., Malvestuto, M., Dozzi, M.V., et al. (2016). Influence of TiO2 electronic structure and strong metal–support interaction on plasmonic Au photocatalytic oxidations. Catal. Sci. Technol. *6*, 3220–3229.

34. Setvin, M., Shi, X., Hulva, J., Simschitz, T., Parkinson, G.S., Schmid, M., Di Valentin, C., Selloni, A., and Diebold, U. (2017). Methanol on Anatase TiO2 (101): Mechanistic Insights into Photocatalysis. ACS Catal. *7*, 7081–7091.

35. Selcuk, S., and Selloni, A. (2016). Facet-dependent trapping and dynamics of excess electrons at anatase TiO2 surfaces and aqueous interfaces. Nature Mater *15*, 1107–1112.

36. Li, W.-K., Gong, X.-Q., Lu, G., and Selloni, A. (2008). Different Reactivities of TiO$_2$ Polymorphs: Comparative DFT Calculations of Water and Formic Acid Adsorption at Anatase and Brookite TiO$_2$ Surfaces. J. Phys. Chem. C *112*, 6594–6596.

37. Naldoni, A., D'Arienzo, M., Altomare, M., Marelli, M., Scotti, R., Morazzoni, F., Selli, E., and Dal Santo, V. (2013). Pt and Au/TiO2 photocatalysts for methanol reforming: Role of metal nanoparticles in tuning charge trapping properties and photoefficiency. Applied Catalysis B: Environmental *130–131*, 239–248.

38. Setvin, M., Aschauer, U., Hulva, J., Simschitz, T., Daniel, B., Schmid, M., Selloni, A., and Diebold, U. (2016). Following the Reduction of Oxygen on TiO2 Anatase (101) Step by Step. Journal of the American Chemical Society *138*, 9565–9571.

# Supporting Information

Defect engineering over anisotropic brookite towards substrate-specific photo-oxidation of alcohols

S. M. Hossein Hejazi[1], Mahdi Shahrezaei[1], Piotr Błoński[1], Mattia Allieta[2], Polina M. Sheverdyaeva[3], Paolo Moras[3], Zdeněk Baďura[1], Sergii Kalytchuk[1], Elmira Mohammadi[1], Radek Zbořil[1,4], Štěpán Kment[1,4], Michal Otyepka[1,5], Alberto Naldoni[1]*, Paolo Fornasiero[6,7]*,

[1]Czech Advanced Technology and Research Institute, Regional Centre of Advanced Technologies and Materials, Palacký University Olomouc, Křížkovského 511/8, 77900 Olomouc, Czech Republic

[2]Ronin Institute Montclair, NJ 07043 USA

[3]Istituto di Struttura della Materia-CNR (ISM-CNR), SS 14, Km 163,5, I-34149, Trieste, Italy

[4]Nanotechnology Centre, Centre of Energy and Environmental Technologies, VŠB–Technical University of Ostrava, 17. listopadu 2172/15, 70800 Ostrava-Poruba, Czech Republic

[5]IT4Innovations, VSB – Technical University of Ostrava, 17. listopadu 2172/15, 708 00 Ostrava-Poruba, Czech Republic

[6]Department of Chemical and Pharmaceutical Sciences, ICCOM-CNR Trieste Research Unit, INSTM-Trieste, University of Trieste, Via L. Giorgieri 1, 34127 Trieste, Italy

[7]Center for Energy, Environment and Transport Giacomo Ciamician - University of Trieste, Italy

*Corresponding authors: alberto.naldoni@upol.cz; pfornasiero@units.it

*Preparation of TiO$_2$ photocatalysts*

Titanium (IV) bis (ammonium lactate) dihydroxide Ti(NH$_4$C$_3$H$_4$O$_3$)$_2$(OH)$_2$ aqueous solution (50 wt.%, Sigma–Aldrich) (TALH) and urea (ACS reagent, Sigma–Aldrich), were used as precursors for the synthesis of TiO$_2$ nanocrystals using a hydrothermal method, according to the procedure previously reported with some modifications [1–4]. A solution containing 45 mL of urea in deionized (DI) water and 5 mL of TALH was stirred until a clear solution was obtained. The solution was afterwards transferred to a 125 mL Teflon lined autoclave and placed in an oil bath at 180°C and stirred at this temperature with 800 rpm for 20 days. The autoclave was then cooled down in air and the precipitate was centrifuged and dispersed by sonication in DI water for several times until the pH of supernatant water became ~7. Finally, the precipitate was dried at 80°C for 12 h. To prepare pure brookite and pure anatase samples, 0.15M and 11.5M urea solution in DI water were used, respectively. Commercial TiO$_2$ brookite was purchased from Sigma-Aldrich (99.99 wt. % purity). To prepare the reduced powders, 20 mg of TiO$_2$ nanopowders were placed in a crucible within a quartz chamber in a tubular furnace (10 °C min$^{-1}$ heating/cooling ramp in N$_2$ flow rate 10 mL min$^{-1}$, 1 h dwell in H$_2$ flow rate 10 mL min$^{-1}$ at predefined temperature. Before starting the heat treatment, the tube furnace was cleaned up increasing the temperature up to 1000°C in air. 1 wt.% platinum nanoparticles were loaded on TiO$_2$ by via photodeposition method. Briefly, 50 mg of TiO$_2$ powder suspended in 25 mL of methanol (ACS reagent, Sigma–Aldrich) and bubbled with Ar for 30 min. Then, a solution of H$_2$PtCl$_6$.6H$_2$O (ACS reagent, Sigma–Aldrich) was added and stirred for 20 min in the dark to favor Pt adsorption of the TiO$_2$ surface. Then the solution was illuminated for 1h using a solar simulator equipped with a 150 W Xe arc lamp and an AM 1.5G filter and calibrated to deliver a power of 100 mW cm$^{-2}$ (1 Sun).

*Characterization*

The morphological analyses of the samples were performed by transmission electron microscopy (TEM) JEM-2100 (JEOL, Tokyo, Japan) at 200 kV of accelerating voltage. For TEM measurements, the samples were dispersed in ethanol by sonication for 5 minutes and then the suspensions were dropped on the copper grid with holey carbon film and dried upon air exposure. The average particle size of brookite nanorods were assessed by analyzing TEM micrographs and by considering at least 100 nanorods. The high resolution transmission electron microscopy (HRTEM) analysis were performed using a HRTEM Titan G2 (FEI) with image corrector on accelerating voltage 300 kV. Images were taken with BM UltraScan CCD camera (Gatan).

X-ray diffraction (XRD) patterns were recorded at room temperature with an Empyrean (PANalytical, Almelo, The Netherlands) diffractometer in the Bragg-Brentano geometry and using Co-$K_\alpha$ radiation (40 kV, 30 mA, $\lambda$ = 0.1789 nm). The diffractometer was equipped with a PIXcel3D detector and programmable divergence and diffracted beam anti-scatter slits. The same amount of powders was placed on a zero-background Si slide. The measurement range was 2θ = 10° - 100°, with a step size of 0.0167° and acquisition time of 4 s per step. Standards SRM640 (Si) and SRM660 (LaB$_6$) were used to evaluate the line position and the instrumental line broadening, respectively. The identification of crystalline phases was performed using the High Score Plus software that includes the PDF-4+ and ICSD databases. Rietveld analysis was performed through the GSAS program [5]. We use the brookite orthorhombic model of Pbca space group with Ti and two O, namely O1, O2, in 8c position (x,y,z) [6]. During the refinement, the background was subtracted using shifted Chebyshev polynomials and the diffraction peak profiles were fitted with a modified pseudo-Voigt function. In the last calculation cycles all the parameters were refined: cell parameters, atomic positional degrees of freedom, isotropic thermal parameters, anisotropic microstrain broadening parameters, background, diffractometer zero point. To evaluate annealing T evolution ionic charge of Ti from the experimental dTi-O1, dTi-O2 of brookite, we calculated Bond Valence Sum (BVS) by using the tabulated parameters [7]. Crystallite size of synthesized brookite samples was estimated through the Williamson-Hall (WH) method [8] employing at least 15 reflections for each calculation. Single peak fitting to extract peak positions and profile parameters was performed through the WinPLOTR Software [9]. The crystallite size of as received and reduced commercial brookite and anatase was calculated from XRD patterns according to the Scherrer equation as follows:

$$D = \frac{K \times \lambda}{\beta \times cos\theta}$$

where, D is the mean size of crystallite, K is the dimensionless shape factor, λ is the x-ray wavelength, β is the full width half maximum intensity (FWHM), and θ is the Bragg angle and considering K=0.9, λ=1.79 Å (Co).

Raman spectra were collected using a DXR Raman spectrometer (Thermo Scientific, Massachusetts, USA). The excitation laser operated at the wavelength of 455 nm. The samples were deposited on a silicon wafer and the laser was focused on its surface and tuned to maximize the signal. The laser power on the sample was set to 0.1 mW cm$^{-2}$ and exposure time was 3 s. The reported Raman spectra were averaged over 512 experimental microscans.

The surface area and pore size analyses were performed by means of $N_2$ adsorption/desorption measurements at 77 K on a volumetric gas adsorption analyzer 3 Flex (Micromeritics, Georgia, USA) up to 0.965 P/P$_0$. Prior the analysis, the sample was degassed under high vacuum (10$^{-4}$ Pa) at 130°C for 12 hours, while high purity (99.999 %) $N_2$ and He gases were used for the measurements. The Brunauer–Emmett–Teller area (BET) was determined with respect to Rouquerol criteria [10] for $N_2$ isotherm.

The ultraviolet-visible diffuse reflectance spectra (UV-Vis DRS) of the fabricated samples were obtained by Specord 250 plus (Analytik Jena, Jena, Germany) spectrophotometer. An integrating sphere was used to collect the spectrum and a Spectralon reference sample was used to measure the background.

X-ray photoelectron spectroscopy (XPS) was carried out with a PHI 5000 VersaProbe II (Physical Electronics, Chanhassen, USA) spectrometer using an Al K$_\alpha$ source (15 kV, 50 W). The obtained data were evaluated with the MultiPak software package (Ulvac-PHI Inc., Chigasaki, Japan). High-resolution spectra of C$1s$ peaks were acquired by setting the pass energy to 23.500 eV and step size to 0.200 eV. The binding energy values were corrected considering the C$1s$ peak at 284.8 eV as a reference. The spectral analysis included Shirley background subtraction and peak deconvolution using Gaussian functions.

Photoluminescence spectroscopy (PL) was performed on an FLS980 fluorescence spectrometer (Edinburgh Instruments, Livingston, United Kingdom) equipped with a R928P photomultiplier (Hamamatsu, Japan), with a 450 W xenon arc lamp as the excitation source for steady-state spectra and an EPL-375 picosecond pulsed diode laser (λ$_{em}$= 372 nm with a pulse width of 66.5 ps, a repetition rate of 10 MHz and an average power of 75 μW, Edinburgh Instruments) in conjunction with a time-correlated single-photon counting system for time-resolved photoluminescence measurements. Spectral correction curves were provided by Edinburgh Instruments. The emission of TRPL spectra were detected at 450 nm. PL decay curves were fitted using a multi-exponential function:

$$I(t) = \sum_{i=1}^{n} B_i \exp\left(-\frac{t}{\tau_i}\right), \sum_{i=1}^{n} B_i = 1,$$

Where, the fit parameter $\tau_i$ represents the decay time constant, $B_i$ represents the normalized amplitude of each component, n is the number of decay times.

The amplitude weighted average decay lifetime $\tau_{ave}$ of the entire PL decay process reads as:

$$\tau_{ave} = \frac{\sum B_i \tau_i^2}{\sum B_i \tau_i}$$

A nitrogen bath cryostat holder Optistat/DNV (Oxford instruments, Abingdon, United Kingdom) was used to control the temperature of sample during measurements. Since the PL emission of $TiO_2$ is weak at room temperature and due to highly scattering nature of a typical nano-$TiO_2$ sample, the stray excitation light could be wrongly assigned as PL signal [11]. To avoid this, the PL spectra were measured at 80 K. Using low temperature condition results in slower non-radiative decay and brighter PL [11], which decreases the effect of stray light scattered by the sample. The powders in solid were pressed between two flat quartz and put into the chamber. $N_2$ and methanol were used to investigate the PL behavior of the $TiO_2$ samples in contact with different environments. The methanol was degassed with argon bubbling for 10 minutes before wetting the sample.

An in-depth analysis of the electronic structure of the samples was carried out at the VUV-Photoemission synchrotron beamline (Elettra, Trieste) at room temperature with a Scienta R-4000 electron spectrometer. The O$1s$ and Ti$2p$ core levels were measured with photon b 650 eV with an instrumental energy resolution of 0.2 eV. The valence band was probed in near-resonant conditions to the Ti $L_{2,3}$ edge, in order to enhance the signal of Ti-related states. A photon energy of 468 eV (energy resolution 0.14 eV) was used to avoid the appearance of spurious Ti$2p$ signal in the region of interest (from 4 eV binding energy up to the Fermi level), due to high harmonics contribution from the beamline.

Electron paramagnetic resonance (EPR) spectra were recorded using a continuous wave X-band JEOL JES-X-320 spectrometer operating at ~9.1 GHz. The EPR spectrometer is equipped with a variable temperature control ES 13060DVT5 apparatus. The cavity Q quality factor was kept above 6000 in all measurements. Highly pure quartz tubes were used (Suprasil, Wilmad, ≤ 0.5 OD) and accuracy on $g$-values was obtained against a $Mn^{2+}$/MgO standard (JEOL standard). For all experiments the same acquisition conditions were kept. The microwave power was set to 1.5 mW, therefore, no power saturation effects was occurring in the EPR traces. The modulation width of 0.7 mT and modulation frequency 100 Hz were used. Experimental temperature was set to 78 K. All spectra were recorded with 30 ms time constant and 2 minutes sweep time with 3 accumulations, to improve signal to noise ratio. HeCd (200 mW) laser with 325 nm wavelength was employed as the UV light source during EPR experiments. EPR envelopes were simulated in Matlab software where the spin-Hamiltonian EasySpin simulation package [12] is implemented. During the measuring the samples in contact with DI water + methanol solution at 80K, the head space of EPR tube was purged by nitrogen to avoid any parasitic oxygen signals coming from the air. The EPR cavity was kept in constant flow of nitrogen to eliminate the formation of ice. The samples were measured in suspension form. For each experiment 10 mg of the powder and 100 μl of DI water/methanol mixture were used. In all experiments, to make spectra more comparable, the position of the EPR tube inside the cavity was kept the same.

$^1$H-NMR (proton nuclear magnetic resonance) spectra were obtained on a JEOL 400 MHz spectrometer in $CD_3OD$, using dimethyl sulfoxide (DMSO) as the internal standard. All the measurements were collected at ambient temperature with a spectral width of 20 ppm, a pulse width of 5.7 μs (90 °), a relaxation delay of 60 s, and 8 scans. Chemical shifts (δ) are expressed in ppm.

C, H and N elemental analyses was performed on a Flash EA 1112 instrument (Thermo Finnigan, North Carolina, USA).

Pt loading on $TiO_2$ was determined by ICP-MS (Agilent 7700x, Agilent, USA) at isotope 105 using He mode and an external calibration. Calibration solutions were prepared from a certified reference material with Pt concentration 100,0 +/- 0,2 mg/L (Analytika Ltd., Czech Republic). A mixture of nitric acid (ACS reagent, 70% , Sigma–Aldrich) and hydrochloric acid (ACS reagent, 37% , Sigma–Aldrich) in a molar ratio of 1:3, was used to digest the Pt. The Pt loading in pristine and reduced brookite nanorods was 0.98 and 0.90%, respectively.

The hydrodynamic diameter of the brookite B700 was measured by Dynamic Light Scattering (DLS) at 23°C, using a Malvern Nano-ZS instrument (Malvern Ltd., Leamington Spa, UK). The light source was a laser 633 nm, 4 mW. The measurement was performed at the beginning (time = 0 h) and after specific times of reaction. The sample was taken from 10 mL of DI water and methanol (1:1 vol) solution with 2mg of dispersed B700-Pt. At time = 0h the solution was ultrasonically dispersed for 10 min.

*Photocatalytic experiments*

Photocatalytic activity was measured in a quartz reactor with 10 mL solution of DI water and methanol (volume ratio 1:1). The same conditions and volume ratio were applied for the photocatalytic reactions with ethanol (99.8 %, BC Chemservic), isopropanol (ACS reagent, 99.8%, Sigma–Aldrich), formaldehyde (36-38%, Penta) and formic acid (99%, Penta). After sealing the reactor with a rubber septum, the photocatalysts were sonicated for 10 min to create a homogeneous and dispersed suspension. Afterwards, the suspension was bubbled with argon for 30 min to remove the unwanted gasses and dissolved oxygen. The samples were irradiated using a solar simulator equipped with a 150 W Xe arc lamp and an AM 1.5G filter and calibrated to deliver a power of 100 mW cm$^{-2}$ (1 sun). A calibrated reference solar cell (Newport, California, USA) was used before and after reaction to check the power of irradiation (1 Sun). Each sample was irradiated for 24 h under continuous stirring before measuring the amount of evolved hydrogen. The test was repeated three times and the average amount of measured hydrogen was reported. The photocatalytic hydrogen was detected with a gas chromatograph GCMS-QP2010 SE (Shimadzu, Kyoto, Japan ) and a TCD (Thermal conductivity detector), using Ar as carrier gas. The temperature of the reaction suspension was measured with a thermocouple and was 23°C both before and after 24 h of irradiation under 1 sun illumination.

To identify the wavelength dependence of AQY, the reactor was illuminated with the wavelengths 316 (1.6 mW cm$^{-2}$), 334 (1.2 mW cm$^{-2}$), 360 (2.8 mW cm$^{-2}$), 369 (8.7 mW cm$^{-2}$), 386 (5 mW cm$^{-2}$) and 402 nm (17.7 mW cm$^{-2}$) using a tunable diode light source of Zahner CIMPS PP201 system. The illuminated surface area was 0.94 cm$^2$ and the power of each wavelength was measured using an external digital power meter Thorlabs PM100D. The sample was illuminated for 1h and the reacting medium was a 1:1 vol/vol% water:methanol mixture. The apparent quantum yield (AQY) was calculated according to the following equation:

$$AQY = \frac{Number\ of\ reacted\ electrons}{Number\ of\ incident\ photons} \times 100 = \frac{2 \times Number\ of\ evolved\ H_2\ molecules}{Number\ of\ incident\ photons}$$

The qNMR (quantitative nuclear magnetic resonance) analysis was used to investigate the rate of methanol consumption during the photocatalytic hydrogen evolution. For this purpose, a 10 mL mixture of DI water: CD$_3$OD (volume ratio 1:1) was prepared. To have a detectable signal, 50 μL of non-deuterated methanol (i.e. CH$_3$OH) was added to the above mixture. The photocatalytic experiment was performed following the same procedure explained above for hydrogen evolution measurements. At the specified times, 600 μL of solution was taken by syringe and centrifuged at 15000 rpm for 30min to separate the catalyst. Then the clear and transparent solution was transferred to a quartz NMR tube and 0.2 μL DMSO was added as internal standard. The NMR spectra of the samples were recorded and the peak area of methanol at 3.32 ppm was compared to the peak area of internal standard to study the rate of photocatalytic methanol consumption over time.

*DFT calculations*

Calculations were carried out by using the DFT-based Vienna Ab Initio Simulation Package (VASP) [13,14]. The projector-augmented-wave (PAW) formalism [15,16] was used to treat the electron–ionic core interactions. A plane-wave basis with a 400 eV energy cutoff was used. Test calculations were also performed with energy cutoff increased to 600 eV. Exchange and correlation effects were treated within a generalized-gradient approximation (GGA) by using Perdew-Burke-Ernzerhof (PBE) functional [17,18]. To counteract the problems of standard density functionals associated with the self-interaction error (SIE) we applied on-site Hubbard corrections [19] to both Ti-*d* and O-*p* states [20] with an effective U parameter of 6 eV. All computations were performed in spin unrestricted manner. Brillouin zone samplings were kept restricted to Gamma point only due to the supercell dimensions being sufficiently large. We modeled the (210) surface of brookite by a bulk-terminated slab of 12 Ti-layers in thickness and 14.38 Å × 15.41 Å of surface area and containing 432 atoms (see the structure shown in Figure S41) [21], and with a vacuum layer of length ~20 Å deployed along the off-planar direction to ward off spurious interactions with the periodic images. Except atoms in the middle part of the slab (area enclosed by green planes in the structure displayed in Figure S41), all other atoms were relaxed until all forces were reduced below 0.025 eV/Å and the change in total energy between successive iteration steps became smaller than 10$^{-6}$ eV.

Several possible sites for oxygen vacancy formation were considered (and denoted by V1–V8 in the structure in Figure S41) and the system was re-optimized. For oxygen vacancy defects present in the bulk region of the slab, atoms in the immediate vicinity of the vacancy were allowed to relax too.

The effect of distortion of TiO$_6$ octahedra on the electronic structure of brookite was also considered. Two models were considered: (i) Several Ti-O axial distances were modified by up to ±0.1 Å and the change in densities of states with respect to an ideal (210) surface of brookite was monitored. (ii) Randomly chosen Ti-O distances were modified within ±0.1 Å and accompanied by a displacement of Ti atoms from its equilibrium positions.

## Supplementary Text

### *Digital pictures of TiO₂ photocatalysts*

Upon reduction under hydrogen atmosphere at different temperatures, the color of synthesized brookite and anatase samples changed from white to gray and black, while the commercial brookite showed a relatively small color change (Table S1), suggesting its non-reducibility, as also confirmed by the other characterizations reported below.

**Table S1.** Digital photographs of as synthesized brookite (B-AS), as-received commercial brookite (CB-AR), as-synthesized anatase (A-AS) and reduced samples at different temperatures. The numbers after abbreviations stand for reducing temperature in °C.

| Sample | | Sample | | Sample | |
|---|---|---|---|---|---|
| B-AS | 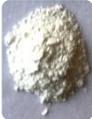 | CB-AR | 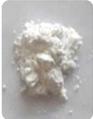 | A-AS | 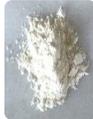 |
| B500 | 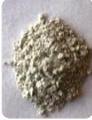 | CB400 | 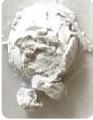 | A400 | 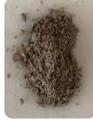 |
| B600 | 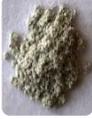 | CB500 | 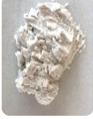 | A500 | 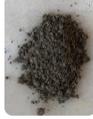 |
| B700 | 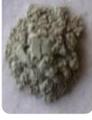 | CB600 | 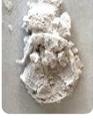 | A600 | 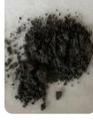 |
| B800 | 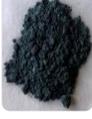 | CB700 | 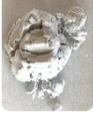 | A700 | 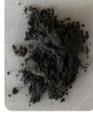 |
| B900 | 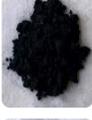 | CB800 | 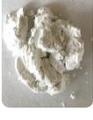 | | |
| B1000 | 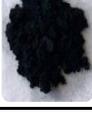 | | | | |

### *CHN elemental analysis*

To exclude the contribution of carbon, nitrogen, and hydrogen impurities in the photocatalytic activity, CHN analysis of TiO₂ samples were carried out before and after reduction under hydrogen atmosphere for the most active photocatalysts. As Table S2 shows, there is no significant difference between carbon, nitrogen, and hydrogen concentration in the samples before and after the thermal treatment in hydrogen.

**Table S2.** CHN analysis of the pristine and the most photoactive sample for brookite, commercial brookite and anatase.

| Sample | C (wt. %) | H (wt. %) | N (wt. %) |
|--------|-----------|-----------|-----------|
| B-AS   | 0.25 ± 0.08 | 0.27 ± 0.03 | 0.12 ± 0.03 |
| B700   | 0.19 ± 0.02 | 0.07 ± 0.01 | 0.03 ± 0.01 |
| CB-AR  | 1.22 ± 0.04 | 0.83 ± 0.02 | 0.07 ± 0.01 |
| CB600  | 0.73 ± 0.05 | 0.08 ± 0.01 | 0.09 ± 0.03 |
| A-AS   | 1.03 ± 0.06 | 1.43 ± 0.09 | 0.11 ± 0.02 |
| A-500  | 0.39 ± 0.06 | 0.16 ± 0.05 | 0.06 ± 0.02 |

*Morphology of $TiO_2$ photocatalysts*

Figure S1A and Figure S1B show the TEM images of B-AS and B700, respectively. The as-synthesized brookite nanorods have an average length of 90 ± 35 nm that after reduction under hydrogen atmosphere at 700°C decreased to 82 ± 28 nm. However, the reduction in high temperature resulted in aggregation of particles due to sintering as well as losing their well-defined facets. Figure S2A and Figure S2B present the TEM images of CB-AR and CB600, respectively. The particle shape of commercial brookite is rounded (average diameter 29 ± 10 nm) in comparison with the synthesized brookite and showed almost no change in shape and a slightly increase in size after reduction (average diameter 32 ± 7 nm). The as synthesized anatase is spherical in shape with average diameter of 6 ± 1 nm (Figure S3) and its shape remained unchanged with slighty increase in diameter after reduction (average diameter 8 ± 2 nm).

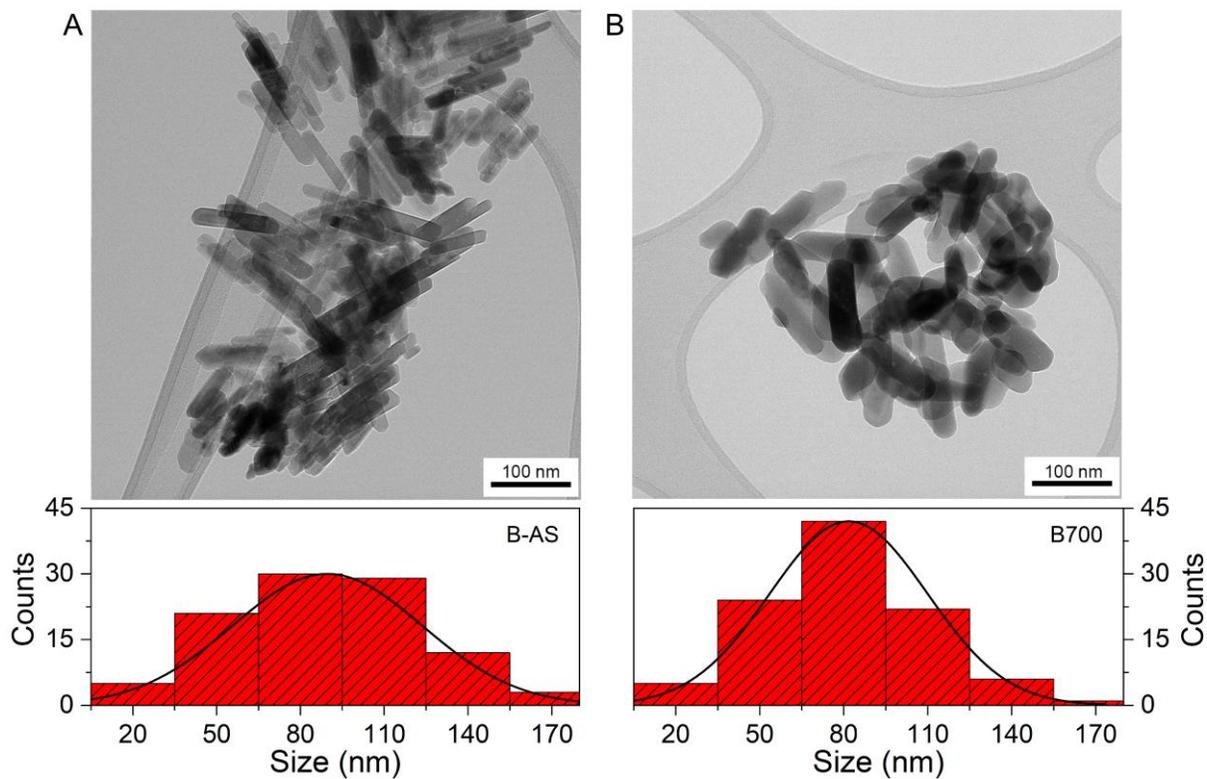

**Figure S1.** TEM images of (A) B-AS and (B) B700 with associated histograms of size distributions (bottom) based on 100 measurements of nanorods length.

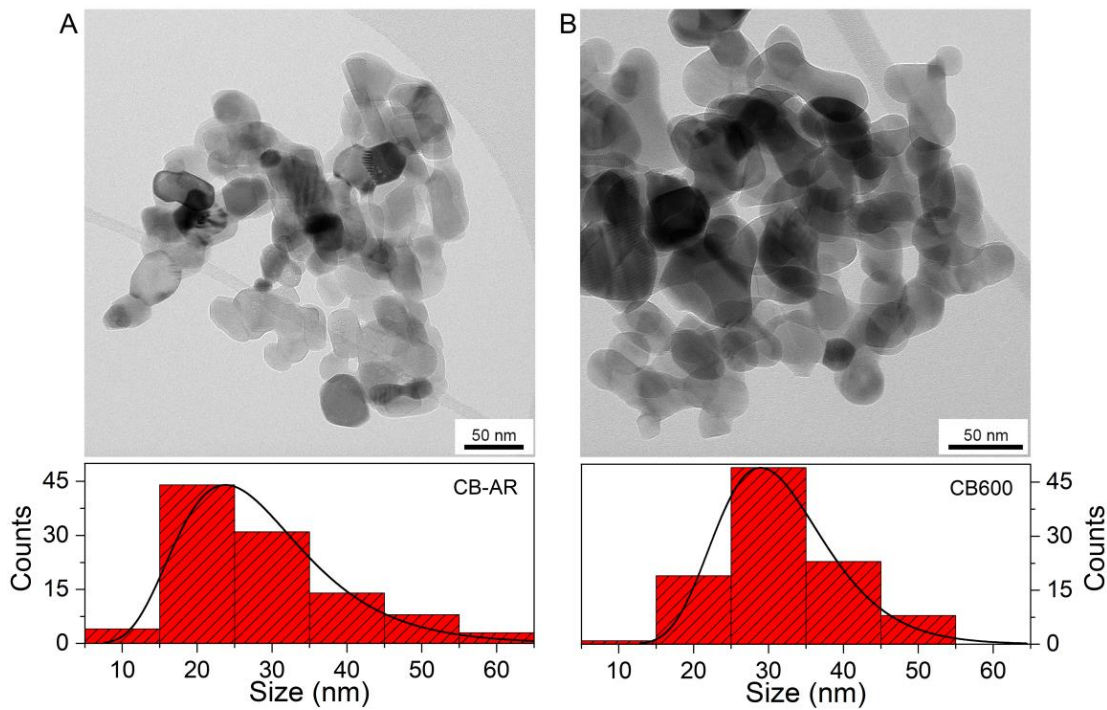

**Figure S2.** TEM images of (A) CB-AR and (B) CB600 with associated histograms of size distributions (bottom) based on 100 measurements of particle diameter.

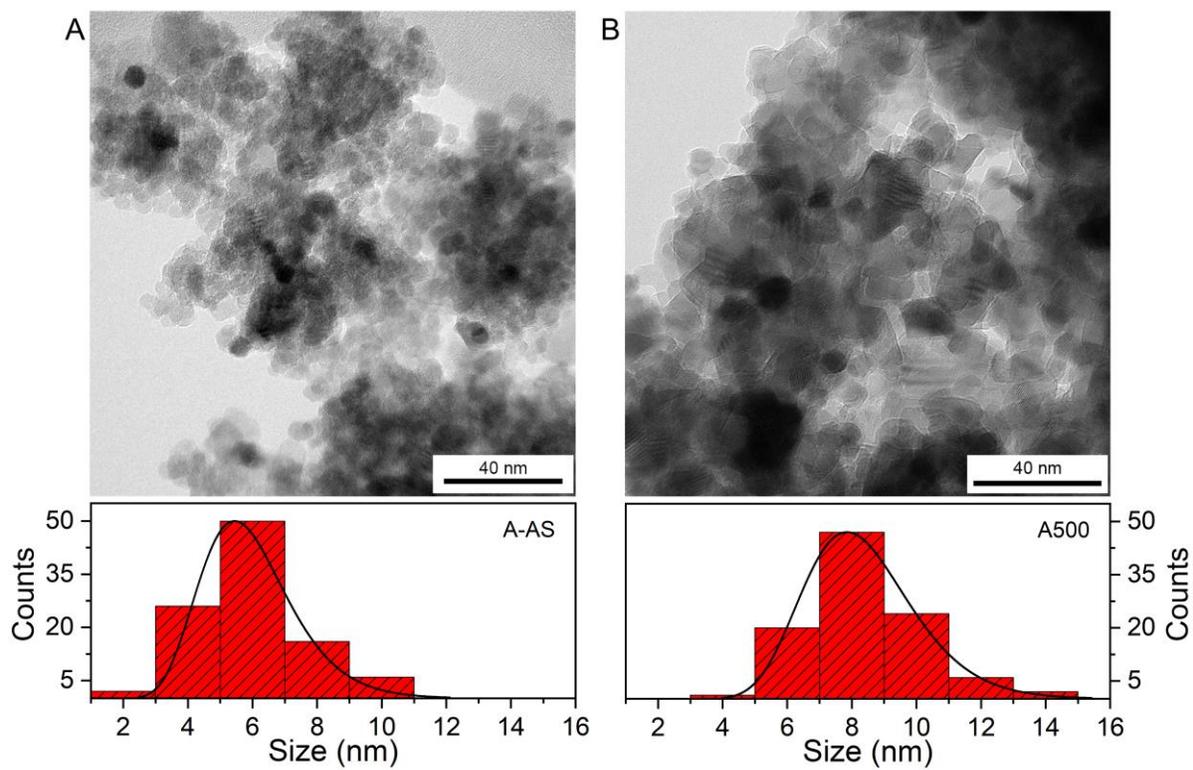

**Figure S3.** TEM images of (A) A-AS and (B) A500 with associated histograms of size distributions (bottom) based on 100 measurements of particle diameter.

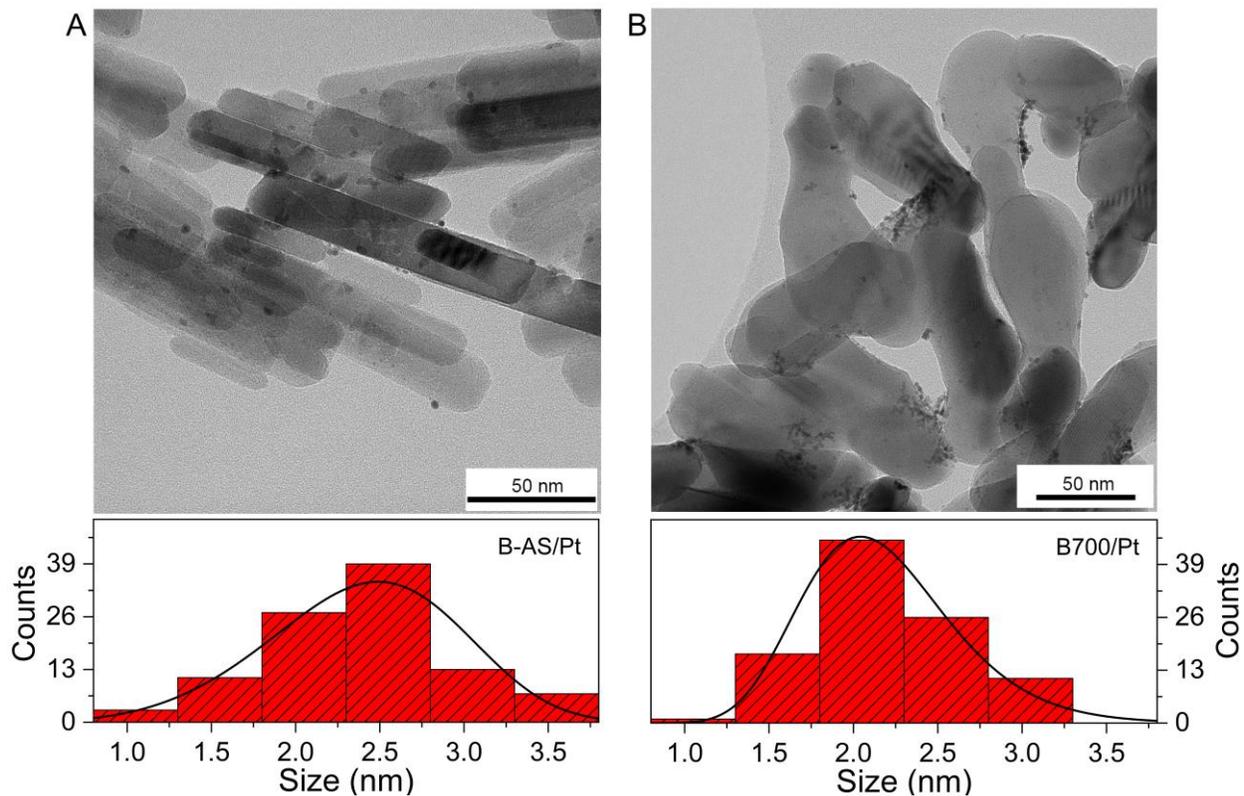

**Figure S4.** TEM images of (A) platinized pristine brookite and (B) platinized reduced brookite at 700°C with associated histograms of size distributions of Pt particles (bottom) based on 100 measurements of particle diameter. The loaded reduced brookite present large Pt aggregates and therefore the size distribution refers to the isolated Pt nanoparticles found in the sample. The photodeposition of Pt results clearly in different results: pristine brookite present isolated homogeneously dispersed Pt nanoparticles, while reduced brookite show mainly large Pt aggregates and some isolated Pt nanoparticles.

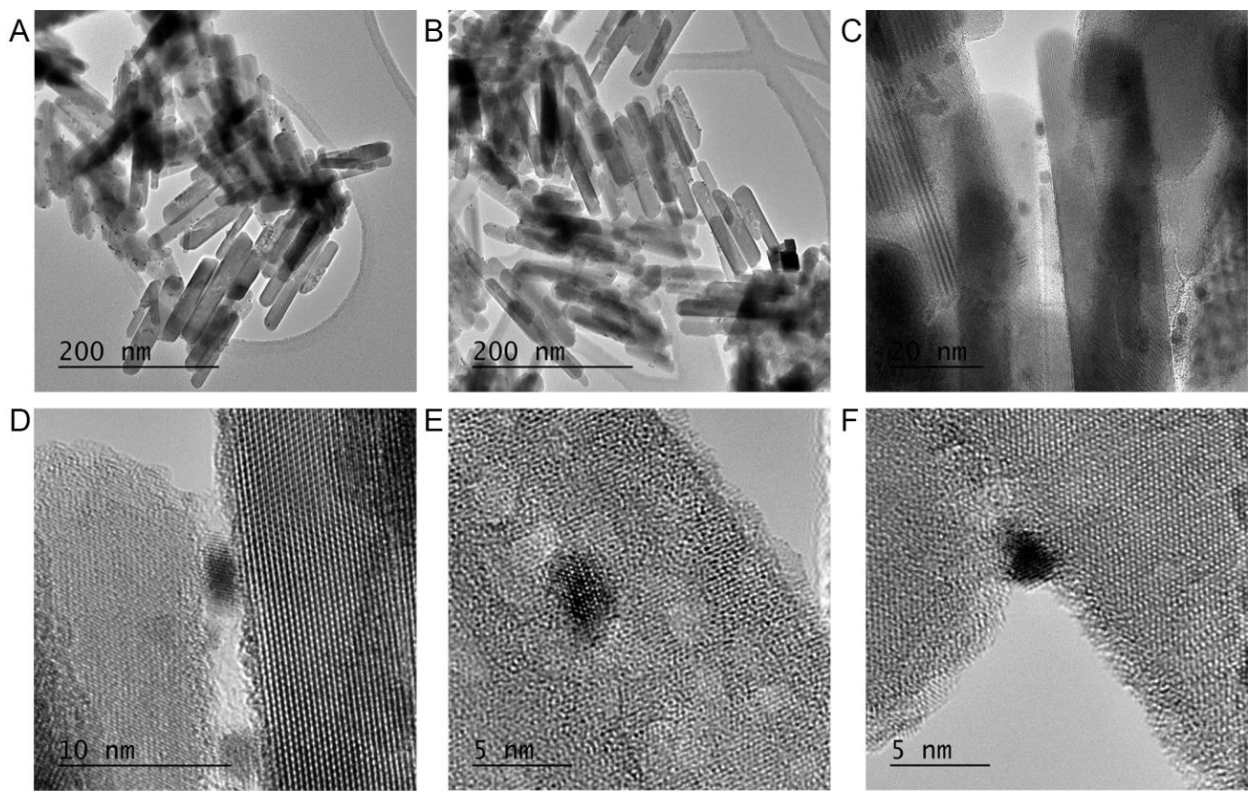

**Figure S5.** (A,B) TEM and HR-TEM (C-F) micrographs of pristine brookite nanorods loaded with 1 wt% Pt showing their homogeneous distribution over the $TiO_2$ matrix. The darker dots are the Pt nanoparticles.

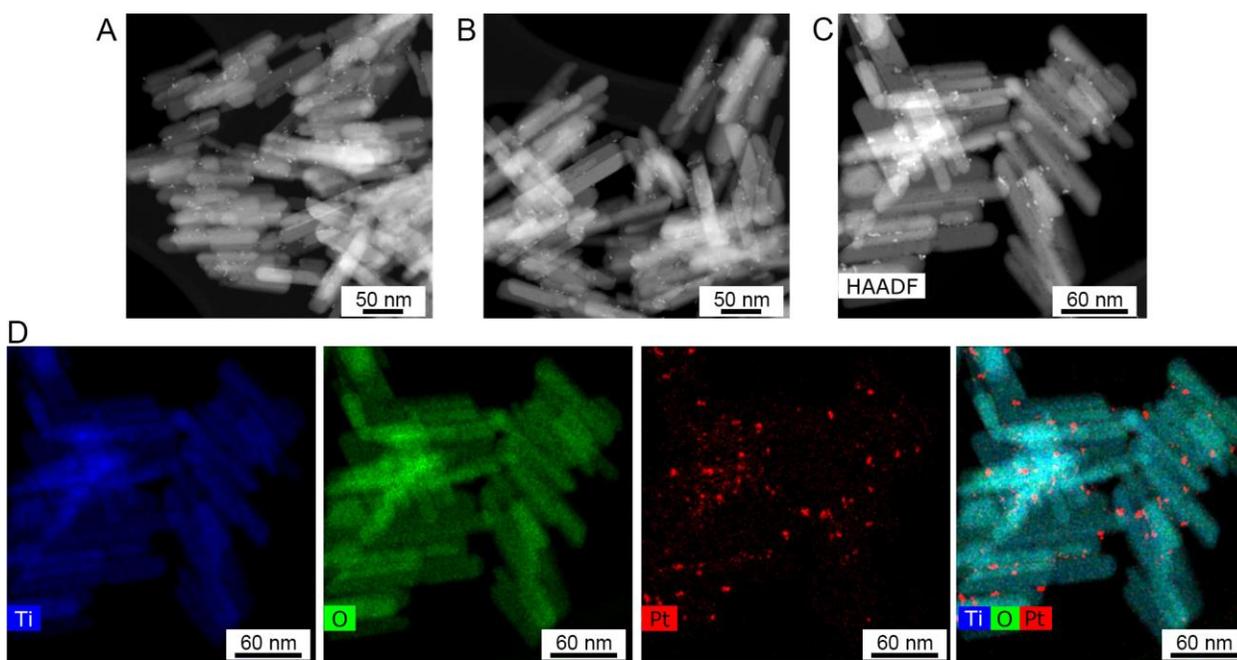

**Figure S6.** (A-C) STEM-HAADF images and (D) elemental EDS mapping of (C) for pristine brookite nanorods loaded with 1 wt% Pt showing their homogeneous distribution over the $TiO_2$ matrix.

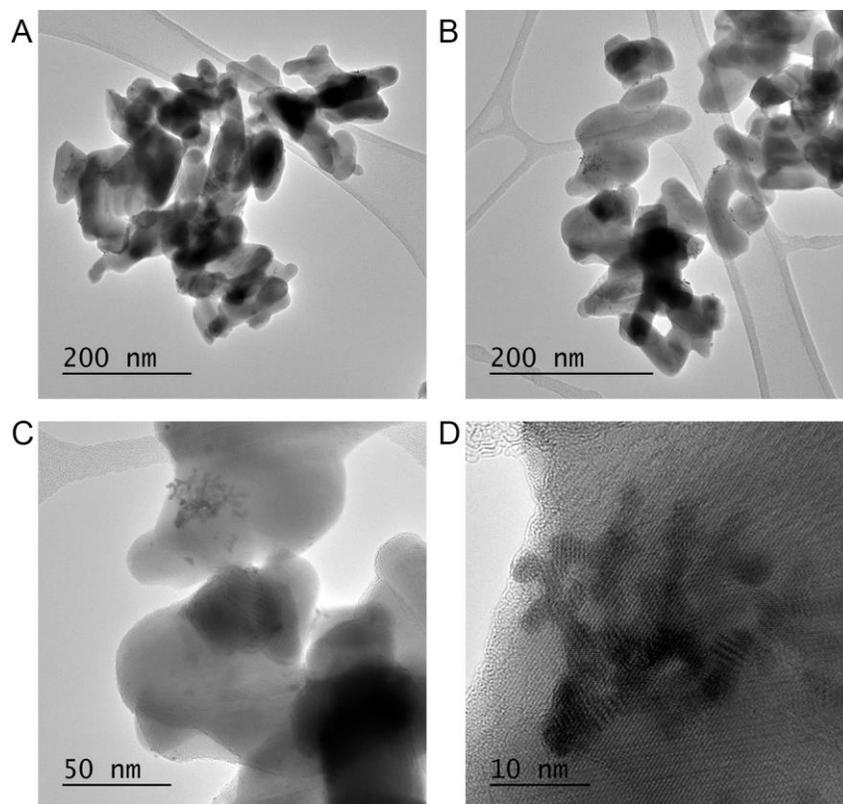

**Figure S7.** (A-C) TEM and HR-TEM (D) of micrographs reduced brookite nanorods (at 700°C) loaded with 1 wt% Pt showing their aggregation over the $TiO_2$ matrix. The darker dots and nanostructured aggregates are made by Pt.

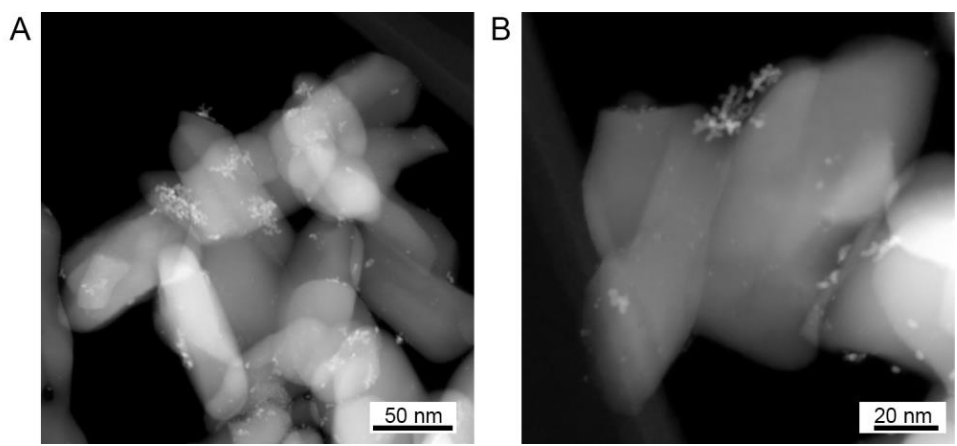

**Figure S8.** (A-C) TEM and HR-TEM (D) of micrographs reduced brookite nanorods (at 700°C) loaded with 1 wt% Pt showing their aggregation over the $TiO_2$ matrix. The Pt nanocrystals are the brighter spots.

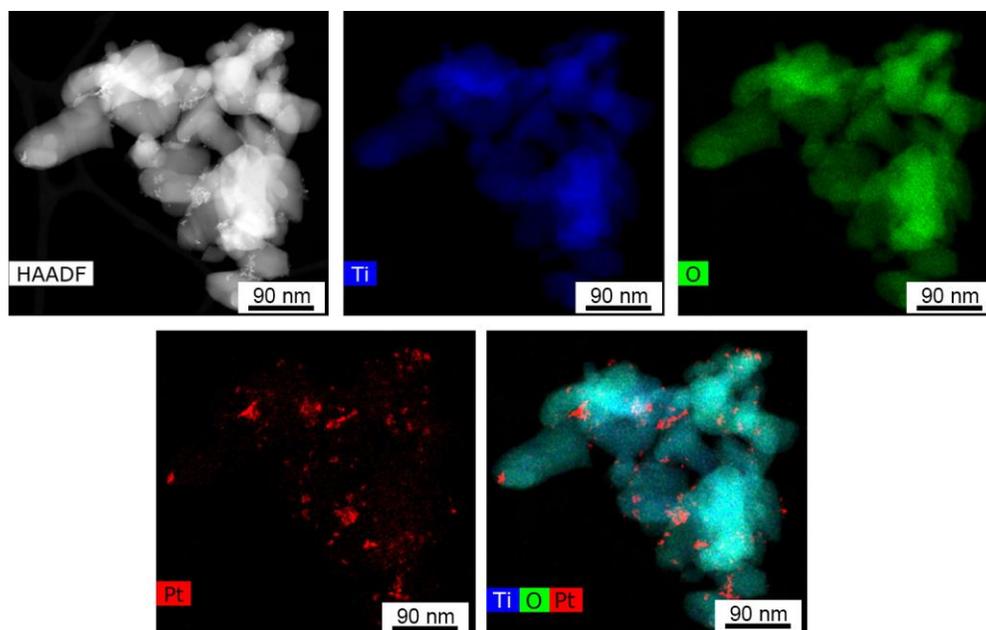

**Figure S9.** STEM-HAADF and elemental mapping images of reduced brookite nanorods (at 700°C) loaded with 1 wt% Pt showing their aggregation over the $TiO_2$ matrix.

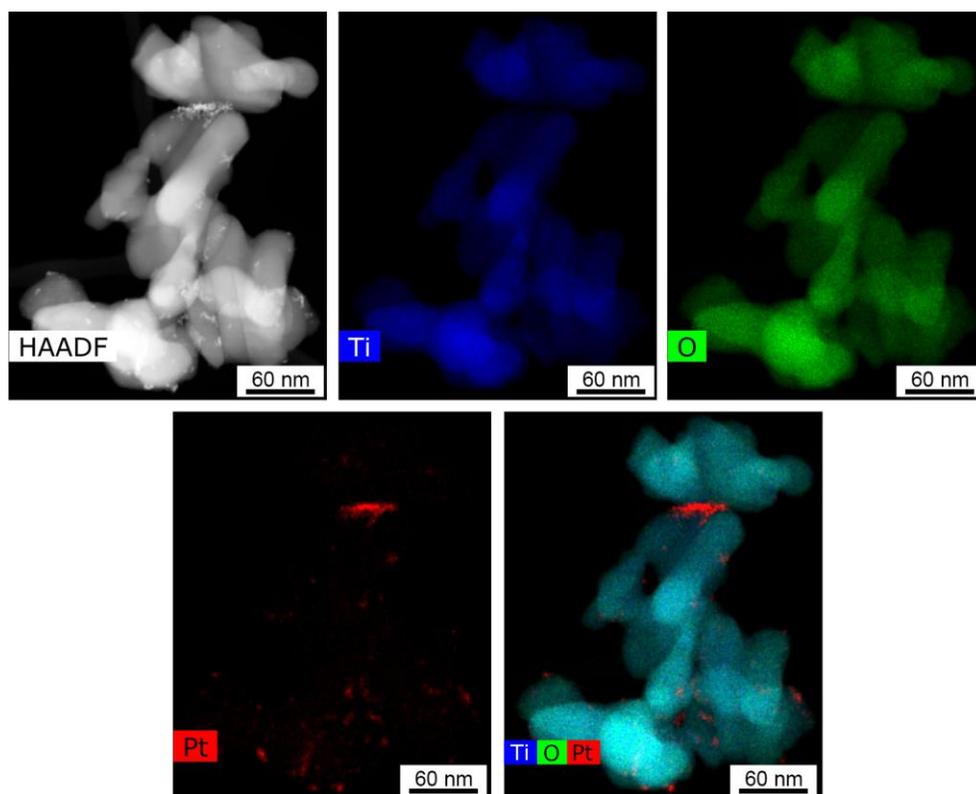

**Figure S10.** STEM-HAADF and elemental mapping images of reduced brookite nanorods (at 700°C) loaded with 1 wt% Pt showing their aggregation over the $TiO_2$ matrix.

Specific surface area measurement

Figure S11 shows the nitrogen sorption isotherm profiles of the samples. In all cases, the BET surface area decreases after reduction under hydrogen atmosphere at high temperature (Table S3).

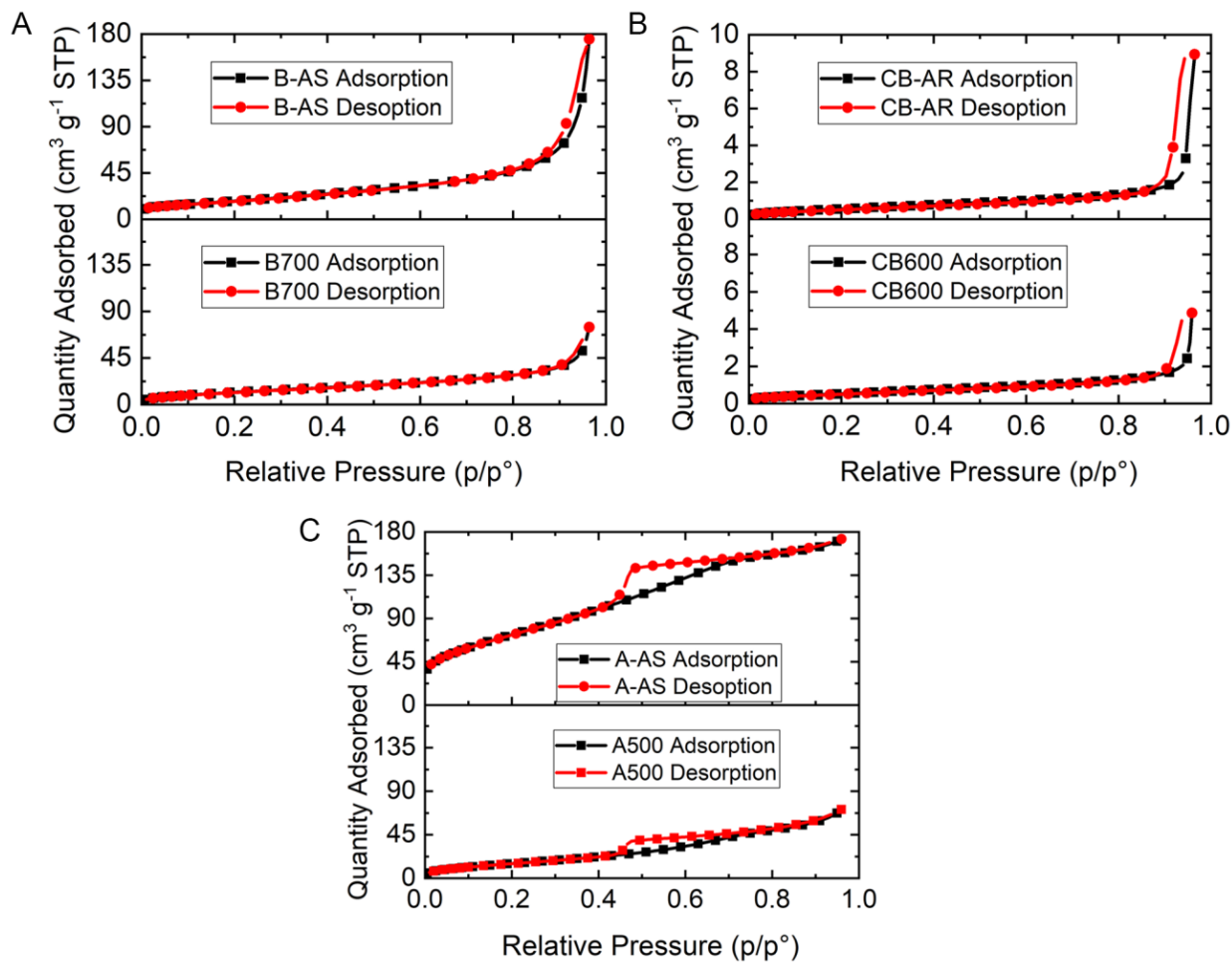

**Figure S11.** $N_2$ adsorption/desorption type IV isotherms (mesoporous solids) at 77K for (A) B-AS and B700, (B) CB-AR and CB600, and (C) A-AS and A500.

**Table S3.** Brunauer−Emmett−Teller (BET) specific surface area for the pristine and the most photoactive sample of each studied phases of $TiO_2$.

| Sample | BET surface area ($m^2\ g^{-1}$) |
|---|---|
| anataseB-AS | 67 |
| B700 | 47 |
| CB-AR | 50 |
| CB600 | 48 |
| A-AS | 273 |
| A-500 | 189 |

*Structural characterization*

*X-ray diffraction*

XRD analysis (Figure S12 and Table S4) revealed that the synthesized brookite is crystalline and stable up to 700°C. The sample treated at 800°C, instead, presented around 91 wt. % of rutile phase, while at 900°C almost all the rutile is transformed to Magnéli phase $Ti_9O_{17}$. At 1000°C, the other Magnéli phases $Ti_4O_7$ in 94 wt. % and $Ti_5O_9$ in 6 wt. % were generated.

The phase stability of commercial brookite is up to 700°C (the same as synthesized brookite) and after that reduction at 800°C induced the complete conversion to rutile (Figure S13 and Table S5). The XRD pattern of as synthesized and reduced anatase (Figure S14) samples indicates that the as synthesized anatase photocatalysts are stable at up to 500°C , while almost full conversion to rutile is obtained at 700°C (Table S6). The average particle sizes obtained using the Scherrer method for commercial brookite and anatase samples are in good agreement with those retrieved from TEM micrographs analysis.

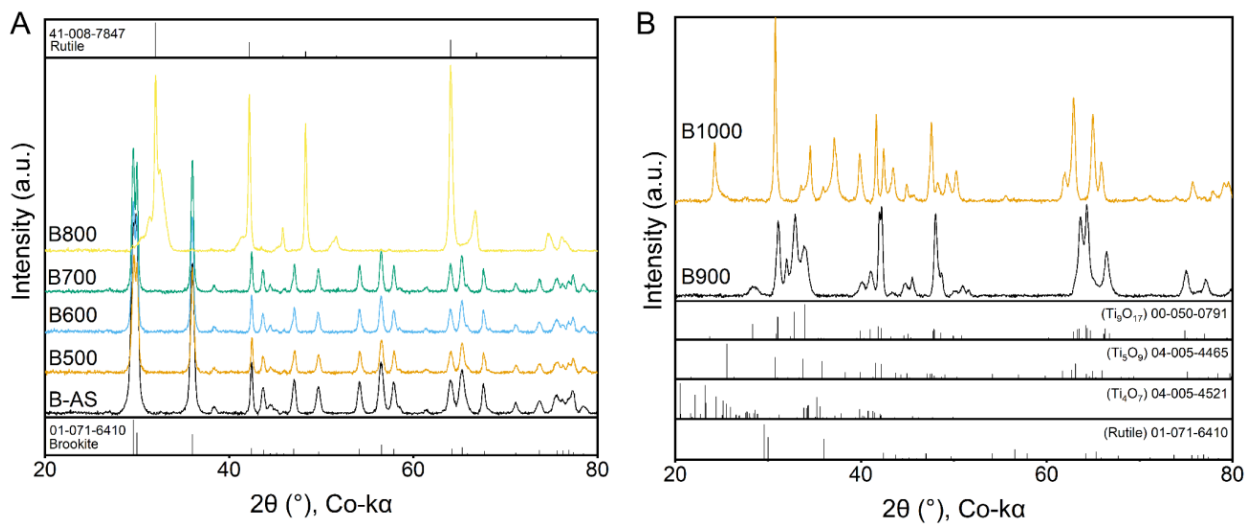

**Figure S12.** (A) XRD patterns of brookite samples and reference patterns for brookite (bottom) and rutile (top). (B) XRD patterns for reduced brookite at 900 and 1000 °C and reference patterns (bottom) for rutile and various magnéli phases ($Ti_xO_{2x-1}$).

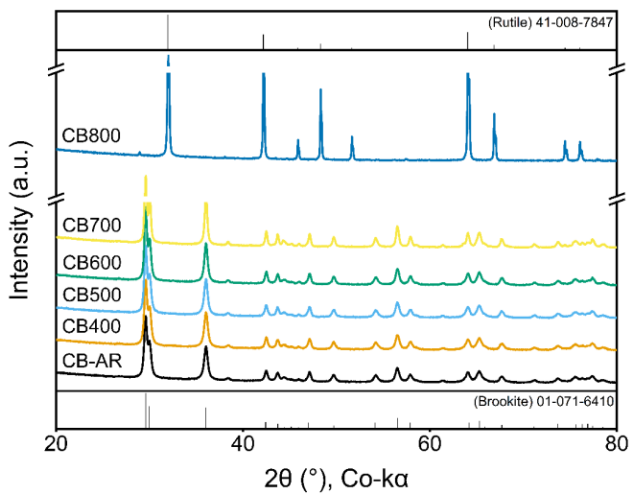

**Figure S13.** XRD patterns for commercial brookite samples and reference patterns for brookite (bottom) and rutile (top).

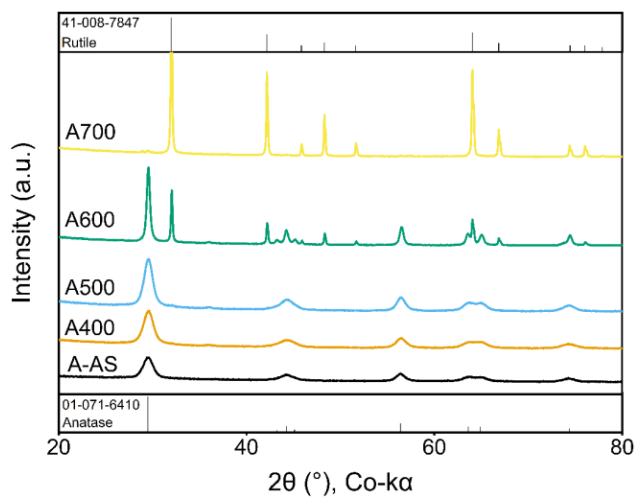

**Figure S14**. XRD patterns for anatase samples and reference patterns for anatase (bottom) and rutile (top).

**Table S4**. Phase composition of brookite samples retrieved from Rietveld refinement of XRD patterns.

| Sample | Brookite (wt.%) | Rutile (wt.%) | $Ti_4O_7$ (wt.%) | $Ti_5O_9$ (wt.%) | $Ti_9O_{17}$ (wt.%) |
|---|---|---|---|---|---|
| B-AS | 100 | - | - | - | - |
| B500 | 100 | - | - | - | - |
| B600 | 100 | - | - | - | - |
| B700 | 100 | - | - | - | - |
| B800 | 9 | 91 | - | - | - |
| B900 | - | 9 | - | - | 91 |
| B1000 | - | - | 94 | 6 | - |

**Table S5.** Rietveld quantitative phase analysis of commercial brookite samples and crystallite size calculated according to the Scherrer method.

| Sample | Brookite (wt.%) | Anatase (wt.%) | Rutile (wt.%) | Crystallite size (nm) |
|---|---|---|---|---|
| CB-AR | 100 | - | - | 27 |
| CB400 | 100 | - | - | 27 |
| CB500 | 100 | - | - | 28 |
| CB600 | 100 | - | - | 31 |
| CB700 | 100 | - | - | 41 |
| CB800 | - | - | 100 | 69 |

**Table S6.** Rietveld quantitative phase analysis of anatase samples and crystallite size calculated according to the Scherrer method.

| Sample | Anatase (wt.%) | Rutile (wt.%) | Crystallite size (nm) |
|---|---|---|---|
| A-AS | 100 | - | 8 |
| A400 | 100 | - | 8 |
| A500 | 100 | - | 13 |
| A600 | 68 | 32 | 24 |
| A700 | - | 100 | 28 |

To investigate more in depth the structural modifications induced by the reduction treatment, we performed detailed Rietveld refinements for the samples treated up to 700°C (Figure S15). All of the patterns are well described by single phase of brookite $TiO_2$ and, within the resolution of our measurements, we did not detect any spurious crystalline phases.

Figure S1A and B shows the reduction temperature (T) dependence of the average particle size, $D$ (nm), and the average strain as obtained using the Williamson-Hall method. We observed a clear decrease of D for T = 400°C followed by an increase of average particle size above this T. On the other hand, the average particle strain weakly decreases at T=400°C keeping roughly the same value at higher T. This phenomenon can be associated to particle sintering since the sudden aggregation of particle can result in an increase of $D$ owing to relieve of intraparticle tension by decreasing the overall particle strain. This is confirmed by the annealing temperature evolution of component of empirical extension of anisotropic microstrain broadening tensor refined by Rietveld refinements (Figure S1C). Indeed, we note a progressive convergence of these parameters to the same values indicating that the evolution of Williamson-Hall strain is explained by the tendency to form more isotropic particle aggregates at high annealing temperature. This morphological observation is in agreement with the TEM micrographs of reduced brookite B700 (Figure S1B), which confirm that brookite nanorods tends to aggregate in isotropic agglomerate upon reduction. The tendency toward powder aggregation can also explain the disagreement between the average nanorod lengths detected by TEM, both before and after reduction, and the average particle size retrieved by Rietveld refinements.

Figure S1D shows the annealing temperature evolution of the refined lattice parameters. Orthorhombic strain is defined as:

$$\eta = \frac{2(a - b)}{(a + b)}$$

where $a$ and $b$ are the unit cell axes. We observed that $\eta$ remains almost unchanged for all the investigated annealing temperature whereas the c-axis shows clear decrease. In other words, this indicates that the annealing temperature does not affect the *ab* plane but induces distortion along the c-axis only. This anisotropic evolution of structural parameters upon reduction is even better outlined by the interatomic distances related to the $TiO_6$ octahedra. $TiO_2$ structure is composed of $TiO_6$ octahedra, each with a titanium atom at its center and oxygen atoms at its corners (Figure S1A). In brookite $TiO_6$ are distorted oxygen atoms in two different positions, namely O1, O2. This results in two groups of distances namely $d_{Ti-O1}$, $d_{Ti-O2}$ (Figure S1B, C) which can be regrouped into two axial and four equatorial different interatomic distances which have been averaged out and shown in Figure S1. Averaged axial distance expand upon increasing temperature, whereas the average equatorial distances show a weak contraction. Both distances tend to the same value and this may indicate that the annealing process induces the $TiO_6$ to be more regular, i.e. isotropic, and less distorted at high temperature. To figure out the effect of reduction treatment on the brookite structure we argue that the reduction of $TiO_2$ at high temperature creates oxygen vacancies ($V_O$, in Kröger–Vink notation) [22,23] and induces $Ti^{3+}$ ions electron trapped in $Ti^{4+}$ lattice sites ($Ti'_{Ti}$) according to the following relation:

$$O_O^x + 2Ti_{Ti}^x + H_2 \rightarrow V_O^{\bullet\bullet} + 2Ti'_{Ti} + H_2O$$

Two negatively charged $Ti'_{Ti}$ species can then couple with one double positive charged $V_O^{\bullet\bullet}$ promoting the formation $<Ti'_{Ti} - V_O^{\bullet\bullet} - Ti'_{Ti}>$ defects. The mutual attraction between them can produce a contraction of the structure to account for electron neutrality, which is fully in agreement with the observed c-axis contraction upon reduction at high temperature. The reduction of Ti valence to produce $Ti'_{Ti}$ species, is compatible with a depletion of ≈ - 0.5% as evidenced by bond valence sum (BVS) analysis, see for B700 (Figure S1A). Further, the correlation between the BVS and c-axis contraction is shown in Figure S1B, showing a nearly linear relationship between these two parameters and thus indicating that more $V_O s$ are induced by the reduction and more the c-axis resulted to be contracted. This highlights that reduction treatment introduced an anisotropic and preferential deformation of the brookite lattice along the c-axis.

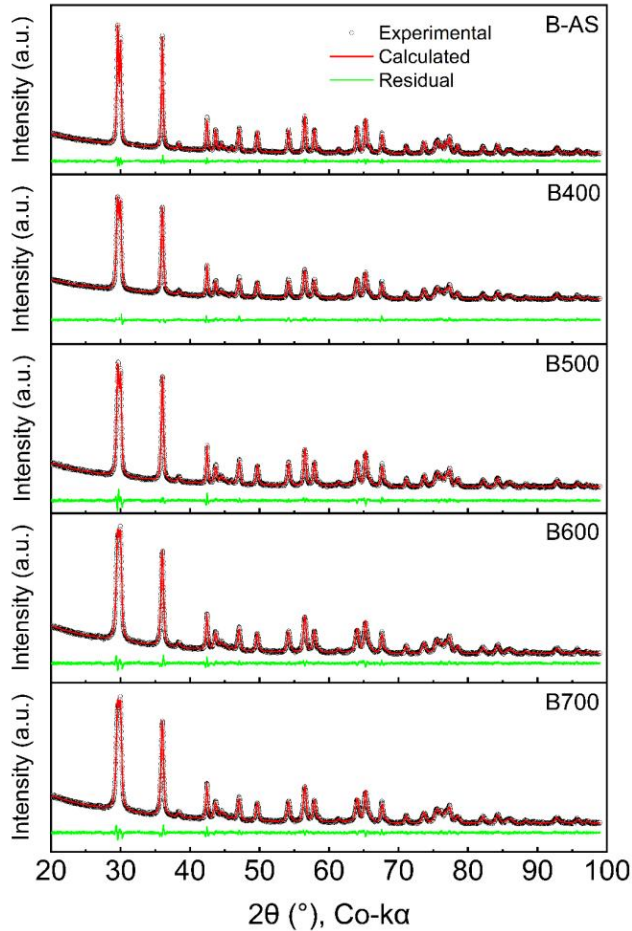

**Figure S15.** Rietveld refinements of brookite samples reduced at different temperatures. Dots are experimental data; continuous lines are the calculated profiles; Rietveld agreement factors [R (F2)] between observed and calculated patterns ranged from 0.06 to 0.08.

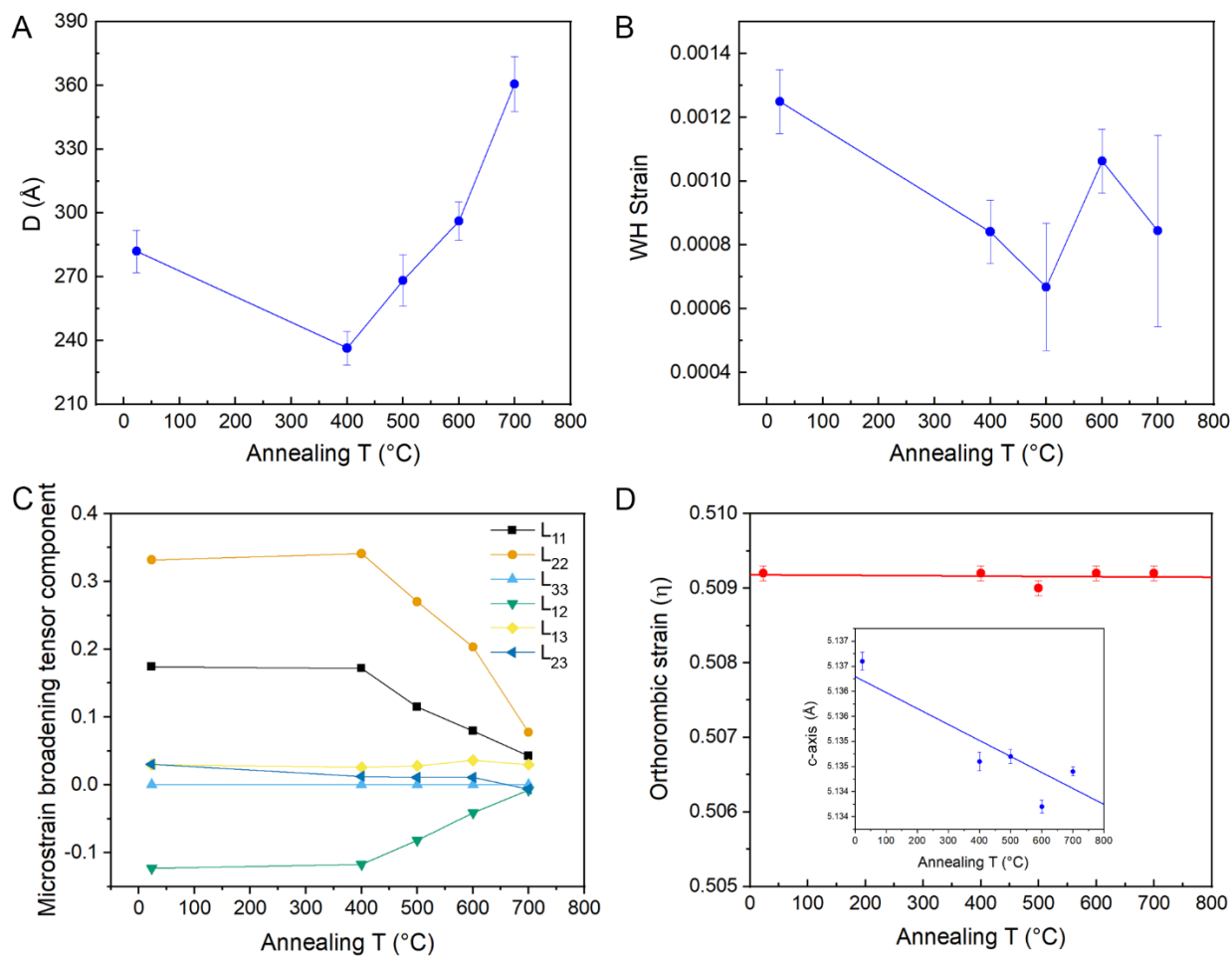

**Figure S16.** Structural parameters retrieved from Rietveld refinements of XRD patterns for as synthesized brookite and brookite reduced at different temperatures (i.e. annealing T in x axis). (A) Reducing temperature dependence of average particle size (D), and (B) crystal structure strain as obtained by Williamson- Hall (WH) method. (C) Annealing temperature evolution of components of empirical extension of anisotropic microstrain broadening tensor (L) refined by Rietveld refinements. Indexes are referring to the following expression: L = $L_{11}h^2+L_{22}k^2+L_{33}l^2+2L_{12}hk+2L_{13}hl+2L_{23}kl$. (D) Annealing temperature dependence of orthorhombic strain (see text) and c-axis (inset).

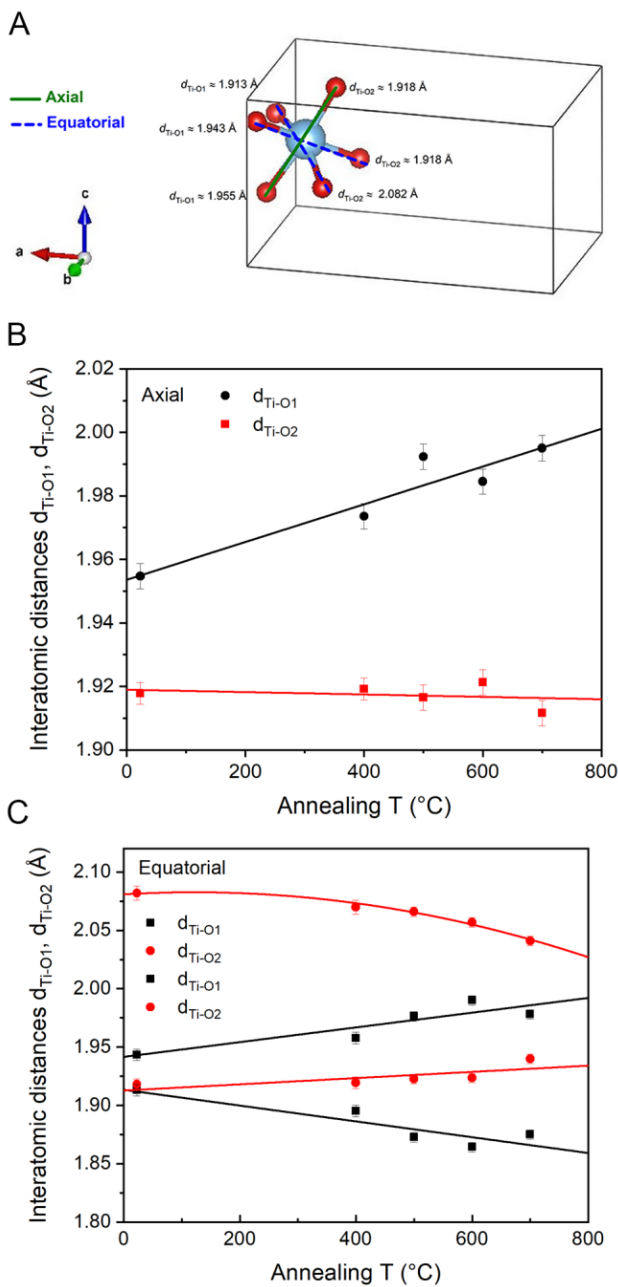

**Figure S17.** (A) Representation of octahedra (TiO$_6$) in TiO$_2$ brookite unit cell (Pbca) showing the arrangement of distorted TiO$_6$ resulting in two groups of distances, namely d$_{Ti-O1}$ and d$_{Ti-O2}$, composed by two axial and four equatorial different interatomic distances. Values refer to structure refined at room temperature. (B) Evolution of axial and (C) equatorial interatomic distances retrieved from Rietveld refinements of XRD patterns for as synthesized brookite and brookite reduced at different temperatures (i.e. annealing T in x axis). Solid lines are guide to the eye.

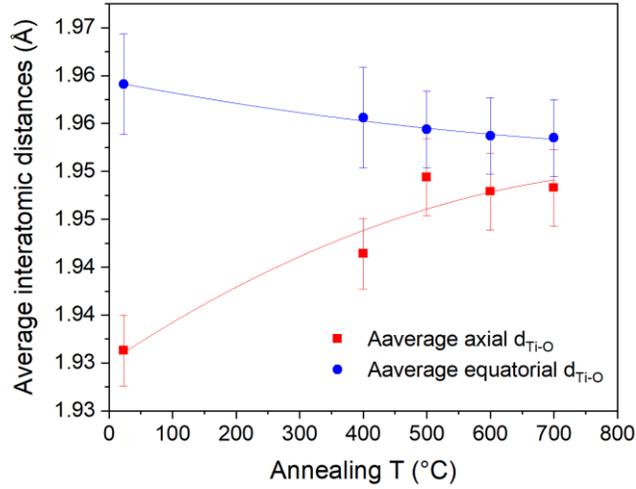

**Figure S18.** Annealing Temperature dependence of averaged axial and equatorial interatomic retrieved from Rietveld refinements of XRD patterns for as synthesized brookite and brookite reduced at different temperatures (i.e. annealing T in x axis).

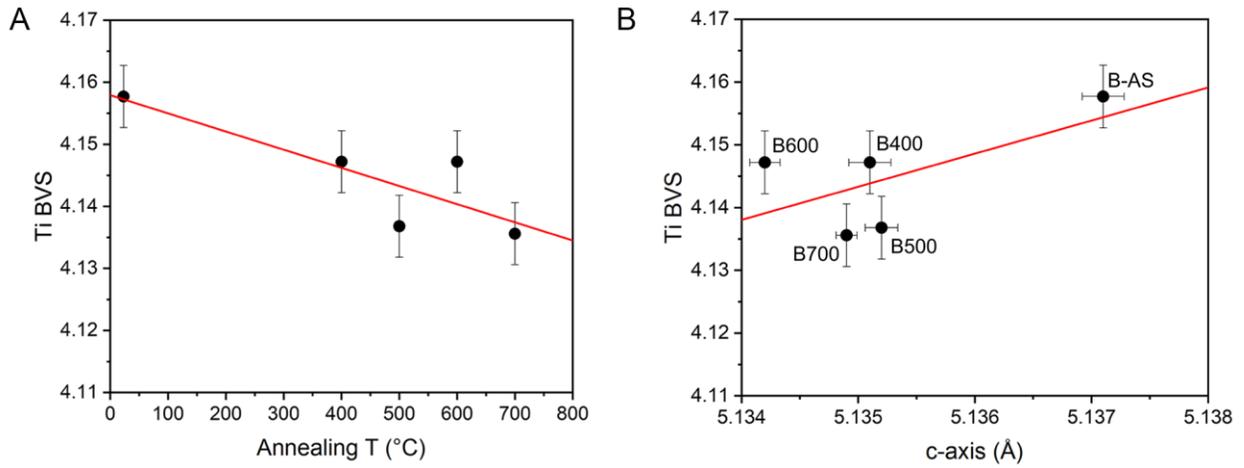

**Figure S19.** (A) Annealing Temperature dependence of Ti Bond Valence Sum (BVS) and (B) correlation between Ti BVS and c-axis contraction retrieved from Rietveld refinements of XRD patterns for as synthesized brookite and brookite reduced at different temperatures (i.e. annealing T in x axis).

*Raman spectroscopy*

The crystal system of brookite is orthorhombic and has eight formula units per unit cell and thirty-six Raman active modes ($9A_{1g}+9B_{1g}+9B_{2g}+9B_{3g}$) [24]. The crystal system of anatase is tetragonal and has two formula units per unit cell and six Raman active modes ($A_{1g}+2B_{1g}+3E_g$) [24]. The crystal system of rutile is tetragonal and has two formula units per unit cell and four Raman active modes ($A_{1g}+B_{1g}+B_{2g}+E_g$) [24].

Raman spectra of synthesized and reduced brookite samples (Figure S20A) confirmed that the synthesized brookite is stable up to 700°C and after that it transforms to rutile, as evidenced by the disappearance of brookite peaks and appearance of rutile peaks in Raman spectra of B700 and B800. The same Raman modes of rutile remained evident also for B900 and B1000, where an additional phase transition to Magnéli phases is completed. The main Raman peak of as synthesized brookite at 147.3 cm$^{-1}$ is blue shifted to 146.0 cm$^{-1}$ after reduction at 700°C. Moreover, there is a peak narrowing in the main peak due to the reduction (FWHM$_{B-AS}$ = 15.82 cm$^{-1}$, FWHM$_{B700}$ = 12.81 cm$^{-1}$).

Raman spectroscopy measurements also revealed that the commercial brookite sample is stable under reducing conditions up to 600°C, after which, rutile Raman fingerprint developed. There is almost no peak shift and no change in FWHM of the main peak (FWHM$_{CB-AR}$ = 16.41 cm$^{-1}$, FWHM$_{CB600}$ = 16.64 cm$^{-1}$) (Figure S20B), suggesting that no significant modifications occurred in the commercial brookite lattice upon reduction at high temperature.

Figure S20C shows, instead, that anatase is stable up to 500°C in our reduction conditions showing also a significant blueshift of the main Raman mode from 142.2 cm$^{-1}$ for as-synthesized anatase to 134.6 cm$^{-1}$ for anatase reduced at 500°C. The same peak narrowed significantly from FWHM$_{A-AS}$ = 30.97 cm$^{-1}$ to FWHM$_{A500}$ = 13.80 cm$^{-1}$.

Several phenomena can give rise to TiO$_2$ Raman peak shifting and broadening (or narrowing) [24–28] as follows: (i) the lattice strain [26], (ii) the crystal size that could regulate the phonon confinement and the Raman scattering (increasing in crystal size results in redshift and peak narrowing) [24], and (iii) the oxygen stoichiometry, depending on the phase of TiO$_2$, could affect the position of Raman peaks and also could modify the FWHM [25]. In the case of synthesized brookite and anatase, we observed a blueshift and peak narrowing of the main Raman peak after reduction. Upon reduction, we detected a reduction of lattice microstrain in brookite nanorods (Figure S20C) as well as an increase in the overall crystal size due to nanorods aggregation. Therefore, we propose that the change in oxygen stoichiometry (as also evidenced by the other characterizations reported) underlies the blue shift and the peak narrowing witnessed for both brookite and anatase [25]. In contrast, in case of commercial brookite neither peak shifting nor peak narrowing (or broadening) were observed. This result confirms that the commercial brookite was not reduced under reduction conditions as already suggested by UV-vis diffuse reflectance spectra analysis, TPR-MS measurements, and also by the color of the samples, which did not change upon reduction treatment (Table S1).

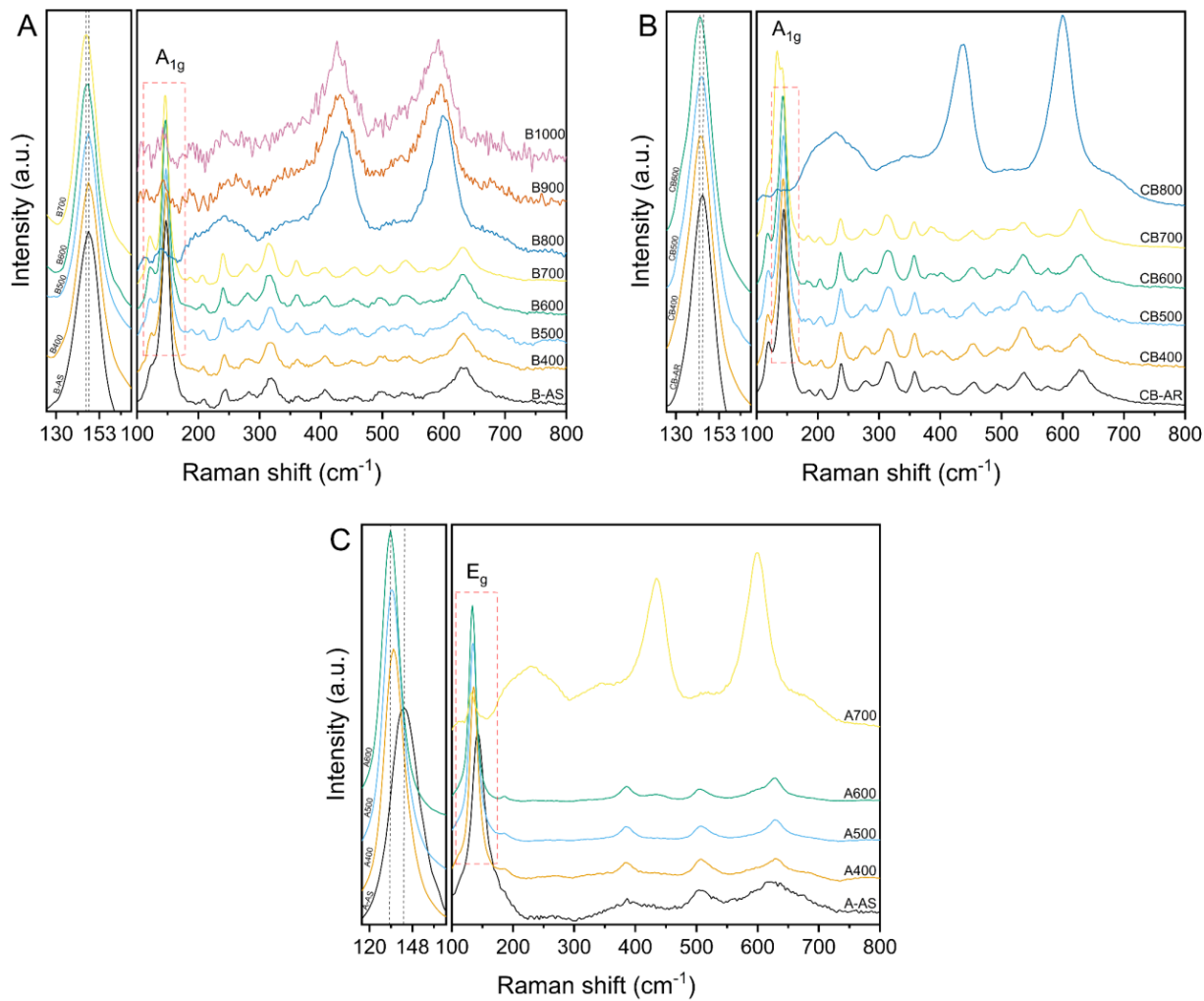

**Figure S20.** Raman spectra of (A) as-synthesized and reduced anisotropic brookite samples, (B) as-received and reduced commercial brookite samples, and (C) as-synthesized and reduced anatase samples. Measurements performed with a 455 nm laser at a power density of 0.1 mW cm$^{-2}$. The left part of each panel is the zoomed view of the main Raman peaks.

*Diffuse reflectance spectroscopy*

The Tauc method [29] is one of the most common procedures for determining the bandgap of materials. This method makes a relationship between absorption coefficient and optical bandgap of material base on the following formula:

$$(\alpha h\nu)^n = (h\nu - E_g)$$

where, $\alpha$ is the absorption coefficient, $h$ is the Plank constant, $\nu$ is the frequency of radiation and $E_g$ is the bandgap of material. The $n$ factor depends on the nature of the electronic transitions and is equal to 2 for $TiO_2$ as it is an indirect band gap semiconductor [30,31]. This method assumes that the scattering component of the reflected irradiation is zero. However, in case of nanopowders with high amount of scattering, this component could not be neglected. Therefore, the Kubelka-Munk theory is used to make an estimation of absorption from reflectance according to the following formula [32]:

$$F(R) = \alpha = \frac{K}{S} = \frac{(1-R)^2}{2R}$$

where, $R$ is the reflectance of the sample with infinite thickness to avoid any contribution of substrate, $K$ and $S$ are the absorption and scattering coefficients, respectively. The bandgap of $TiO_2$ can be retrieved using the Tauc method as follows:

$$(F(R)h\nu)^{1/2} = (h\nu - E_g)$$

From the Tauc plot, the x-axis intersection point of the tangent to the linear increase of light absorption in the Tauc plot gives the band gap energy. This method is accurate and used here for determining the bandgap energy of pristine $TiO_2$ materials, while for reduced $TiO_2$ powders the baseline method was employed to calculate the bandgap [32].

Furthermore, we analyzed the Urbach tail in absorption spectra, as it originates optical transitions involving intragap states related to defects [33–35]. The Urbach energy can be calculated as follow [36,37]:

$$\alpha = \alpha_0 + \exp\left(\frac{E}{E_u}\right)$$

where $\alpha$ is the absorption coefficient, $E$ is the photon energy equal to $h\nu$ and $E_u$ is the Urbach energy, which can be retrieved by the reciprocal of the slope of the linear part of the curve.

Figure S21 shows the absorption spectra of as synthesized and reduced anisotropic brookite samples. As expected, most of the UV light is absorbed owing to the wide band gap of brookite, which was found to be comprised between 3.32 and 3.36 eV (Table S7) for all samples containing only the brookite phase (B-AS, B500, B600, B700). A large shift in the absorption edge of B800 and B900 was observed due to the phase change to rutile, with a corresponding decrease of bandgap values around 3.1 eV. The sample B1000 absorbs almost the entire spectrum and it is not possible to calculate any bandgap. This can be ascribed to a phase composition including several semimetallic (Magnéli) phases [38]. The absorption of anisotropic brookite samples in the visible region increases by increasing the temperature of reduction, which is predictable from the color of powders, and it can be assigned to the introduction of oxygen vacancies and $Ti^{3+}$ electronic states. However, visible light has no influence in the photocatalytic activity of the reduced brookite, as discussed in photocatalysis section reported above. Furthermore, in all of the samples, a change in the Urbach tail due to the reduction at different temperatures was observed. It varied from 69 to 115 meV passing from B-AS to B700, suggesting an increased population of point defects in $TiO_2$.

Reflectance spectra for commercial brookite and synthesized anatase (with isotropic crystal shape) are reported in Figure S22 and Figure S23. In addition, Table S8 and Table S9 provide the results obtained from analysis of bandgap and Urbach energy of commercial brookite and anatase, respectively. It is apparent from the results that in all cases the bandgap underwent to only a slightly modification after reduction, remaining around 3.3 and 3.2, for commercial brookite and anatase, respectively. After phase transformation to rutile (at 800°C in commercial brookite and at 600°C in anatase), bandgap values decreased and Urbach energy increased. Furthermore, no change in color of the commercial brookite samples was observed after reduction at different temperatures (Table S1), while in anatase samples the color change from white to gray and black. The variations of Urbach energy is in agreement with this observation: for commercial brookite it remained at a stable value of ~55 meV before and after reduction, while for anatase Urbach energy increased from 115 (A-AS) to 205 meV (A500). These results further confirmed that commercial brookite was hardly reducible, while defects could be introduced in the synthesized anatase.

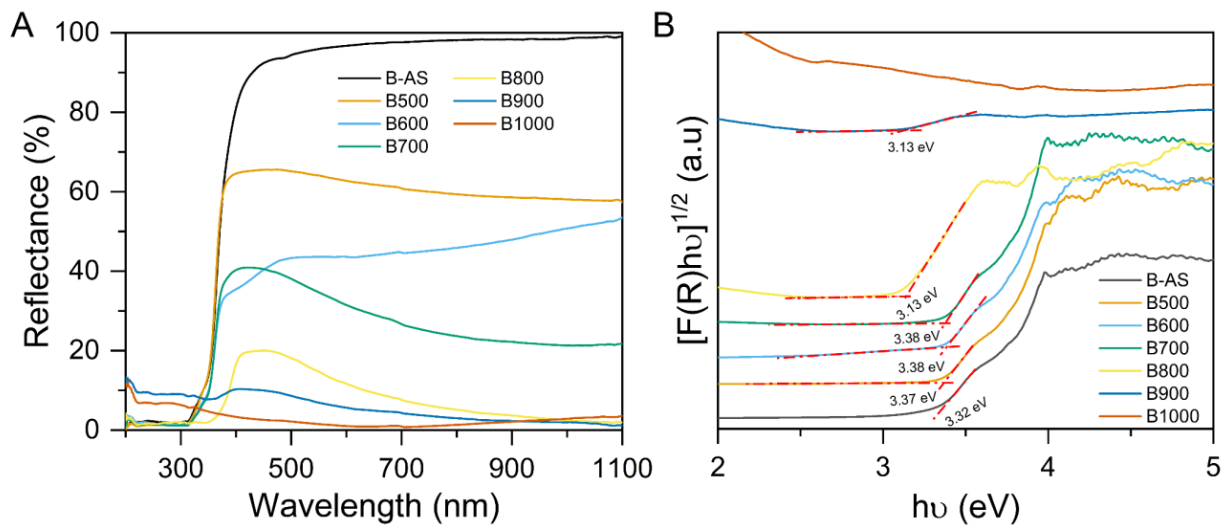

**Figure S21.** (A) Diffusive reflectance spectra and (B) Tauc plots of as synthesized and reduced anisotropic brookite samples.

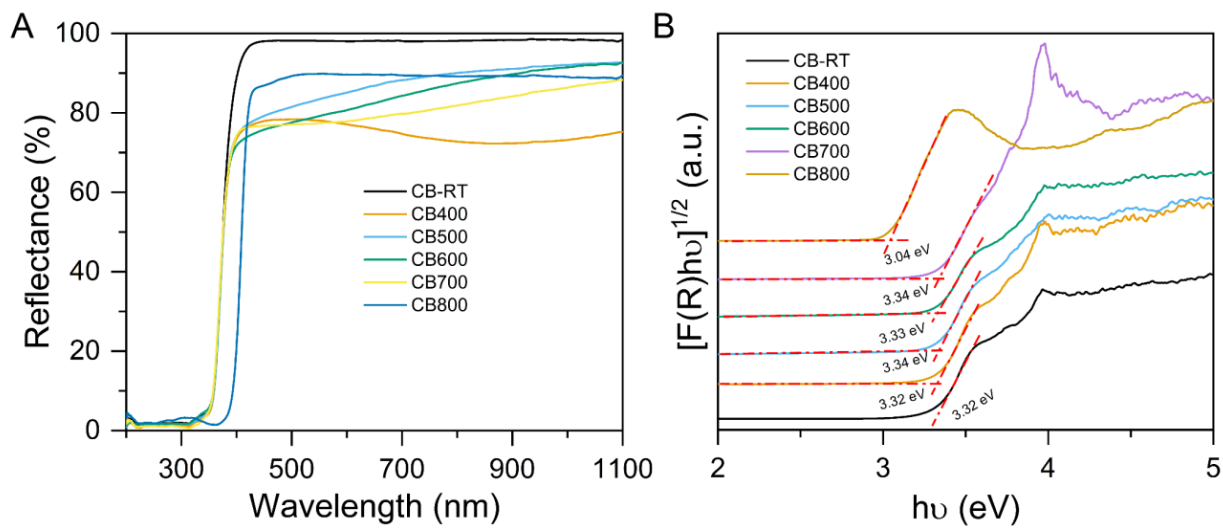

**Figure S22.** (A) Diffusive reflectance spectra and (B) Tauc plots of as received and reduced commercial brookite samples.

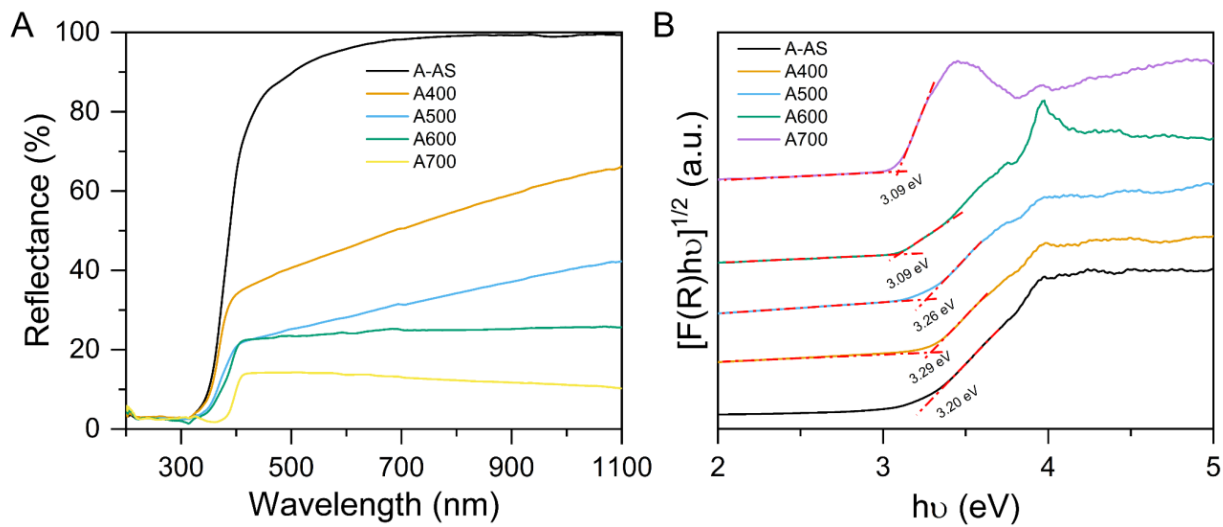

**Figure S23.** (A) Diffusive reflectance spectra and (B) Tauc plots of as synthesized and reduced anatase samples.

**Table S7.** Optical bandgap and Urbach energy of anisotropic brookite samples.

| Sample | Optical bandgap (eV) | Urbach energy (meV) |
|---|---|---|
| B-AS | 3.32 | 69 |
| B500 | 3.37 | 67 |
| B600 | 3.38 | 104 |
| B700 | 3.38 | 115 |
| B800 | 3.13 | 155 |
| B900 | 3.13 | - |
| B1000 | - | - |

**Table S8.** Optical bandgap and Urbach energy of commercial brookite samples.

| Sample | Optical bandgap (eV) | Urbach energy (meV) |
|---|---|---|
| CB-AR | 3.32 | 58 |
| CB400 | 3.32 | 55 |
| CB500 | 3.34 | 54 |
| CB600 | 3.32 | 56 |
| CB700 | 3.34 | 39 |
| CB800 | 3.04 | 36 |

**Table S9**. Optical bandgap and Urbach energy of anatase samples.

| Sample | Optical bandgap (eV) | Urbach energy (meV) |
|--------|----------------------|---------------------|
| A-AS   | 3.20                 | 115                 |
| A400   | 3.29                 | 154                 |
| A500   | 3.26                 | 205                 |
| A600   | 3.09                 | 252                 |
| A700   | 3.09                 | 100                 |

*Methanol photoreforming*

In order to demonstrate the better photo-oxidation acitivty toward methanol of reduced brookite, we performed a series of photocatalytic experiments detecting (1) the methanol consumption rate through $^1$H-NMR spectroscopy and (2) the hydrogen evolution rate with GC for platinized B-AS and B700 (Figure SS24-S25 and Figure 1 of the main text).

We did not detect any trace of formaldehyde nor formic acid in both liquid phase (through NMR analysis) and gas phase (through GC analysis), suggesting that methanol oxidation proceeded toward $CO_2$, as confimed by the high reactivity of brookite samples toward the oxidation of formaldehyde and formic acid solutions (Figure S27).

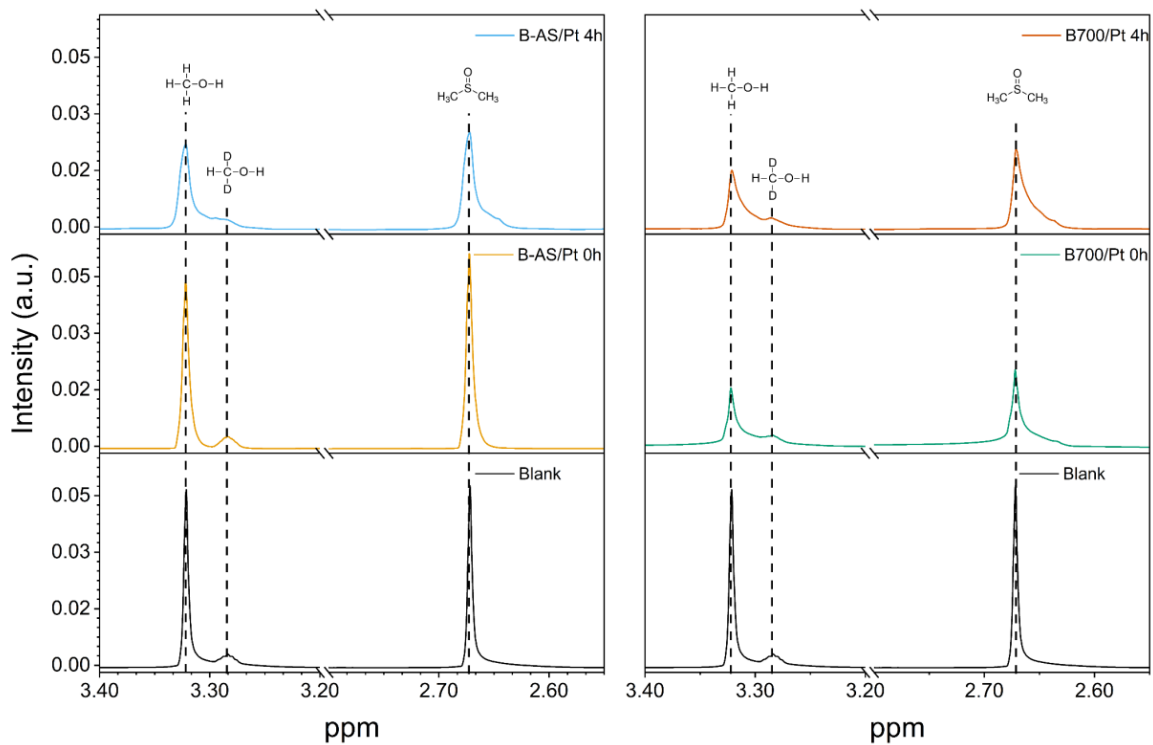

**Figure S24.** $^1$H-NMR spectra of the solutions after photocatalysis at representative reaction times for B-AS/Pt and B700/Pt samples. The blank samples were recorded by using solutions containing $CD_3OD$, DI water, and DMSO (($CH_3$)$_2$SO) as the internal standard. In the other samples, an aliquot of 50 μL of $CH_3OH$ was added to the solution and its photocatalytic consumption rate was measured by assessing the area of its characteristic NMR peak. $CHD_2OH$ is the impurity presents in the deuterated methanol.

**Table S10**. Relative NMR integration values of MeOH signal to internal standard (DMSO) at different reaction times for the pristine (B-AS/Pt) and reduced brookite (B700/Pt) samples.

| Time (h) | B-AS/Pt | B700/Pt |
|---|---|---|
| -[a] | 1.04[a] | 0.92[a] |
| 0[b] | 1.03[b] | 0.92[b] |
| 2 | 1.03 | 0.88 |
| 4 | 1.02 | 0.85 |
| 6 | 1.02 | 0.83 |
| 8 | 1.01 | 0.83 |
| 10 | 1.02 | 0.82 |
| 16 | 0.99 | 0.78 |
| 24 | 0.96 | 0.75 |

[a]Blank measurements with no photocatalyst, but in the presence of the reagents *d4*-MeOH (5 mL), $H_2O$ (5 mL), MeOH (50 μL), and DMSO (2 μL).
[b]Results were obtained after addition of the photocatalysts (B-AS-Pt and B700-Pt) to the solutions and stirring for 30 min in the absence of light to allow for reaching the adsorption/desorption equilibrium.

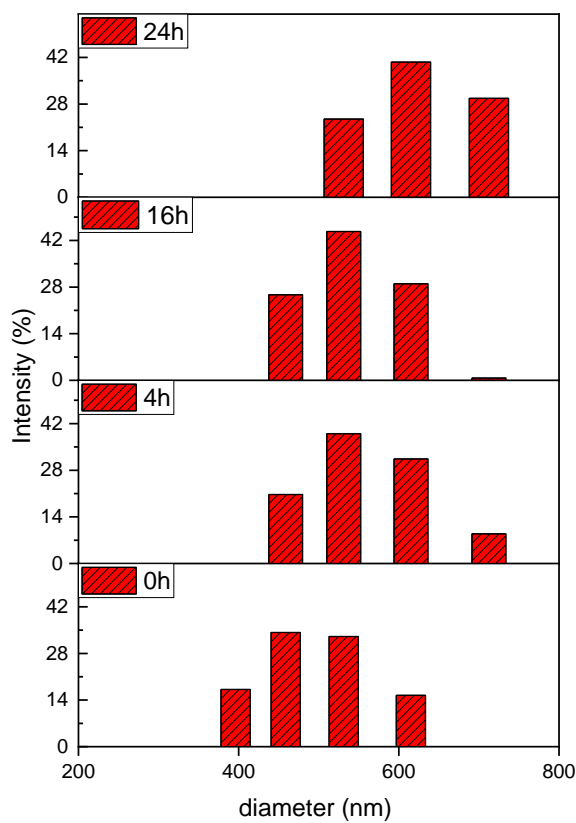

**Figure S25.** Dynamic Light Scattering (DLS) measurements analysis showing the hydrodynamic diameter evolution of the B700/Pt catalysts during the reaction.

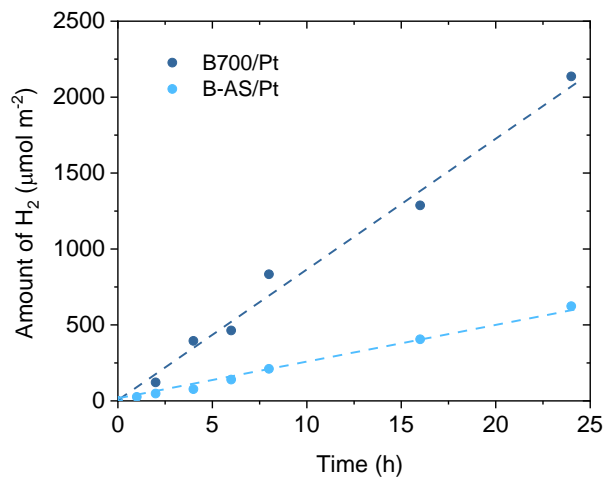

**Figure S26.** Hydrogen evolution kinetics for B-AS/Pt and B700/Pt under one sun illumination.

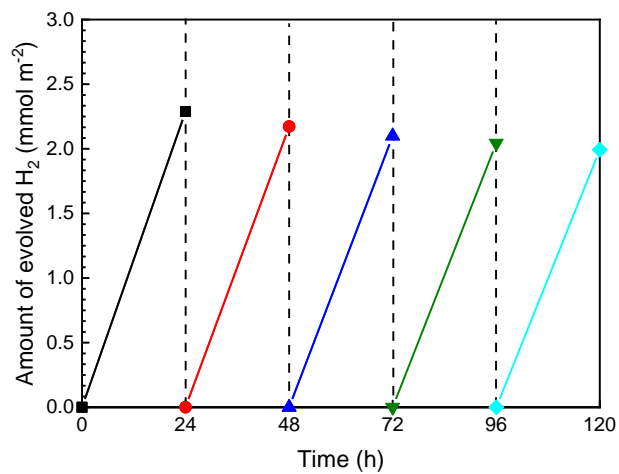

**Figure S27**. Hydrogen amount evolved from five consecutive photocatalytic cycles for B700/Pt during methanol photoreforming under 1 Sun illumination.

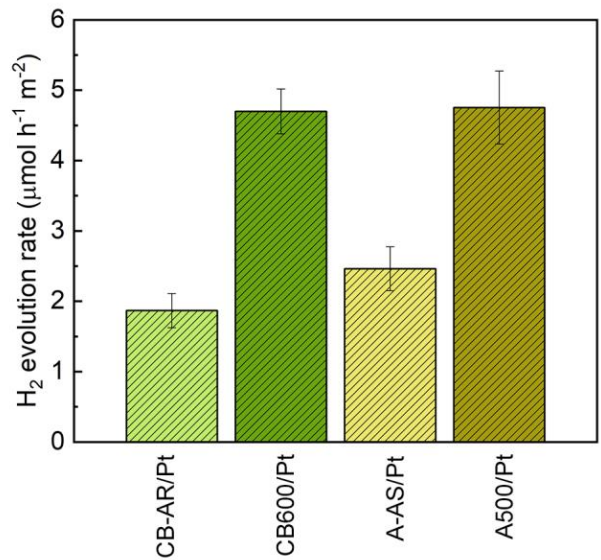

**Figure S28.** Specific hydrogen evolution rate obtained for samples loaded with 1 wt.% Pt by normalizing the photocatalytic rates by BET specific surface area for the as-received (CB-AR/Pt) and the most active reduced commercial brookite (CB600/Pt), and as-synthesized (A-AS/Pt) and the most active reduced anatase (A500/Pt).

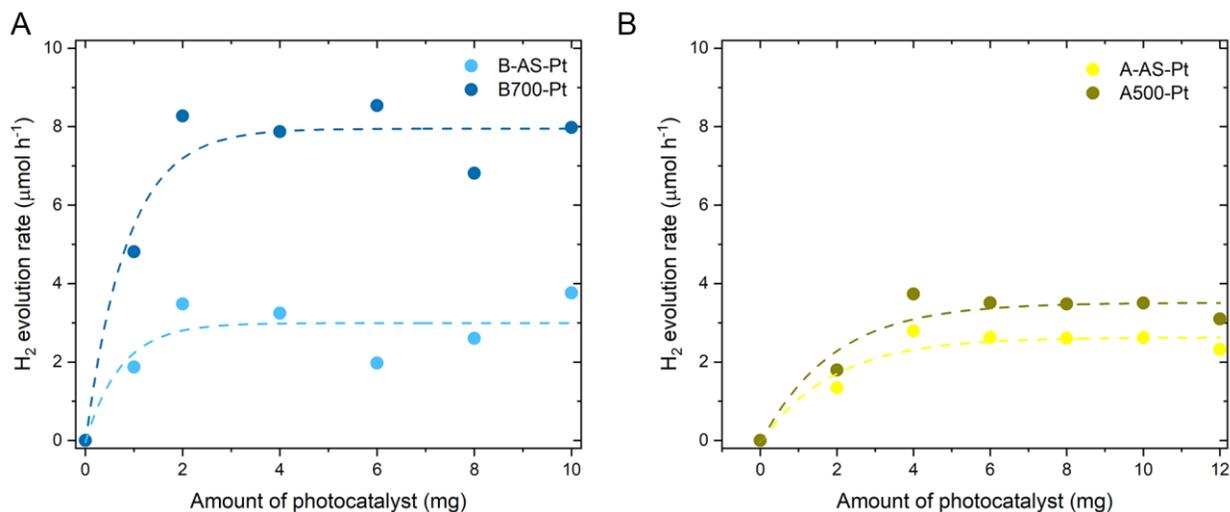

**Figure S29.** Photocatalytic hydrogen evolution rate optimized per mass of used photocatalyst for (A) for B-AS/Pt and B700/Pt; (B) A-AS/Pt and A500/Pt under one sun illumination.

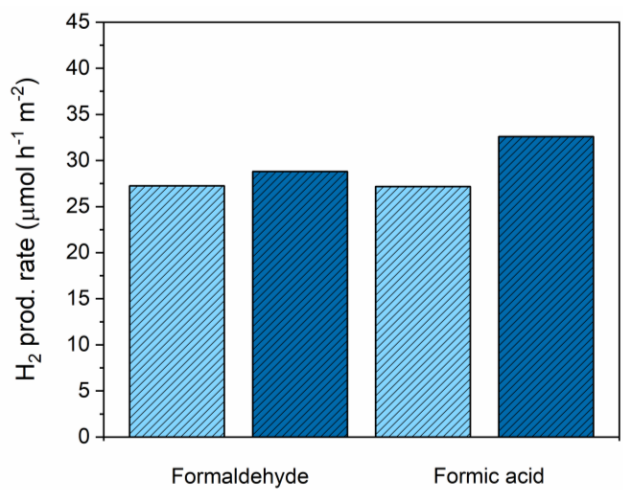

**Figure S30.** (A) Hydrogen production rate from different intermediates of methanol oxidation (formaldehyde and formic acid) of as-synthesized (light blue) and reduced at 700°C (dark blue) brookite samples loaded with 1 wt.% Pt. These experiments confirmed GC analysis of the gas phase (i.e., we detected only carbon dioxide) and NMR analysis of the liquid phase after reaction (i.e. we did not detect any signal related to formaldehyde nor formic acid) indicating that the methanol photo-oxidation proceeded to carbon dioxide, due to reactivity of brookite nanorods toward oxidation of the intermediates (formaldehyde and formic acid) of methanol oxidation. Photolysis of formaldehyde under AM 1.5G 1sun illumination produced a very small hydrogen production rate, namely, ~85 nmol m$^{-2}$ h$^{-1}$.

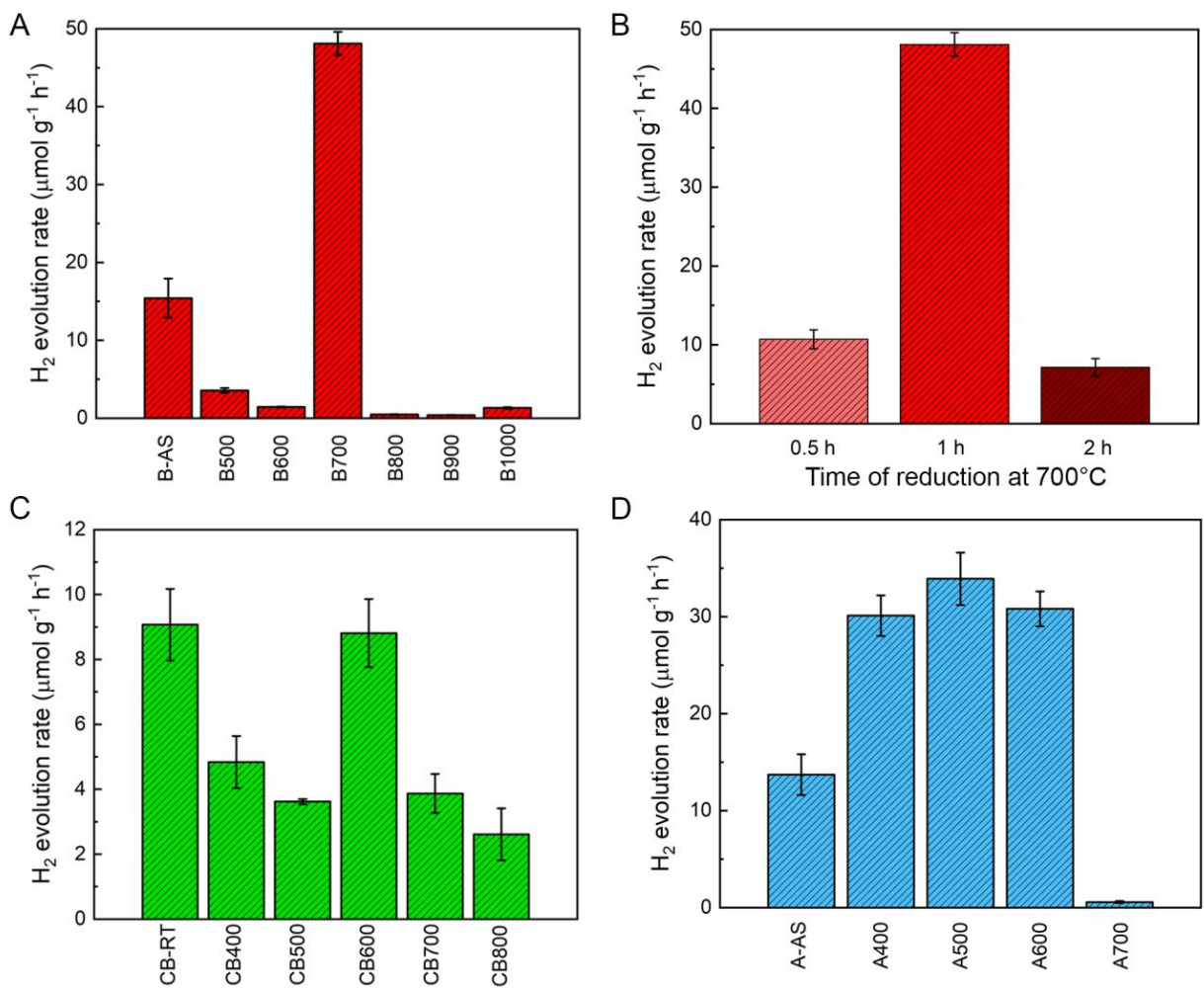

**Figure S31.** Co-catalyst free rate of hydrogen evolution of TiO$_2$ samples tested under 1 Sun illumination for 24 h and reduced under pure hydrogen for 1 h: (A) brookite nanorods, (B) brookite nanorods reduced at 700°C for different times, (C) commercial brookite, and (D) synthesized anatase. We report the specific H$_2$ evolution rate only for the most performing sample for each series in Figure S29.

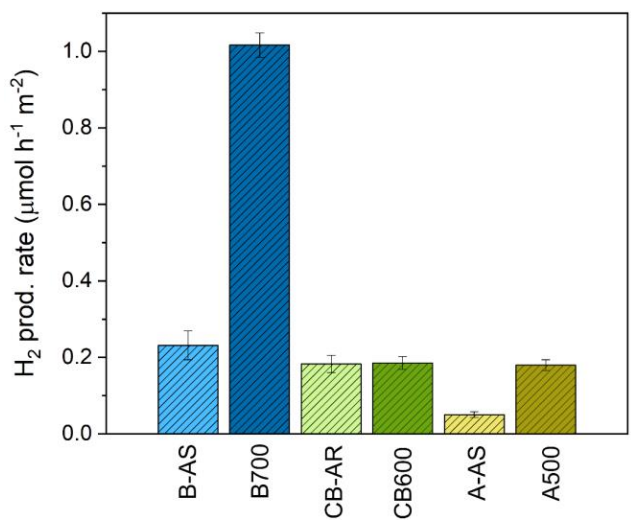

**Figure S32.** Specific hydrogen evolution rate obtained by normalizing the photocatalytic rates by BET specific surface area for as-synthesized (B-AS) and the most active reduced brookite (B700), as-received (CB-AR) and the most active reduced commercial brookite (CB600), and as-synthesized (A-AS) and the most active reduced anatase (A500).

The specific photocatalytic $H_2$ evolution rates expressed in µmol m$^{-2}$ h$^{-1}$ shown in Figure S29 underly that nanorods of reduced brookite (B700) have a specific photocatalytic activity that is 4.4-, 5.6-, and 5.7-fold the ones of as synthesized brookite (B-AS), reduced commercial brookite (CB600), and reduced anatase (A500), respectively. These results demonstrate that higly active sites for photocatalysis are formed in B700 upon reduction under hydrogen atmosphere. Interestingly, when we performed $H_2$ evolution experiments with B700 in $H_2O$/methanol under 1 sun light irradiation and applying a longpass optical filter to cutoff λ ≥ 380 nm, i.e. cutting optical excitation above bandgap energy, we did not detect any $H_2$ after 24 h of reaction. This finding suggests that the intragap electronic states due to the introduction of defects in $TiO_2$ upon reduction (see materials characterization and DFT calculations below) do not directly participate to the photocatalytic activity in the $H_2$ evolution reaction. Further, experiments performed under full 1 sun illumination and using $Na_2S$ (0.1 M) instead of methanol as a hole scavanger, did not produce any $H_2$ suggesting that the $H_2$ evolution activity of our reduced brookite is regulated by its selective methanol photo-oxidation ability. This is supported by the photocatalytic experiments carried out using different alcohols (Figure 1 main text).

*Electron paramagnetic resonance spectroscopy*

In order to identify the nature of spin containing sites and their contribution to photocatalysis we used electron paramagnetic resonance (EPR) spectroscopy. Three samples have been tested, which differ in the annealing temperature in hydrogen atmosphere, B-AS, B500, and B700. We investigated these samples in powder form and in situ under photocatalytic conditions by using a water/methanol (MeOH) matrix.

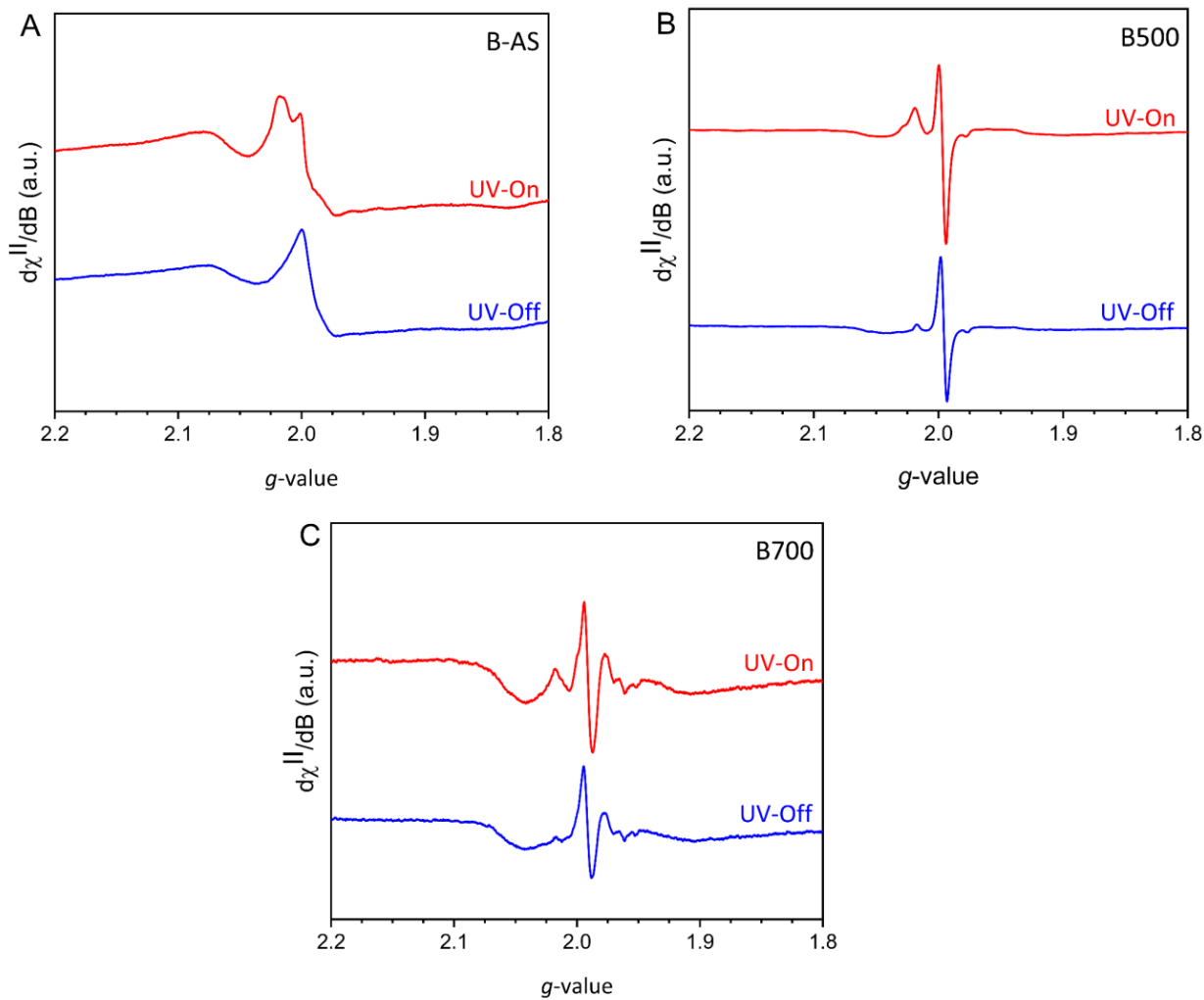

**Figure S33.** X-band EPR spectra (9.1 GHz, T = 77 K) of (A) B-AS, (B) B500, and (C) B-700 recorded in powder form and measured in dark (blue lines) and under UV irradiation (red lines).

The EPR spectrum of B-AS in powder form shown in Figure SA exhibits a broad resonance, centered at $g \sim 2.0$. This signal arises from delocalized defects and presence of significant strains in the crystalline lattice, due to the synthetic procedure pursued. Following irradiation, a new resonant line appears in the spectrum, around $g = 2.017$. This signal originates from holes/oxygen-based radicals. From the double integration of the EPR signal recorded in light and dark conditions, we calculated an increase of about 25.6% in the total number of spin contacting sites after illumination. The observed increase in spin concentration indicates that fast electron/hole recombination processes are here hindered; thus, the photogenerated states are rather stable.

The EPR trace of B500 in powder form, on the contrary, is very strong and it is dominated by a sharp resonant line at $g=1.997$ (Figure S30B). This resonance arises from localized lattice embedded $Ti^{3+}$ sites in an octahedral field. Upon the UV irradiation, we witness an increase of the signal in the region of the spectrum associated to oxygen-based radicals at $g=2.017$. In this case, the double integrated EPR signal shows after UV light exposure and increase in the total number of spins of 13.7%. Therefore, the photoexcited states are here less stable than in B-RT and fast $e^-/h^+$ recombination processes occur in this sample.

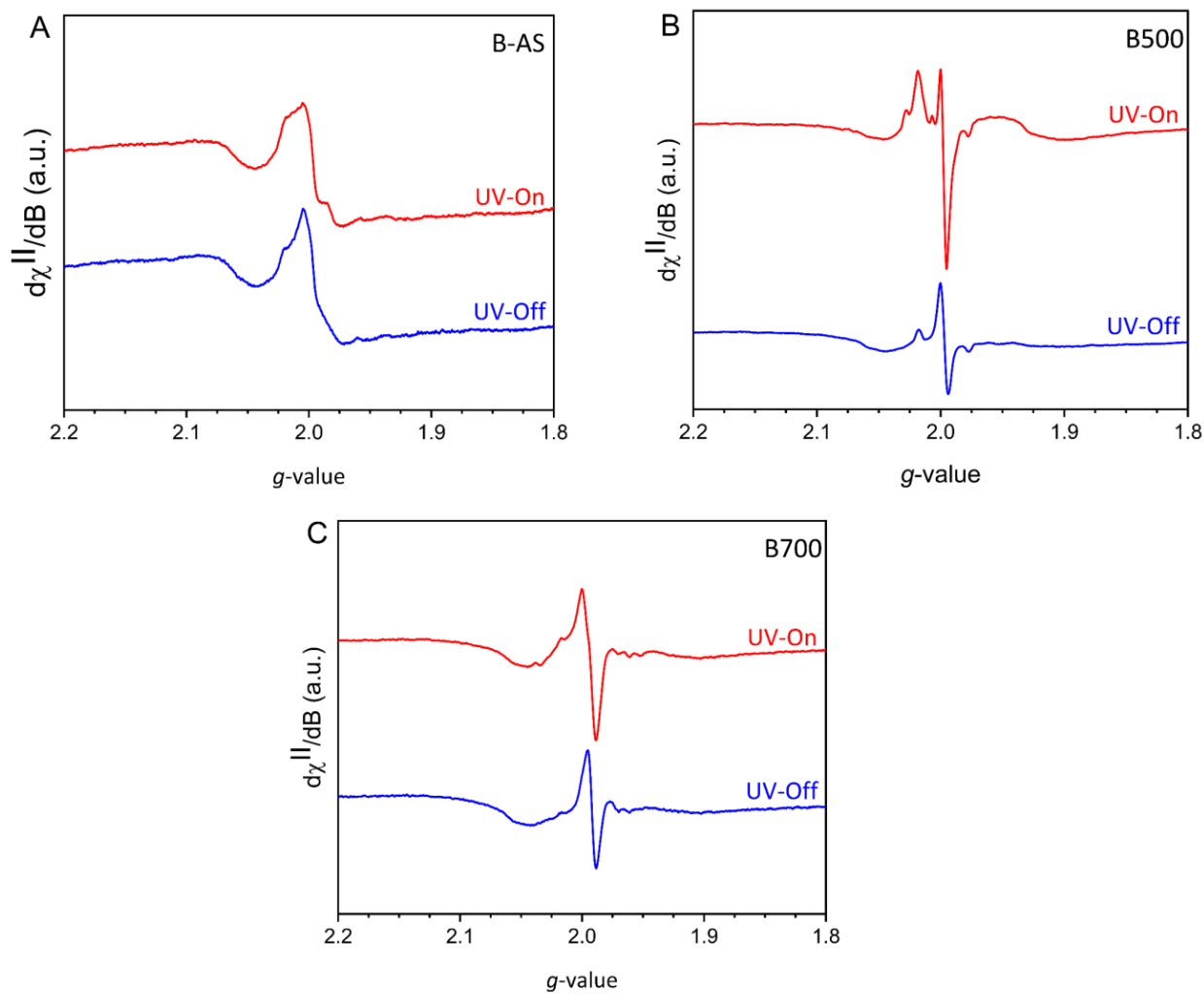

**Figure S34.** X-band EPR spectra (9.1 GHz, T = 77 K) of three samples studied (A) B-AS, (B) B500, and (C) B700 dispersed in the frozen solution of DI water and methanol (1:1, volume ratio) recorded in dark (blue lines) and under UV irradiation (red lines).

The sample B700 in powder form in the dark (Figure SC) gives a weaker EPR signal compared to B500, although being qualitatively similar. The major difference between B700 and B500 appears, in the former, to be linked with a larger distribution in crystal field strain (octahedral to rhombic) associated to the $Ti^{3+}$ spins due to structural defects. Upon UV irradiation, an increase of 56.6% in spin concentration was observed. This behavior is interpreted as due to the high stability of photoexcited states and slow recombination processes in the material, i.e. in B700 in powder form we observed the highest increase of paramagnetic species accumulating under irradiation.
It is worthy to point out that the most efficient sample in hydrogen production and methanol consumption is B700, which gives indeed the weakest intensity in the EPR powder spectrum among the series here shown. Therefore, the number of spins recorded by EPR do not directly correlate with the system reactivity and its overall efficiency in the catalytic process.

To unveil in more detail the reasons underlying the different efficiency in the hydrogen production recorded in this series of tested materials, we performed a set of experiments under in situ photocatalytic conditions. In our experiments, 10 mg of materials were dispersed in 100 µL of solution of deionized (DI) water and MeOH (50:50). Under dark conditions the EPR fingerprints in frozen solutions of B-AS, B500, and B700 (Figure S32A-E) appear very similar to those observed in their correspondent powder forms in contact with $N_2$. However, upon irradiation, significant differences in the ability of MeOH molecules to quench the photogenerated holes emerged. In particular, while in B-AS and B700 the interaction of MeOH molecules with photogenerated $h^+$ seems more effective, in B500 is not, as validated from the appearance of a strong peak at $g=2.017$ associated to accumulation of holes. Therefore, the decreased catalytic efficiency of B500 compared to B-AS and B700 arises from combination of two factors, i) fast electron/hole quenching during photoexcitation and ii) when holes are formed they do not tend to react with MeOH molecules, which translates into a lower probability of successful delivery of electrons for hydrogen production. From time-resolved PL measurements, however, the average lifetime of charge carriers is similar for both B500 (1.9 ns) and B700 (2.0 ns), suggesting how the reaction between holes and methanol is a determining factor addressing the photocatalytic activity trend.

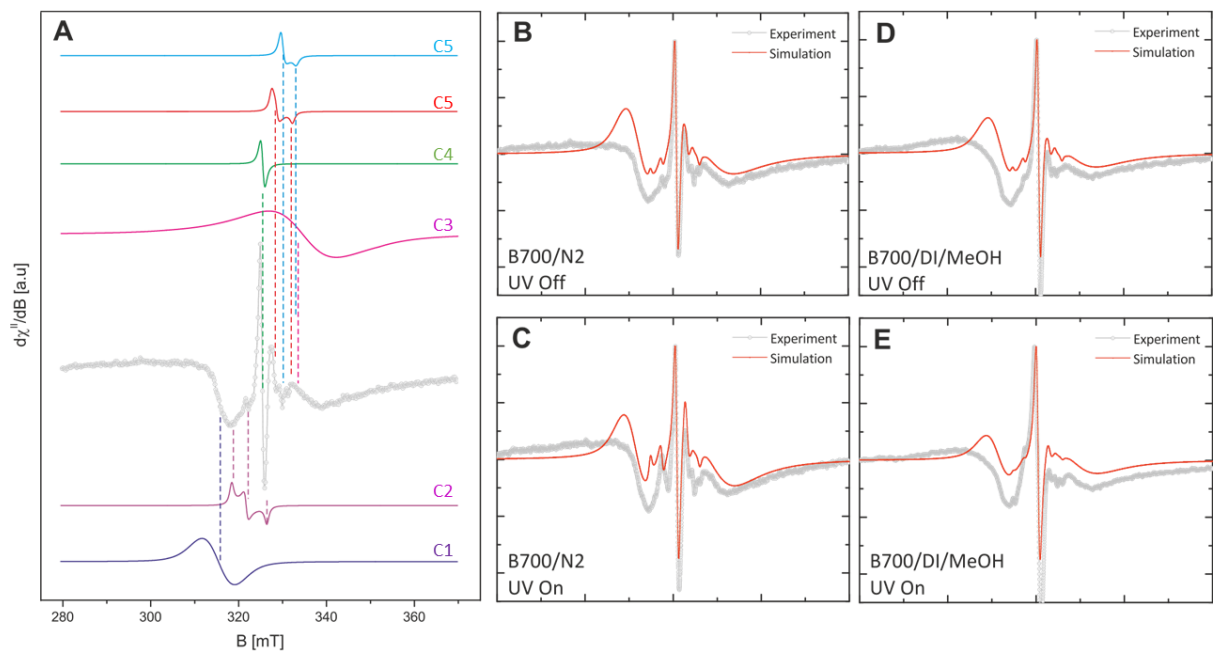

**Figure S35.** (A) an illustrative EPR envelope of B700 in contact with $N_2$ at 80K (grey spectrum) together with individual components (C1-C6) derived from simulation and discussed below, (B-E) comparisons of EPR spectra of B700 recorded in different conditions together with theoretically proposed model.

The simulation and deconvolution of EPR spectra for samples B700 in light and dark conditions in contact with N2 (in powder form) and DI water: MeOH (1:1, volume ratio) at 80K is shown in Figure S32. By deconvolution of EPR spectra we simulated the possible model of paramagnetic species present in the sample B700, which can be applied on all related spectra just by varying the proportions of individual components. We assigned five main species commented bellow.

Component C1: $g=2.059$ related to peroxyl radicals (Ti-O-O$^{\bullet}$) on the surface.
Component C2: $S=1/2$ system associated with a trapped hole (O-centre) in the lattice with $g_x=2.038$, $g_y=2.017$ and $g_z=1.999$.

S37

<u>Component C3:</u> surface exposed $Ti^{3+}$ centres which are highly disordered, and which give wide resonant line at $g=1.935$.
<u>Component C4:</u> sharp isotropic signal at $g=1.992$, we assume that this signal is associated with interstitial $Ti^{3+}$ sides in pseudo octahedral positions.
<u>Component C5:</u> typical axial signal arising from $Ti^{3+}$ in regular lattice positions with $g_{x,y}=1.978$ and $g_z=1.952$.
<u>Component C6:</u> $Ti^{3+}$ in the lattice located in some sublayer under the surface, where are lattice parameters slightly shortened and therefore *g* value is smaller comparing to C5, to be precise $g_{x,y}=1.963$ and $g_z=1.944$.

*Photoluminescence (PL) and time-resolved photoluminescence (TRPL) spectroscopy*

Photoluminescence spectroscopy is a powerful technique for studying the radiative recombination of electrons and holes that populate intragap energy positions and that are due to the presence of structural defects in semiconducting materials [39]. Photoluminescence spectroscopy is often used in the literature to reconcile charge carrier recombination phenomena occurring in $TiO_2$ and its photocatalytic activity [40–42]. At first approximation, an higher PL intensity and lower PL lifetime underlie a higher electron and hole recombination rate and consequently a lower photocatalytic activity. However, the way radiative recombination processes may bring insight to photocatalytic activity also relates to: (i) the position of structural defects into the $TiO_2$ lattice (i.e. surface, subsurface, bulk) that, in turn, influences the energy distribution of electronic states due to defects (and therefore also the energy distribution of PL spectra), and (ii) how PL intensity and energy distribution change upon exposure to molecules employed in the photocatalytic tests as a hole scavenger (i.e. methanol).

To examine the luminescence behavior of our $TiO_2$ samples, we measured PL map spectra using several excitation wavelengths from 258 to 590 nm (4.8 eV down to 2.1 eV) and monitoring PL emission in the energy range from 364 to 730 nm (3.4 eV down to 1.7 eV). The samples were measured initially in the solid state at the temperature 80 K in contact with $N_2$ atmosphere. The PL maps for as synthesized brookite and B700 (the most active photocatalyst) are reported in the main text (Figure 2), while PL maps for B500, B600 and B800 are illustrated in Figure S36. From these PL maps, it is evident that PL intensity and energy distribution changed in each sample depending on the reduction temperature, suggesting that recombination centers (i.e. defects) re-organized and moved within the brookite lattice upon reduction at increasing temperatures [23,43]. A similar phenomenon was observed for each set of samples investigated, namely commercial brookite (Figure S37) and anatase (Figure S38). Within the synthesized brookite series, the most intense PL signal was recorded for B700 ≈ B-AS > B500 > B600 > B800. According to the XRD results, the sample B800 is already transformed to rutile phase and for this reason show a very weak PL in the investigated emission energy range. It is interesting to note that B-AS presents PL signal only when excited with above-bandgap photons, while B700 highlights a complex PL signal even for below-bandgap excitations, demonstrating that reduction modified the electronic structure of reduced brookite. In order to investigate more in depth this aspect, we deconvoluted the PL spectra of synthesized and reduced brookite (Figure S39) obtained using an excitation wavelength 340 nm (3.6 eV, above-bandgap). The spectra were fitted by 4 components peaking at 450, 490, 545 and 620 nm (2.75, 2.53, 2.27 and 2.0 eV, respectively). In each case, the peak position and the full width at half maximum (FWHM) of the deconvoluted peaks were kept the same and just the intensity of peaks was free to change.

Generally, it has been shown that anatase and brookite exhibit distinct PL bands in the visible region of the electromagnetic spectrum [44]. The well-accepted interpretative model [39] for these emissions include the presence of the main visible emission is composed of a green component ("type 1 PL" or "green PL"), at higher energies, due to recombinations between shallowly trapped electrons (or conduction band electrons) and deeply trapped holes, and a red component ("type 2 PL" or "red PL"), at lower energies, due to recombinations between valence band holes and deeply trapped electrons [45–49]. Spectra deconvolution reported in Figure S39 show that the weight of different PL components in our brookite samples significantly changes upon treatment in hydrogen at increasing temperatures. Importantly, the most intense PL components for B-AS are those at 2 and 2.27 eV, while for B700 (i.e. the most active photocatalyst), the higher energies PL components become dominant. The experimentally determined valence band photoemission spectra recorded at synchrotron using photon energy in resonance with the titanium x-ray absorption edge (Figure 3A in the main text) show that in case of B-AS, the native oxygen vacancies present in the brookite induced a distribution of intragap states acting as deep electronic traps, while for B700 a more defective structure induced a valence band tailing, offering therefore electronic states that may host deeply/shallowly trapped holes. These findings demonstrates that the PL emissions observed for our samples are in agreement with the proposed mechanism about "green and red PL". The different PL behavior in B-AS and B700 therefore remarks that the reduction treatment introduce different defects that have a different PL signatures and influence in different way the photocatalytic activity of brookite samples. These results reflect those of Vequizo et al. [50] who also found that the presence of an appropriate depth of the traps can effectively contribute to enhance the overall photocatalytic activity of $TiO_2$. They found that in case of as synthesized brookite, the moderate depth of electron traps in comparison with the anatase and the rutile phase with shallow and deep electron traps, respectively, help brookite to be active for the both oxidation and reduction reactions. In our case, instead, we provide evidences that recombinations centers in reduced brookite are those defects that regulate the photo-oxidation reaction during $H_2$ evolution from methanol/$H_2O$ photoreforming. Further, PL maps recorded in the presence of methanol (Figure 2 in the main text) show for both B-AS and B700 that the radiative recombinations are almost completely quenched in these conditions, confirming the proposed PL mechanism and the fact that during photocatalysis holes are mainly employed for the photo-oxidation reaction.

Figure S40 shows the TRPL of as-synthesized (received) and reduced samples, while Table S10 shows the fitted parameters employed to calculating the average electron life time of each sample. In all cases, the amplitude of the slow component ($B_1$) is higher than the amplitude of the fast component ($B_2$). The average electron lifetime ($\tau_{ave}$) of reduced samples, in all cases, is less than that of the corresponding pristine samples. This behavior of reduced $TiO_2$ samples suggests that after reduction, the defective centers are responsible for a faster charge carrier recombination [51]. Importantly, the results from TPRL measurements demonstrates that the improved photocatalytic activity of our reduced $TiO_2$ (B700, A500, CB600) is not due to an improved photo-induced charge separation, as shown previously in other $TiO_2$ systems [52,53]. In contrast, in this case it is regulated by other parameters such as the photo-reactivity of the heterogeneous catalytic sites formed around oxygen vacancies and comprised of several Ti atoms sharing extra charge (see DFT calculations below) and lattice distortions (see XRD analysis).

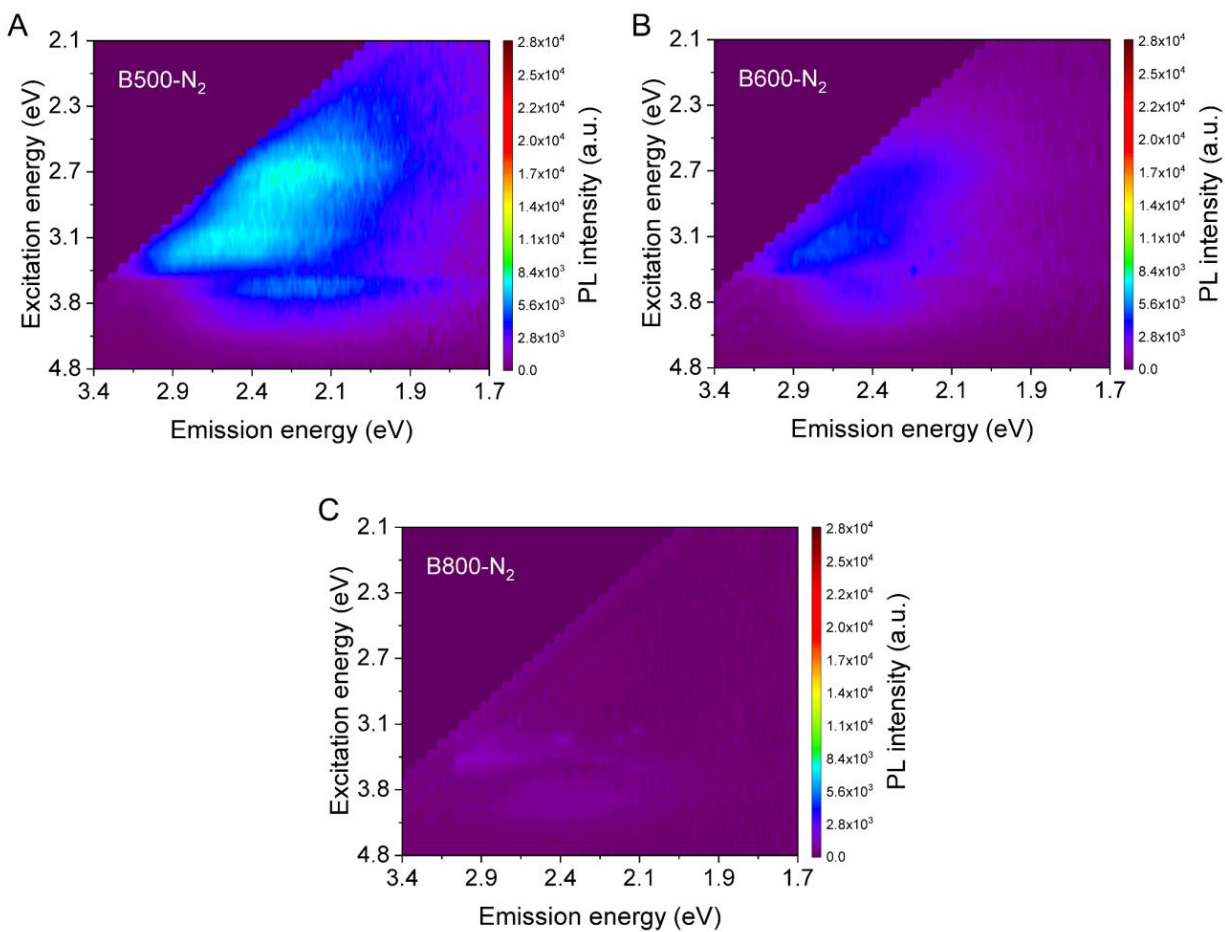

**Figure S36.** PL maps for (A) B500, (B) B600, and (C) B800. The measurement temperature was 80 K under $N_2$ atmosphere.

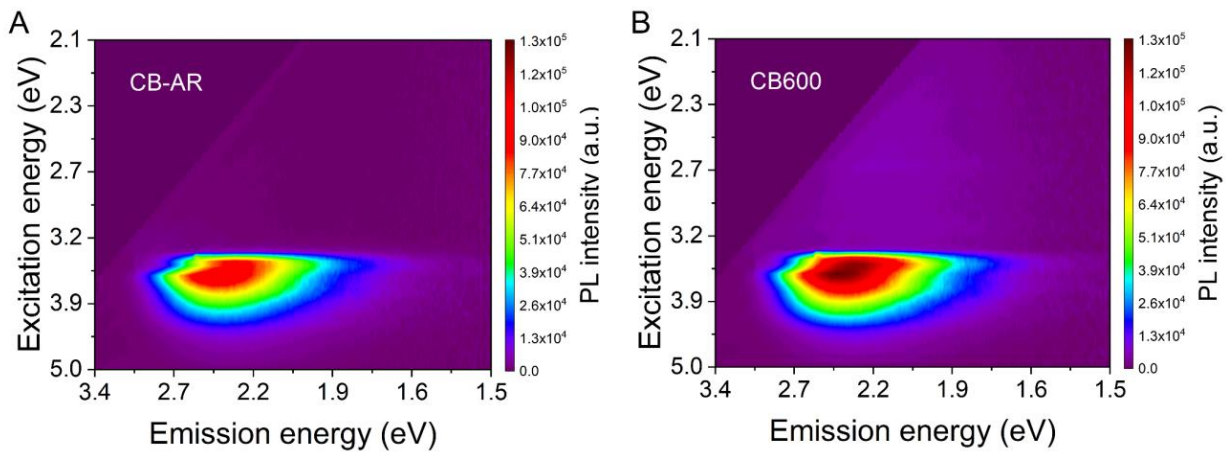

**Figure S37.** PL maps for (A) as-received commercial brookite and (B) reduced commercial brookite at 600°C. The measurement temperature was 80 K under $N_2$ atmosphere.

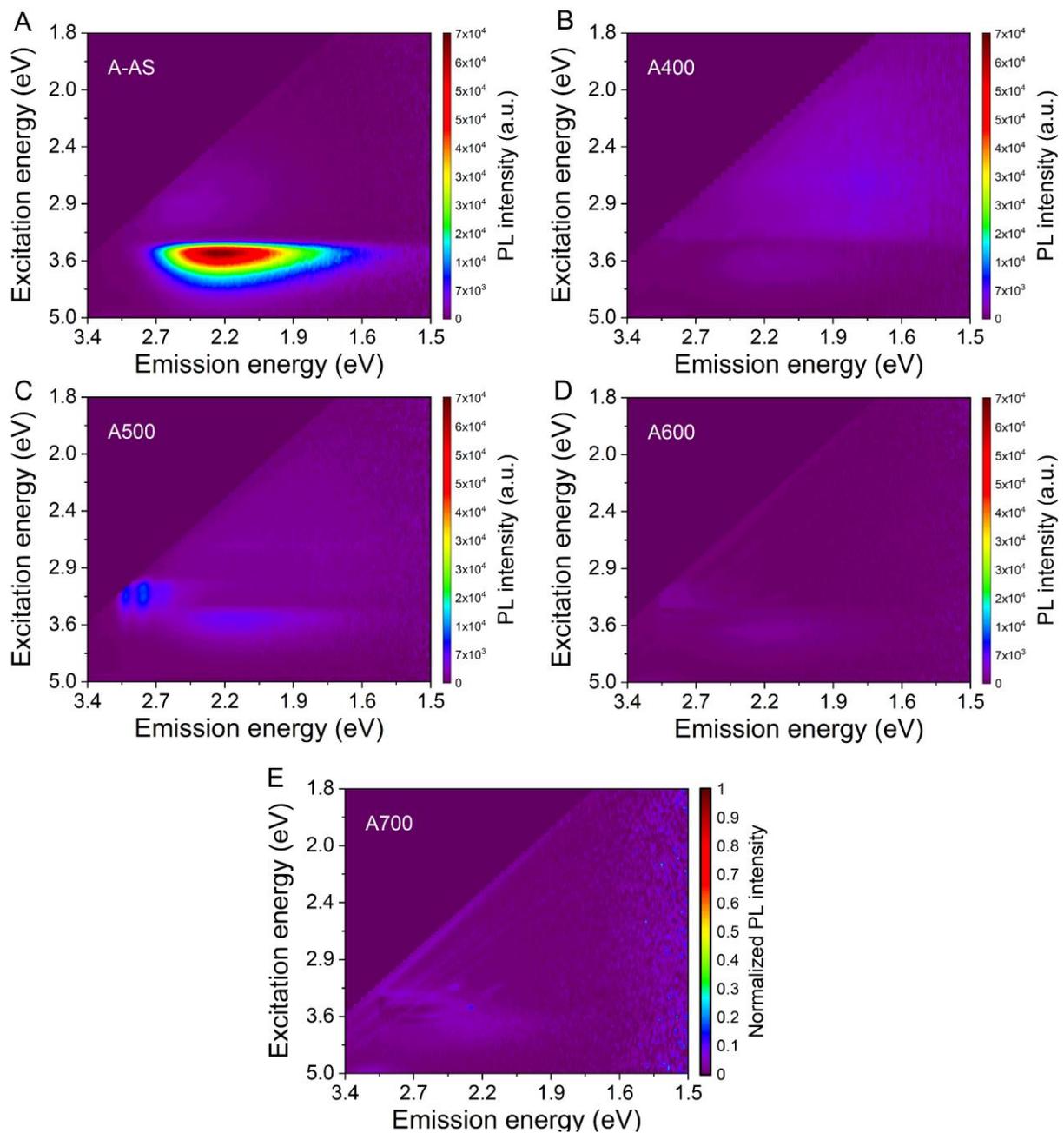

**Figure S38.** PL maps for (A) as-synthesized anatase, and reduced anatase at (B) 400°C, (B) 500°C, (B) 600°C, and (B) 700°C. The measurement temperature was 80 K under $N_2$ atmosphere.

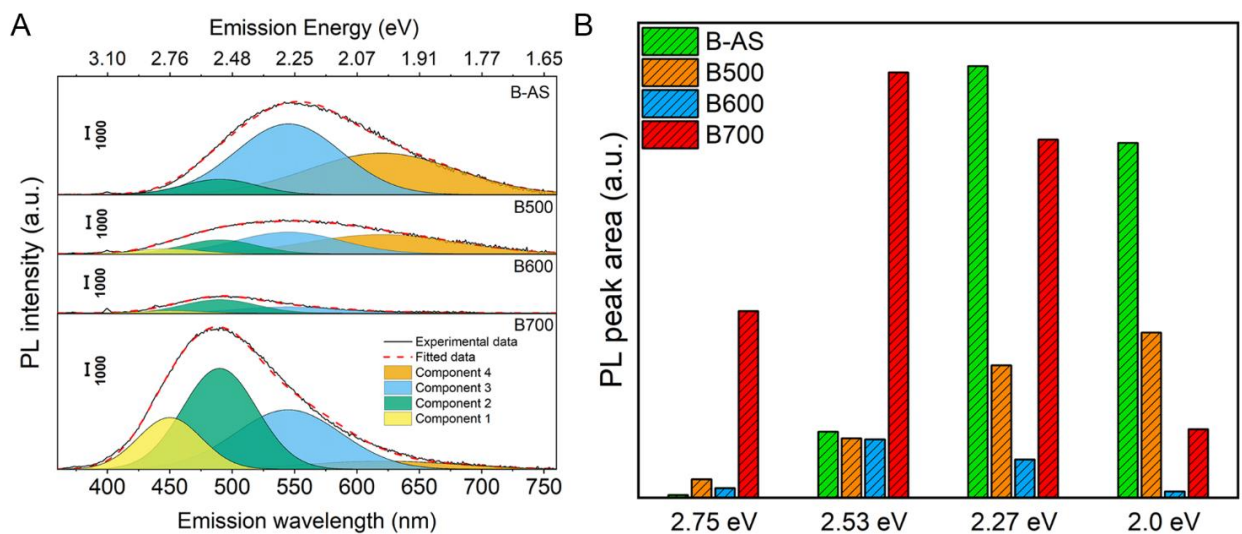

**Figure S39.** (A) Deconvoluted PL spectra for as-synthesized and reduced brookite samples, using an excitation energy 3.6 eV. (B) PL peak area related to each component retrieved from deconvolution and centered at 2, 2.27, 2.53, 2.75 eV. Both peak energy and width of each component were kept constant in each sample.

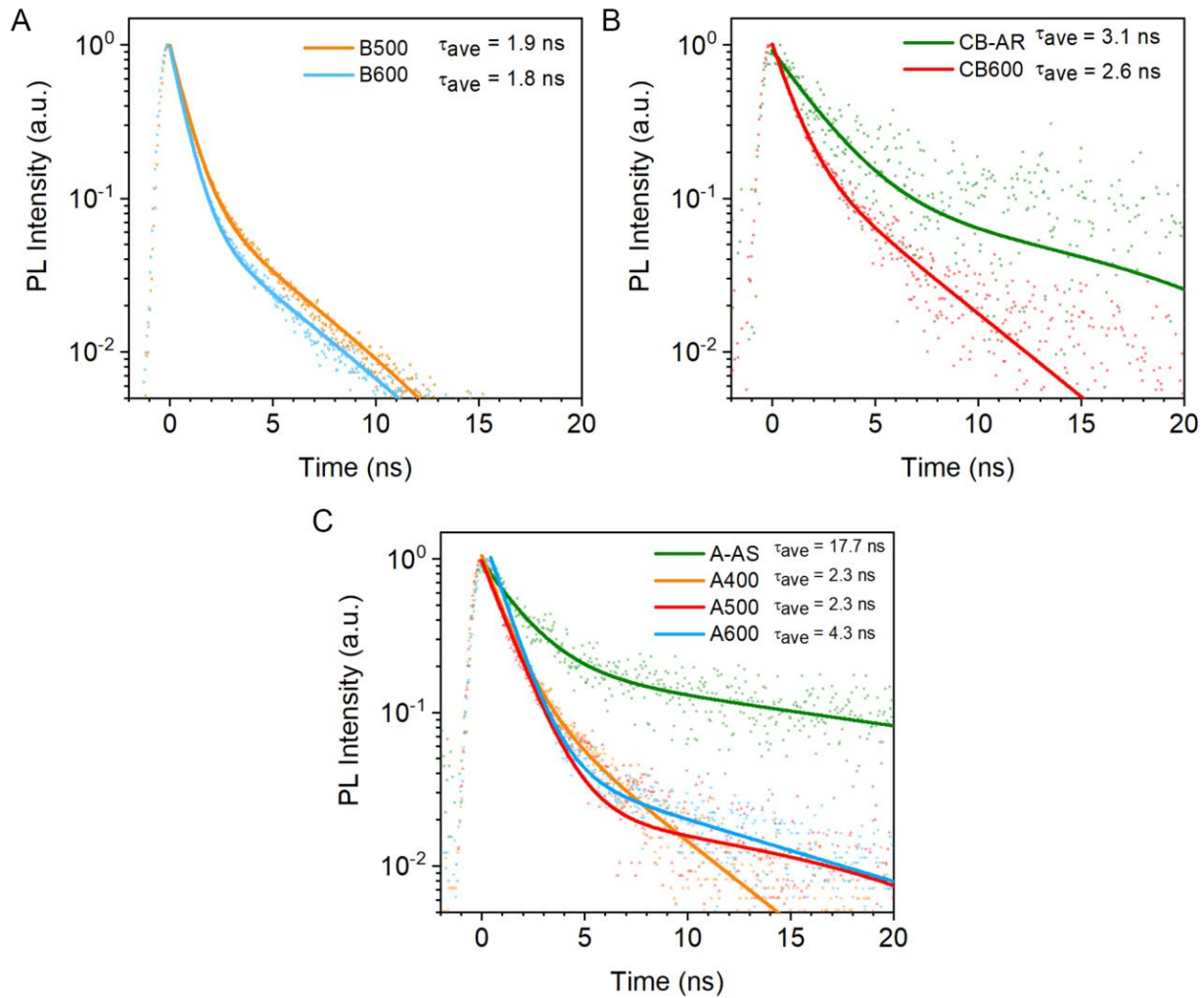

**Figure S40.** Time–resolved photoluminescence spectra of (A) brookite reduced at 500°C and 600°C, (B) as-received and reduced commercial brookite at 600°C, and (C) as-synthesized and reduced anatase at 400, 500, and 600°C. The dots are experimental data and the solid lines are the fitted curve. The $\tau_{ave}$ is the average electron lifetime extracted from the fitting.

**Table S10.** Time–resolved photoluminescence fitted parameters for as-synthesized (as-received) and reduced samples at different temperatures.

| Sample | $B_1$, % | $\tau_1$, ns | $B_2$, % | $\tau_2$, ns | $\tau_{ave}$, ns |
|--------|------|------|------|------|------|
| B-AS   | 85   | 1.2  | 15   | 8.6  | 5.3  |
| B500   | 87   | 0.7  | 13   | 3.6  | 1.9  |
| B600   | 91   | 0.6  | 9    | 3.7  | 1.8  |
| B700   | 90   | 0.8  | 10   | 4.0  | 2.0  |
| A-AS   | 80.2 | 1.8  | 19.8 | 22.7 | 17.7 |
| A400   | 85.1 | 1.1  | 14.9 | 4.1  | 2.3  |
| A500   | 84.7 | 0.9  | 15.3 | 4.1  | 2.3  |
| A600   | 95.4 | 1.0  | 4.6  | 10.8 | 4.3  |
| CB-AR  | 100.0| 3.1  | -    | -    | 3.1  |
| CB600  | 77.1 | 0.9  | 22.9 | 3.9  | 2.6  |

*Electronic characterization*

*Photoemission spectroscopy*

We examined the electronic state of elements at the surface of the as-synthesized (received) and the best reduced samples using an XPS laboratory source. Figure S compares the XPS survey spectra of samples before and after reduction treatment. An important finding from these survey spectra is that the samples are not contaminated during the reduction process, as already confirmed by CHN analysis (Table S2), since there is no difference between total XPS survey before and after process. This observation supports our hypothesis that all of the changes in photoactivity of the samples are due to the reduction process and not due to materials contamination. The C*1s* peak at 284.8 eV was used as a reference for the energy scale to compensate the charging effect of the samples (all spectra were shifted according to this reference). However, the observed difference between XPS spectra of Ti*2p* and O*1s* for pristine and reduced samples were not significant. A possible explanation for this might be that the amount of changes in the lattice of $TiO_2$ induced through reduction are below the detection limit of XPS [38] or the reduced species like $Ti^{3+}$ can be easily oxidized by exposure to the ambient air [54]. Since, no difference was detected in XPS analysis between the samples before and after reduction, we used synchrotron-based XPS (VUV-Photoemission beamline, Elettra) to study them in more detail. Figure S39 shows the synchrotron-based XPS Ti*2p* spectra of the pristine and the most photoactive sample of brookite. There is no evidence of an increase in $Ti^{3+}$ species after reduction. Additionally, synchrotron-based XPS study of O*1s* orbital of the pristine and the most photoactive sample of brookite (Figure S40) reveals about 18% increase in OH groups on the surface of $TiO_2$ after reduction. This supports the water dissociation on superficial defects created upon reduction treatment, see more details in the mechanism part below.

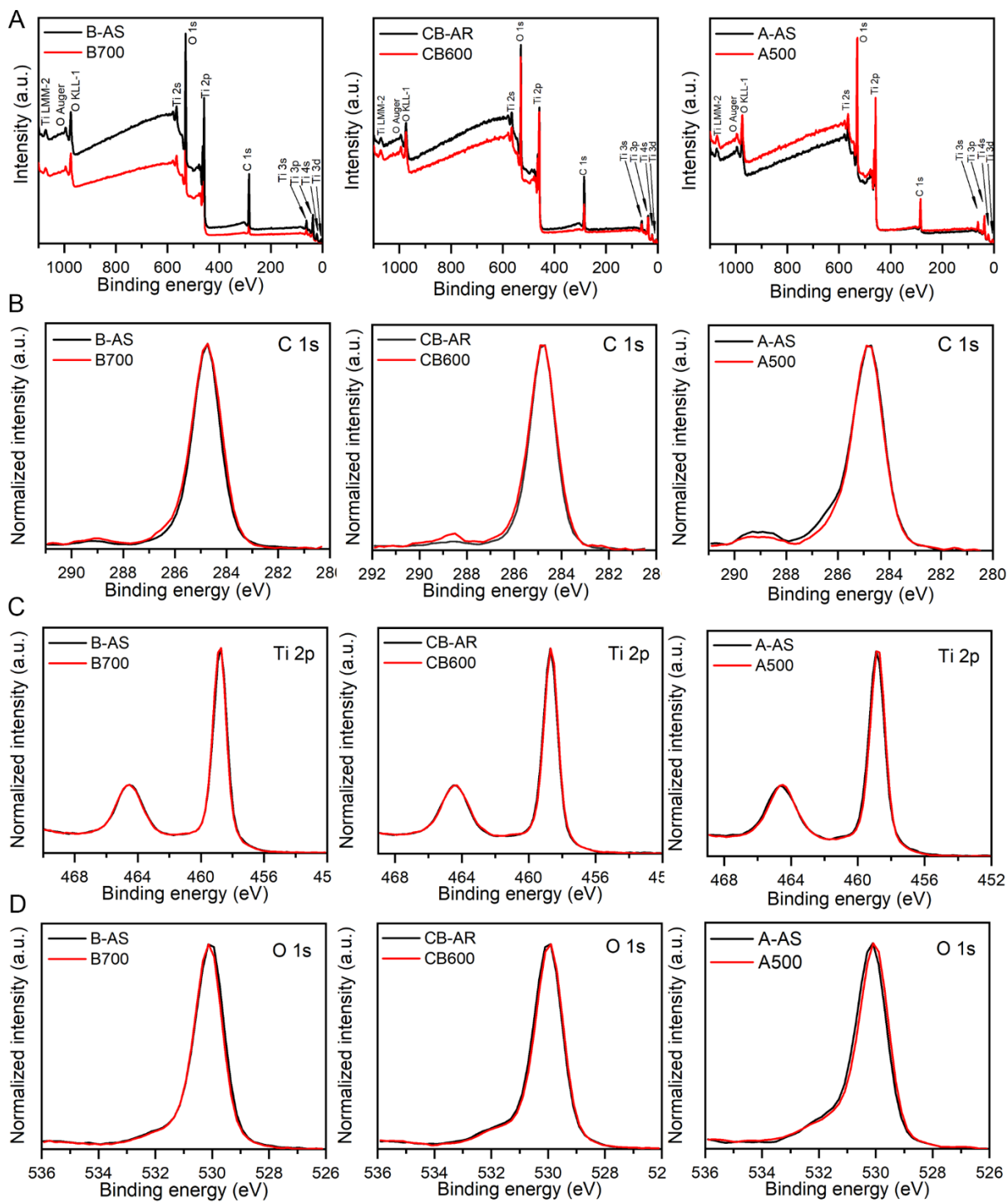

**Figure S41.** (A) XPS survey spectra of the pristine and the most photoactive sample of brookite (left), commercial brookite (middle) and anatase(right) (B) High resolution XPS spectra of the pristine and the most photoactive sample of brookite (left), commercial brookite (middle) and anatase.(right) of (B) C*1s*, (C) Ti*2p* and (D) O*1s* orbitals.

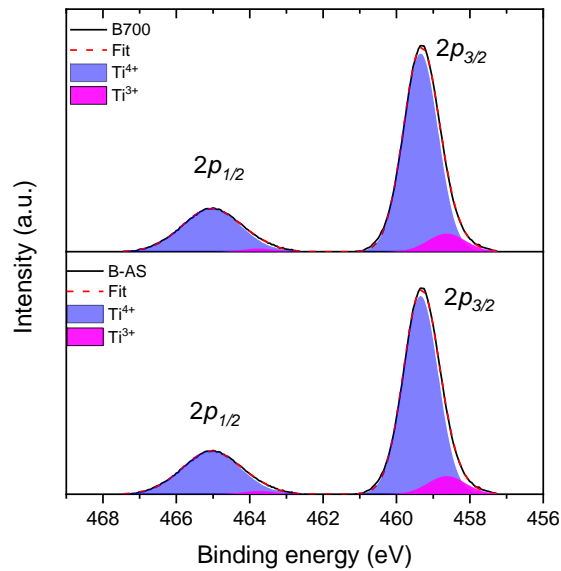

**Figure S42.** Synchrotron-based XPS spectra in the Ti *2p* region of pristine brookite (B-AS), the most photoactive brookite sample (B700).

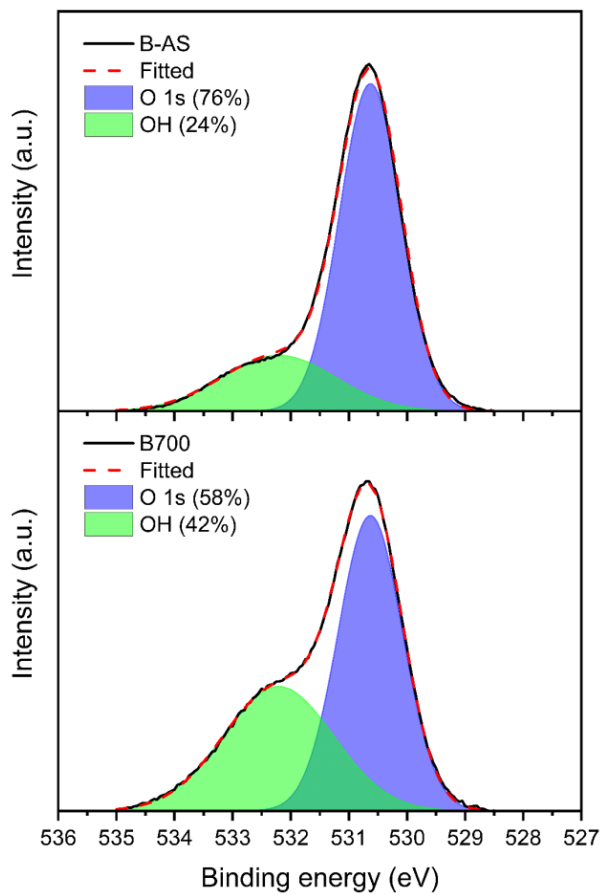

**Figure S43.** Synchrotron-based XPS spectra of the pristine (top) and the most photoactive sample of brookite (bottom), O*1s* orbital.

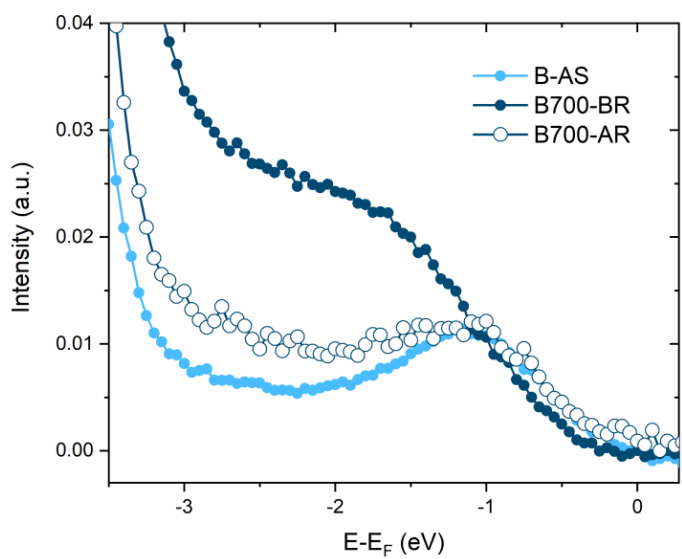

**Figure S44.** Synchrotron-based photoemission spectra around the valence band (VB) region for the pristine brookite B-AS (light blue, full circles), the reduced brookite before reaction B700-BR (dark blue, full circles), and the reduced brookite after 24 h of photocatalytic reaction B700-AR (dark blue, empty circles).

*Density functional theory calculations*

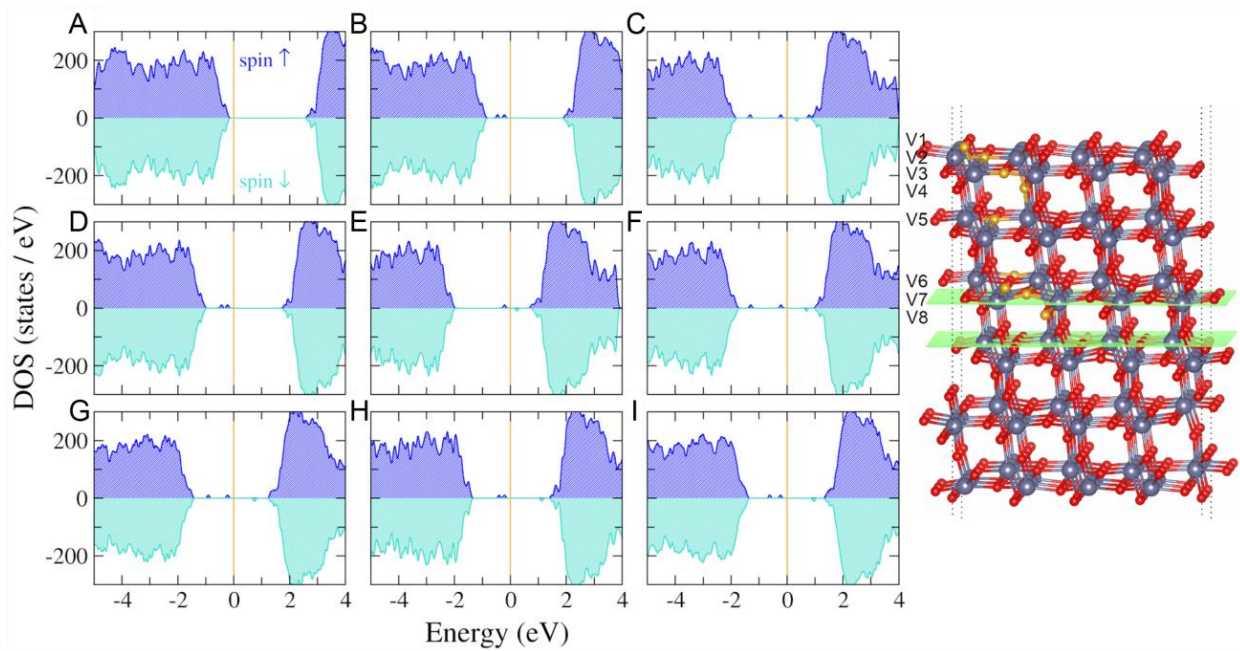

**Figure S45.** Spin-resolved density of states (DOS) plots of brookite $TiO_2$ supercell exposing the (210) surface. The simulated $TiO_2$ structure is shown on the right side of the panel with Ti atoms plotted in grey, O in red, and O vacancies indicated in orange. The middle part of the slab corresponds to the bulk region of $TiO_2$ enclosed by green planes, and the supercell's boundaries are indicated by the dotted lines. (A) DOS for a perfect slab, (B–I) is for the same slab including one O vacancy denoted by V1–V8 in the structure on the right. All plots are zeroth to the Fermi level. DOS of V2 and V4 were calculated for a fixed geometry to prevent the migration of the vacancy to the on-surface V1 position. The DOS plots clearly show that varying the lattice position (surface, subsurface, bulk) of an oxygen vacancy results in the formation of intragap electronic states with different energy and features.

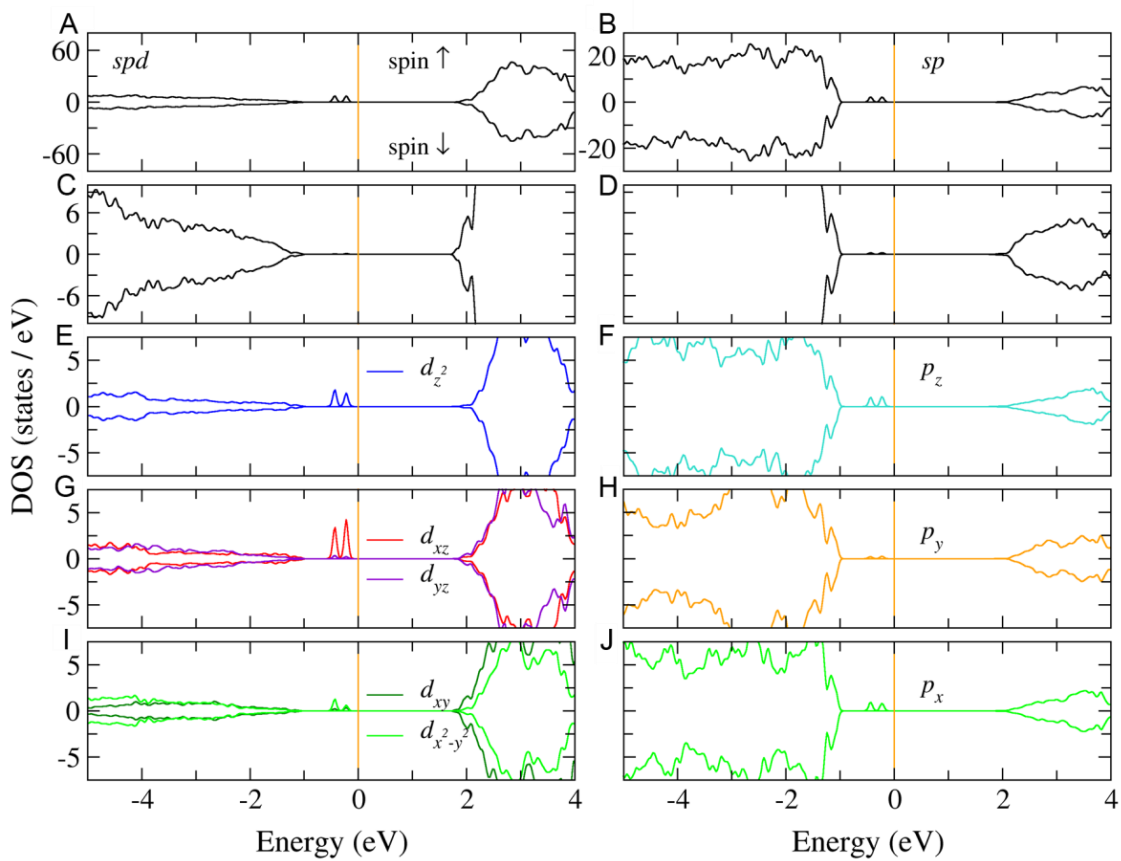

**Figure S46.** (A) DOS of the surface Ti bilayer and (B) the corresponding O atoms with the missing O atom at V3 position. (C) DOS of subsurface Ti bilayer and (D) the corresponding O atoms. (E–J) Orbital resolved partial DOS for the surface Ti bilayer (E, G, I) and O atoms (F, H, J). All plots are zeroth to EF. It is interesting to note how intragap states are composed by hybridized electronic contributions belonging both to Ti and O atoms.

*Mechanism of methanol photo-oxidation*

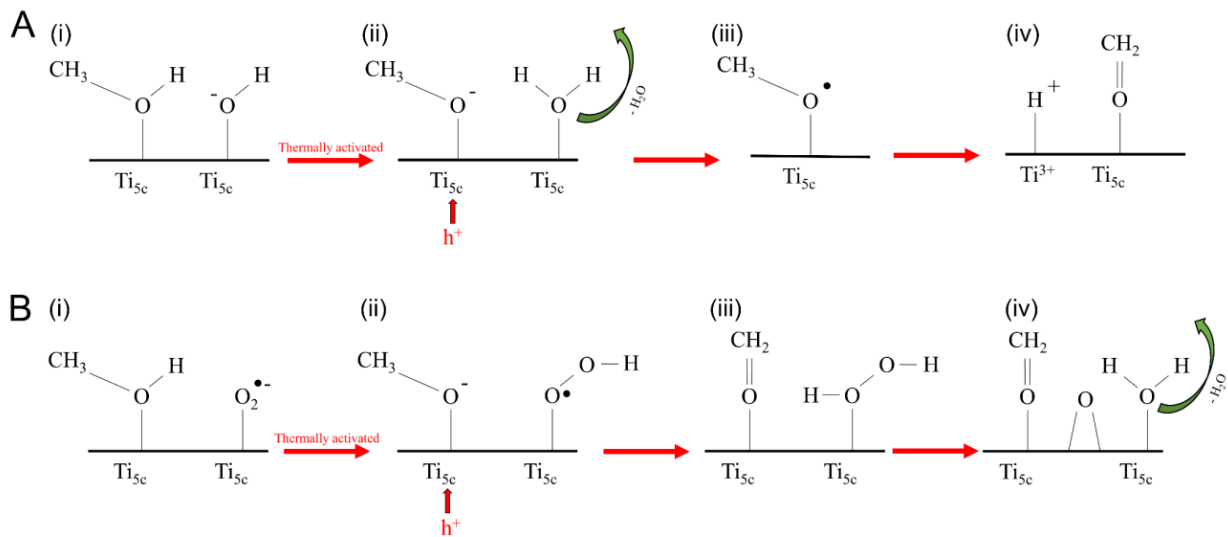

**Figure S47.** Methanol oxidation mechanism at the TiO$_2$ surface. (A) Methanol activation by terminal OH$^-$. In step (i) → (ii) the OH reacts with methanol producing H$_2$O and the methoxy group. In (ii) → (iii) the methoxy group accepts a hole from TiO$_2$ substrate (or donates an electron to TiO$_2$) and converts to methoxy radical, while the water molecule is released from the TiO$_2$ surface. In (iii) → (iv) the methoxy radical decomposes to adsorbed formaldehyde and proton ion. (B) Methanol activation by coadsorbed O$_2$. In step (i) → (ii) The superoxide activates the methanol and to methoxy radical and produces the peroxide radical. Then in step (ii) → (iii) such radicals are converted to adsorbed formaldehyde and hydrogen peroxide, respectively. In step (iii) → (iv) the hydrogen peroxide decomposes to water (then released from the TiO$_2$ surface) and a bridging oxygen dimer [55].

*References*

1. Kandiel, T.A., Feldhoff, A., Robben, L., Dillert, R., and Bahnemann, D.W. (2010). Tailored titanium dioxide nanomaterials: Anatase nanoparticles and brookite nanorods as highly active photocatalysts. Chem. Mater. *22*, 2050–2060.

2. Zhao, H., Liu, L., Andino, J.M., and Li, Y. (2013). Bicrystalline $TiO_2$ with controllable anatase-brookite phase content for enhanced $CO_2$ photoreduction to fuels. J. Mater. Chem. *1*, 8209–8216.

3. Beltram, A., Romero-Ocaña, I., Josè Delgado Jaen, J., Montini, T., and Fornasiero, P. (2016). Photocatalytic valorization of ethanol and glycerol over $TiO_2$ polymorphs for sustainable hydrogen production. Appl. Catal. A: Gen *518*, 167–175.

4. Naldoni, A., Montini, T., Malara, F., Mróz, M.M., Beltram, A., Virgili, T., Boldrini, C.L., Marelli, M., Romero-Ocaña, I., Delgado, J.J., et al. (2017). Hot electron collection on brookite nanorods lateral facets for plasmon-enhanced water oxidation. ACS Catal. *7*, 1270–1278.

5. Larson, A.C., and Von Dreele, A.C. (2004). General structure analysis system (GSAS).

6. Weyl, R. (1977). Zeitschrift fuer Kristallographie.

7. http://www.iucr.org/resources/data/datasets/bond-valenceparameters.

8. Williamson, G.K., and Hall, W.H. (1953). X-ray line broadening from filed aluminium and wolfram. Acta metall. *1*, 22–31.

9. Roisnel, T., and Rodríguez-Carvajal, J. (2001). WinPLOTR: A windows tool for powder diffraction pattern analysis. Mater. Sci. Forum *378–381*, 118–123.

10. Rouquerol, F., Rouquerol, J., Llewellyn, P., Maurin, G., Sing, K., (1999). Adsorption by powders and porous solids (Elsevier).

11. Rex, R.E., Knorr, F. J., and McHale, J.L. (2013). Comment on "Characterization of oxygen vacancy associates within hydrogenated $TiO_2$: A positron annihilation study". J. Phys. Chem *117*, 7949–7951.

12. Stoll, S., and Schweiger, A. (2006). EasySpin, a comprehensive software package for spectral simulation and analysis in EPR. J. Magn. Reson. *178*, 42–55.

13. Kresse, G., and Furthmuller, J. (1996). Efficient iterative schemes for ab initio total-energy calculations using a plane-wave basis set. Phys. Rev. B Condens. Matter Mater. Phys. *54*, 11169–11186.

14. Kresse, G., and Furthmüller, J. (1996). Efficiency of ab-initio total energy calculations for metals and semiconductors using a plane-wave basis set. Comput. Mater. Sci. *6*, 15–50.

15. Blöchl, P.E. (1994). Projector augmented-wave method. Phys. Rev. B Condens. Matter Mater. Phys. *50*, 17953–17979.

16. Kresse, G., and Joubert, D. (1999). From ultrasoft pseudopotentials to the projector augmented-wave method. Phys. Rev. B Condens. Matter Mater. Phys. *59*, 1758–1775.

17. Perdew, J.P., Burke, K., and Ernzerhof, M. (1996). Generalized gradient approximation made simple. Phys. Rev. Lett. *77*, 3865–3868.

18. Perdew, J.P., Burke, K., and Ernzerhof, M. (1997). Generalized gradient approximation made simple (Errata). Phys. Rev. Lett. *78*, 1396–1396.

19. Dudarev, S.L., Botton, G.A., Savrasov, S.Y., Humphreys, C.J., and Sutton, A.P. (1998). Electron-energy-loss spectra and the structural stability of nickel oxide: An LSDA+U study. Phys. Rev. B *57*, 1505–1509.

20. Morgan, B.J., and Watson, G.W. (2009). Polaronic trapping of electrons and holes by native defects in anatase $TiO_2$. Phys. Rev. B *80*, 2331021–2331024.

21. Holmström, E., Ghan, S., Asakawa, H., Fujita, Y., Fukuma, T., Kamimura, S., Ohno, T., and Foster, A.S. (2017). Hydration structure of brookite $TiO_2$ (210). J. Phys. Chem *121*, 20790–20801.

22. Naldoni, A., Allieta, M., Santangelo, S., Marelli, M., Fabbri, F., Cappelli, S., Bianchi, C.L., Psaro, R., and Dal Santo, V. (2012). Effect of nature and location of defects on bandgap narrowing in black $TiO_2$ nanoparticles. J. Am. Chem. Soc. *134*, 7600–7603.

23. Naldoni, A., Altomare, M., Zoppellaro, G., and Liu, N. (2019). Photocatalysis with reduced $TiO_2$ : from black $TiO_2$ to cocatalyst- free hydrogen production. ACS Catal. *9*, 345–364.

24. Bersani, D., Lottici, P.P., and Ding, X. (1998). Phonon confinement effects in the Raman scattering by $TiO_2$ nanocrystals. Appl. Phys. Lett. *72*, 73–75.

25. Parker, J.C., and Siegel, R.W. (1990). Calibration of the Raman spectrum to the oxygen stoichiometry of nanophase $TiO_2$. Appl. Phys. Lett. *57*.

26. Fauchet, I.H.C. and P.M. (1986). The effects of microcrystal size and shape of the one phonon raman spectra of crystalline semiconductor. Solid State Commun. *58*, 739–741.

27. Venkatasubbu, G.D., Ramakrishnan, V., Sasirekha, V., Ramasamy, S., and Kumar, J. (2014). Influence of particle size on the phonon confinement of $TiO_2$ nanoparticles. J. Exp. Nanosci. *9*, 661–668.

28. Fisica, D., Infm, C., Chimica, S., Parma, U., and Scienze, V. (1993). Raman scattering characterization of gel-derived titania glass. J. Mater. Sci. *28*, 177–183.

29. Tauc, J., Grigorovici, R., Vancu, A. (1966). Optical properties and electronic structure of amorphous germanium. Phys. Stat. Sol. *627*, 627–637.

30. Escobedo-Morales, A., Ruiz-López, I.I., Ruiz-Peralta, M.deL., Tepech-Carrillo, L., Sánchez-Cantú, M., and Moreno-Orea, J.E. (2019). Automated method for the determination of the band gap energy of pure and mixed powder samples using diffuse reflectance spectroscopy. Heliyon *5*, 1–19.

31. Mohajernia, S., Andryskova, P., Zoppellaro, G., Hejazi, S., Kment, S., Zboril, R., Schmidt, J., and Schmuki, P. (2020). Influence of $Ti^{3+}$ defect-type on heterogeneous photocatalytic $H_2$ evolution activity of $TiO_2$. J. Mater. Chem. A *8*, 1432–1442.

32. Makuła, P., Pacia, M., and Macyk, W. (2018). How to correctly determine the band gap energy of modified semiconductor photocatalysts based on UV-Vis spectra. J. Phys. Chem. Lett. *9*, 6814–6817.

33. Urbach, T. (1953). The long-wavelength edge of photographic sensitivity and of the electronic absorption of solids. Phys. Rev. *92*, 1324.

34. Rai, R.C. (2013). Analysis of the Urbach tails in absorption spectra of undoped ZnO thin films. J. Appl. Phys. *113*, 153508.

35. John, S., Soukoulis, C., Cohen, M.H., and Economou, E.N. (1986). Theory of electron band tails and the Urbach optical-absorption edge. Phys. Rev. Lett. *57*, 1777–1780.

36. Akshay, V.R., Arun, B., Mandal, G., and Vasundhara, M. (2019). Visible range optical absorption, Urbach energy estimation and paramagnetic response in Cr-doped $TiO_2$ nanocrystals derived by a sol–gel method. Phys. Chem. Chem. Phys. *21*, 12991–13004.

37. H. Tang, F. Levy, H. Berger, P.E.S. (1995). Urbach tail of anatase $TiO_2$. Phys. Rev. B *52*, 7771–7774.

38. Mohajernia, S., Andryskova, P., Zoppellaro, G., Hejazi, S., Kment, S., Zboril, R., Schmidt, J., and Schmuki, P. (2020). Influence of $Ti^{3+}$ defect-type on heterogeneous photocatalytic $H_2$ evolution activity of $TiO_2$. Journal of Materials Chemistry A *8*, 1432–1442.

39. Mascaretti, L., Russo, V., Zoppellaro, G., Lucotti, A., Casari, C.S., Kment, Š., Naldoni, A., and Li Bassi, A. (2019). Excitation wavelength- and medium-dependent photoluminescence of reduced nanostructured $TiO_2$ films. J. Phys. Chem. C *123*, 11292–11303.

40. Liqiang, J., Yichun, Q., Baiqi, W., Shudan, L., Baojiang, J., Libin, Y., Wei, F., Honggang, F., and Jiazhong, S. (2006). Review of photoluminescence performance of nano-sized semiconductor materials and its relationships with photocatalytic activity. Sol. Energy Mater Sol. Cells *90*, 1773–1787.

41. Gerosa, M., Bottani, C.E., Caramella, L., Onida, G., Di Valentin, C., and Pacchioni, G. (2015). Defect calculations in semiconductors through a dielectric-dependent hybrid DFT functional: The case of oxygen vacancies in metal oxides. J. Chem. Phys. *143*.

42. Shi, J., Chen, J., Feng, Z., Chen, T., Lian, Y., Wang, X., and Li, C. (2007). Photoluminescence characteristics of $TiO_2$ and their relationship to the photoassisted reaction of water/methanol mixture. J. Phys. Chem. C *111*, 693–699.

43. Mercado, C.C., Knorr, F.J., McHale, J.L., Usmani, S.M., Ichimura, A.S., and Saraf, L. V. (2012). Location of hole and electron traps on nanocrystalline anatase $TiO_2$. J. Phys. Chem. C *116*, 10796–10804.

44. Vequizo, J.J.M., Kamimura, S., Ohno, T., and Yamakata, A. (2018). Oxygen induced enhancement of NIR emission in brookite $TiO_2$ powders: Comparison with rutile and anatase $TiO_2$ powders. Phys. Chem. Chem. Phys. *20*, 3241–3248.

45. Pallotti, D.K., Passoni, L., Maddalena, P., Di Fonzo, F., and Lettieri, S. (2017). Photoluminescence mechanisms in anatase and rutile $TiO_2$. J. Phys. Chem. C *121*, 9011–9021.

46. Mercado, C., Seeley, Z., Bandyopadhyay, A., Bose, S., and McHale, J.L. (2011). Photoluminescence of dense nanocrystalline titanium dioxide thin films: Effect of doping and thickness and relation to gas sensing. ACS Appl. Mater. Interfaces *3*, 2281–2288.

47. Knorr, F.J., and McHale, J.L. (2013). Spectroelectrochemical photoluminescence of trap states of nanocrystalline $TiO_2$ in aqueous media. J. Phys. Chem. C *117*, 13654–13662.

48. Knorr, F.J., Mercado, C.C., and McHale, J.L. (2008). Trap-state distributions and carrier transport in pure and mixed-phase $TiO_2$: Influence of contacting solvent and interphasial electron transfer. J. Phys. Chem. C *112*, 12786–12794.

49. Mercado, C.C., McHale, J.L. Defect photoluminescence of $TiO_2$ nanotubes. (2010). MRS Online Proceedings Library *1268*, 310.

50. Vequizo, J.J.M., Matsunaga, H., Ishiku, T., Kamimura, S., Ohno, T., and Yamakata, A. (2017). Trapping-induced enhancement of photocatalytic activity on brookite $TiO_2$ powders: Comparison with anatase and rutile $TiO_2$ powders. ACS Catal. *7*, 2644–2651.

51. Yan, Y., Han, M., Konkin, A., Koppe, T., Wang, D., Andreu, T., Chen, G., Vetter, U., Morante, J.R., and Schaaf, P. (2014). Slightly hydrogenated $TiO_2$ with enhanced photocatalytic performance. J. Mater. Chem. A *2*, 12708–12716.

52. Wu, Z., Cao, S., Zhang, C., and Piao, L. (2017). Effects of bulk and surface defects on the photocatalytic performance of size-controlled $TiO_2$ nanoparticles. Nanotechnology *28*.

53. Alsalka, Y., Hakki, A., Schneider, J., and Bahnemann, D.W. (2018). Co-catalyst-free photocatalytic hydrogen evolution on $TiO_2$: Synthesis of optimized photocatalyst through statistical material science. Appl. Catal. B *238*, 422–433.

54. Mohajernia, S., Hejazi, S., Mazare, A., Nguyen, N.T., and Schmuki, P. (2017). Photoelectrochemical $H_2$ generation from suboxide $TiO_2$ nanotubes: visible-light absorption versus conductivity. Chem. Eur. J. *23*, 12406–12411.

55. Setviń, M., Aschauer, U., Scheiber, P., Li, Y.F., Hou, W., Schmid, M., Selloni, A., and Diebold, U. (2013). Reaction of $O_2$ with subsurface oxygen vacancies on $TiO_2$ anatase (101). Science *341*, 988–991.